\documentclass[aps,prx,reprint,superscriptaddress,footinbib]{revtex4-2}

\DeclareUnicodeCharacter{2212}{-}

\usepackage[utf8]{inputenc}
\usepackage{csquotes}
\usepackage{xcolor}
\usepackage{graphicx} 
\usepackage{epsfig}
\usepackage{epstopdf}
\usepackage{braket}
\usepackage[normalem]{ulem}
\usepackage[colorlinks]{hyperref}
\usepackage{bm}
\usepackage{siunitx}
\DeclareSIUnit\gauss{G}
\DeclareSIUnit\rad{rad}
\usepackage{amsmath}
\usepackage{bbold}
\usepackage{orcidlink}

\def\-{\raisebox{.75pt}{-}}

\usepackage{ulem}

\usepackage{placeins}

\begin{document}

\title{Experimental distributed quantum sensing in a noisy environment}

\author{J.~Bate\,\orcidlink{0000-0002-3570-5102}}
\affiliation{Universit\"at Innsbruck, Institut f\"ur Experimentalphysik, Technikerstr. 25, 6020 Innsbruck, Austria}
\author{A.~Hamann\,\orcidlink{https://orcid.org/0000-0002-9016-3641}}
\affiliation{Universit\"at Innsbruck, Institut f\"ur Theoretische Physik, Technikerstr. 21a, 6020 Innsbruck, Austria}
\author{M.~Canteri\,\orcidlink{0000-0001-9726-2434}}
\affiliation{Universit\"at Innsbruck, Institut f\"ur Experimentalphysik, Technikerstr. 25, 6020 Innsbruck, Austria}
\author{A.~Winkler\,\orcidlink{0000-0002-9459-027X}}
\affiliation{Universit\"at Innsbruck, Institut f\"ur Experimentalphysik, Technikerstr. 25, 6020 Innsbruck, Austria}
\author{Z.~X.~Koong\,\orcidlink{0000-0001-8066-754X}}
\affiliation{Universit\"at Innsbruck, Institut f\"ur Experimentalphysik, Technikerstr. 25, 6020 Innsbruck, Austria}
\author{V.~Krutyanskiy\,\orcidlink{0000-0003-0620-4648}}
\affiliation{Universit\"at Innsbruck, Institut f\"ur Experimentalphysik, Technikerstr. 25, 6020 Innsbruck, Austria}
\author{W.~Dür\,\orcidlink{0000-0002-0234-7425}}
\affiliation{Universit\"at Innsbruck, Institut f\"ur Theoretische Physik, Technikerstr. 21a, 6020 Innsbruck, Austria}
\author{B.~P.~Lanyon\,\orcidlink{0000-0002-7379-4572}}
\affiliation{Universit\"at Innsbruck, Institut f\"ur Experimentalphysik, Technikerstr. 25, 6020 Innsbruck, Austria}
\email[Correspondence should be sent to ]{ben.lanyon@uibk.ac.at}

\date{\today}

\begin{abstract}

The precision advantages offered by harnessing the quantum states of sensors can be readily compromised by noise.
However, when the noise has a different spatial function than the signal of interest, recent theoretical work shows how the advantage can be maintained and even significantly improved. 
In this work we experimentally demonstrate the associated sensing protocol, using trapped-ion sensors.  
An entangled state of multi-dimensional sensors is created that  isolates and optimally detects a signal, whilst being insensitive to otherwise overwhelming noise fields with different spatial profiles over the sensor locations. The quantum protocol is found to outperform a perfect implementation of the best comparable strategy without sensor entanglement. 
While our demonstration is carried out for magnetic and electromagnetic fields over a few microns, the technique is readily applicable over arbitrary distances and for arbitrary fields, thus present a promising application for emerging quantum sensor networks.

\end{abstract}

\maketitle

Measuring physical quantities with ever higher precision lies at the heart of the natural sciences and is the key
to many technological applications.
Quantum sensors can achieve Heisenberg-limited scaling, where the standard deviation of the parameter estimate scales as $1/N$, while classical sensors are limited by the standard quantum limit, with standard deviation scaling as $1/\sqrt{N}$, where $N$ is the number of sensors used~\cite{PhysRevD.23.1693,PhysRevLett.72.3439, PhysRevA.47.5138, PhysRevA.54.R4649, giovannettiAdvancesQuantumMetrology2011, RevModPhys.90.035005, DEMKOWICZDOBRZANSKI2015345}.
In scenarios in which a local field strength or frequency should be measured, there are many demonstrations of quantum enhanced precision e.g., for electric and magnetic fields~\cite{Roos2006, Gilmore2021, Taylor2008}, gravitational waves~\cite{Aasi2013}, and momentum changes~\cite{ref1, Hempel2013, Wolf2019}.
In many cases, a spatially-distributed signal is of interest, such as a difference in field strength or frequency. 
In such \emph{distributed sensing} scenarios, quantum sensor networks could also offer Heisenberg-limited scaling~\cite{proctorMultiparameterEstimationNetworked2018,eldredgeOptimalSecureMeasurement2018,qianHeisenbergscalingMeasurementProtocol2019,shettellGraphStatesResource2020,rubioQuantumSensingNetworks2020,bringewattProtocolsEstimatingMultiple2021}, as well as unconditional privacy~\cite{shettellPrivateNetworkParameter2022,bugalhoPrivateRobustStates2024,hassaniPrivacyNetworksQuantum2024}, and the first experimental investigations have been done~\cite{Liu2021, Kim2024, PhysRevX.11.031009, nicholElementaryQuantumNetwork2022, hainzer_correlation_2024, maliaDistributedQuantumSensing2022}. 
However, in general, noise jeopardizes 
accessing the scaling advantage of both local and distributed quantum sensing as the targeted precision is increased~\cite{demkowicz-dobrzanskiAdaptiveQuantumMetrology2017a,zhouAchievingHeisenbergLimit2018a, PhysRevLett.106.130506, Omran2019, Bao2024}. 

\begin{figure}[h!]
	\begin{center}
       \includegraphics[width=1\columnwidth]{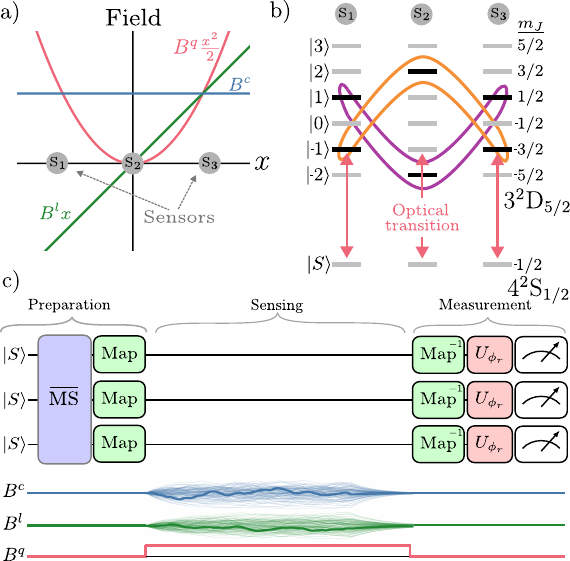}
		\caption{
   		\label{fig:concept} Experimental demonstration with trapped-ion sensors. \textbf{(a)} Three sensors in a line are exposed to spatially-constant ($B^c$), linear ($B^lx$) and quadratic ($B^qx^2/2$) fields. \textbf{(b)} Encoding of sensor state 
       $\ket{\psi_{(1,-2,1)}^{\mathrm{SWD}}} = (\ket{1,-2,1}+\ket{-1,2,-1})/\sqrt{2}$ into superpositions of the $3^2\textrm{D}_{5/2}$ manifolds of three $^{40}$Ca$^+$ ions. 
        \textbf{(c)} Protocol. $\overline{\mathrm{MS}}$ is equivalent to an entangling M{\o}lmer-S{\o}rensen gate on the optical transition. `Map' moves electron population to the $3^2\textrm{D}_{5/2}$ manifold. After state preparation, fluctuating noise ($B^c$ and $B^l$) and signal ($B^q$) fields are turned on. $U_{\phi_r}$ is described in the text.}
		\vspace{-5 mm}
	\end{center}
\end{figure}

Quantum error correction offers a powerful future approach to overcome noise in quantum sensing~\cite{kesslerQuantumErrorCorrection2014,arradIncreasingSensingResolution2014,demkowicz-dobrzanskiAdaptiveQuantumMetrology2017a, sekatskiQuantumMetrologyFull2017,zhouAchievingHeisenbergLimit2018a,faistTimeEnergyUncertaintyRelation2023}.
However, 
when 
noise and signal have a different spatial dependence, a simpler approach has been proposed ~\cite{sekatskiOptimalDistributedSensing2020} based on decoherence-free subspaces (DFSs)~\cite{helstrom1969quantum,PhysRevLett.79.3306} that is equally as powerful~\cite{hamannOptimalDistributedMultiparameter2024} and far less resource intensive. Here, the choice of initial entangled state established between distributed sensors allows optimal estimations of any desired targeted signal, whilst being resilient against fluctuations in noise fields. 
Such DFSs 
lead to a quantum advantage even in the presence of strong noise, and have been 
theoretically constructed 
for general signals and noise sources \cite{sekatskiOptimalDistributedSensing2020,wolkNoisyDistributedSensing2020,hamannApproximateDecoherenceFree2022,hamannOptimalDistributedMultiparameter2024}. 
Those results show how to generalize techniques that exploit e.g., two-sensor states to avoid decoherence due to spatially-constant noise~\cite{Roos2006, landiniPhasenoiseProtectionQuantumenhanced2014, nicholElementaryQuantumNetwork2022}.

In this manuscript, we experimentally demonstrate the SWD (Sekatski-Wölk-Dür) protocol~\cite{sekatskiOptimalDistributedSensing2020}. 
As an example, we consider the task of sensing the strength of a field that varies quadratically in space, whilst being insensitive to fluctuations in the strength of fields with both constant and gradient profiles. 
The sensors are three atomic $^{40}$Ca$^+$ ions, \SI{4.9}{\micro\meter} apart.
The performance of our implementation of the optimal entangled sensor state is compared with the best comparable scheme without entanglement. 

We consider the scenario in which
three sensors are positioned equidistantly along one dimension. 
The total scalar field strength at position $x$ is given by $B(x) = B^c + B^l x + B^q x^2/2$ (Figure~\ref{fig:concept}a), where $B^{j}$ are the coefficients of three linearly-independent field components, corresponding to the first three orders of a Taylor expansion. 
Each of the $L_i$ eigenstates (or `levels') of sensor $i$ ($\ket{s^i_k}$) experience an energy shift proportional to the total field strength at the sensor position $\big(B(x_i)\big)$. 
Specifically, the Hamiltonian of sensor $i$ is given by $H(x_i) = \hbar\kappa\sum_{k = 1}^{L_i}B(x_i)s_k^i\ket{s_k^i}\bra{s_k^i}$, where $\kappa$ is a constant determining the sensor-field interaction strength, and each $s_k^i$ is a real number proportional to the sensitivity of the $k$-th eigenstate. 
The strength of each field component at the three sensor positions $x = (-1, 0, 1)$ is described by the component-vectors $\bm{f}^c=B^c(1,1,1), \bm{f}^l=B^l(-1,0,1)$ and $\bm{f}^q=B^q(1,0,1)$, where the $i$th element is the field strength of the corresponding field component at sensor $i$. 

The SWD protocol~\cite{sekatskiOptimalDistributedSensing2020} requires preparing sensors into entangled states of the form $\ket{\psi_{\bm{s}}^{\mathrm{SWD}}}=(\ket{s_1,s_2,s_3}+\ket{-s_1,-s_2,-s_3})/\sqrt{2}$, where $\bm{s}=(s_1,s_2,s_3)$. Each $s_i$ is constrained such that $\ket{s_i}$ is an eigenstate of the sensor Hamiltonian $H(x_i)$.
Exposure to any three field components, described by component-vectors $\{\bm{f}^1, \bm{f}^2, \bm{f}^3\}$, for a time $t$ causes the state to develop a relative phase $e^{-2i\kappa t \sum_{j = 1}^3 \bm{s}\cdot\bm{f}^j}$. 
The SWD protocol describes how to choose $\bm{s}$ such that the phase evolves due to any one of the field components ($\bm{f}^j$) at the maximal rate ($\max_{\bm{s}}|\bm{s}\cdot\bm{f}^j|$)~\cite{SuppMat}, while the other field components generate no phase ($\bm{s}\cdot\bm{f}^{j'\neq j} = 0$). 
For example, up to a scaling factor in $\bm{s}$, 
the optimal state to sense $B^q$ while being insensitive to $B^c$ and $B^l$
is $\ket{\psi_{(1, -2, 1)}^{\mathrm{SWD}}}$. 
That state exists in a DFS with respect to constant and gradient fields, meaning that it is insensitive to fluctuations in the field strengths $B^c$ and $B^l$. $\ket{\psi_{(1, -2, 1)}^{\mathrm{SWD}}}$ is a GHZ-type entangled state~\cite{Greenberger_1989} of two-level sensors, but with differing sensitivities to the surrounding fields.

We encode sensor states into the metastable $3^2\textrm{D}_{5/2}$ manifold of each of three $^{40}$Ca$^+$ ions, following  $\ket{s_i}{=}\ket{3^2\textrm{D}_{5/2, m_J = (s_i-1/2)}}$ (Figure \ref{fig:concept}b). 
The sensor states couple to magnetic fields via the linear Zeeman effect: the frequency shift of sensor state $\ket{s_i}$  is given by $\kappa B(x_i)(s_{i}+1/2)/(2\pi)$, where $\kappa = g_{5/2}\mu_b/\hbar=(2\pi)\SI{0.0168}{\hertz\per\pico\tesla}$. 
A spatially-constant principle magnetic field of $\SI{4.14686(1)e-4}{\tesla}$ is applied at all times, yielding a level splitting of $2\pi \times \SI{6.96486(2)}{\mega\hertz}$.
Each experimental shot of the SWD protocol (and its classical counterpart) consists of three steps: state preparation, sensing and measurement (Figure~\ref{fig:concept}c). 
State preparation begins with Doppler cooling, sideband cooling and optical pumping that prepares each ion into the state $\ket{S}=|4^2\text{S}_{1/2}, m_j = -1/2\rangle$ and the ground state of the axial center of mass motional mode. 
Next, a M{\o}lmer-S{\o}rensen logic gate (MS-gate)~\cite{PhysRevLett.82.1835} followed by a $\pi/2$ pulse, both driven by a \SI{729}{\nano\meter} laser on the $\ket{S}\leftrightarrow\ket{-1}$ transition, generates the state $\ket{G}=\left(\ket{S,S,S}+\ket{-1,-1,-1}\right)/\sqrt{2}$. 
A final sequence of laser pulses map $\ket{G}$ to $\ket{\psi_{(1, -2, 1)}^{\mathrm{SWD}}}$~\cite{SuppMat}.

During sensing, the ions are exposed to fluctuating magnetic fields generated by two pairs of current-carrying coils mounted outside the vacuum chamber. Each pair is driven by a separate noisy voltage generator that produces currents fluctuating at acoustic frequencies. One pair---referred to as the constant coils (CC)---is operated in Helmholtz-like configuration: the currents flow in the same direction and generate a spatially-constant field $B^c$ varying in the range $\pm\SI{1.02(5)}{\mu\tesla}$ at the positions of the ions. 
The other coil pair---referred to as the gradient coils (GC)---is in anti-Helmholtz-like configuration with counter propagating currents. The GC coil is used to generate a spatial-gradient field $B^l$ which varies in the range $\pm\SI{0.905(3)}{\milli\tesla\per\meter}$: corresponding to field differences in the range $\pm\SI{8.88(2)}{\nano\tesla}$ between the outer ions. 
The GC pair also creates a spatially-constant field which does not affect interpretations of the results. The scalar fields of the SWD protocol correspond to the projections of the vector fields, generated by the coil pairs, onto the principle magnetic field axis. The fluctuating fields are switched off during 
state preparation and measurement, which corrupt those steps. 
To avoid this, one could use logic gates that incorporate dynamical decoupling~\cite{PhysRevLett.110.263002}. 


A laser imprints an effective spatially-quadratic field across the ion sensors, shifting the frequency of the $\ket{-2}$ state of all ions in proportion to the laser intensity via the AC-Stark effect.
The laser's effect on the Hilbert space, spanned by the bold states in Figure \ref{fig:concept}b, is equivalent to a spatially-quadratic field across the ions: shifting  the relevant energy gap of the middle sensor but not those of the outer sensors. 
The following calibrated quadratic field strengths are applied to the sensors: $B^q_{\mathrm{cal}} = [ 0.0, 2.1, 4.7, 7.6, 9.5, 11.9, 15.2]~\SI{}{\pico\tesla\per\square\micro\meter}$~\cite{SuppMat}, corresponding to laser-induced frequency shifts of $\ket{-2}$ by up to $\SI{12.3}{\hertz}$ (an  effective magnetic field change of the central ion by \SI{365}{\pico\tesla} with respect to the outer ions). 

The measurement step begins with \SI{729}{\nano\meter} laser pulses that map the sensor's states from the $3^2\textrm{D}_{5/2}$ manifold to qubit superpositions of $\ket{S}$ and $\ket{-1}$ for each ion. Next, a $\pi/2$-pulse realizes the global rotation $U_{\phi_r} = e^{-i(\pi/4)(\sigma_X\cos{\phi_r} + \sigma_Y\sin{\phi_r})}$, where $\phi_r$ is the angle of rotation, and $\sigma_J = \sum_{l = 1}^3{\sigma_j^l}$, with $\sigma_j^l$ being the Pauli $j$ operator acting on ion $l$. 
Finally, the state of each ion-qubit is measured via the standard electron shelving method on all ions simultaneously.  
One experimental shot is now complete.
Each experimental shot is repeated $n$ times, and from those $n$ shots an estimate $P_{\mathrm{est}}$ for the parity 
$P=\langle U_{\phi_r}^{\dag}\sigma^1_z\sigma^2_z\sigma^3_z U_{\phi_r}\rangle$ is calculated. 
For the initial state $\ket{\psi_{(1, -2, 1)}^{\mathrm{SWD}}}$, the parity has the form 
\begin{equation}
\label{equation:parity}
P(B^q, \phi_r)=A\cos(\omega B^q + 3 \phi_r + \phi_0),
\end{equation}
where $A=1$, $\omega = 2\kappa td^2$, $d = \SI{4.9}{\micro\meter}$ is the distance between neighboring ions and $\phi_0$ is a phase offset discussed later. An estimate for the quadratic field strength $B_{\mathrm{est}}^q$ is given by $B_{\mathrm{est}}^q=\hat{B^q}(P_{\mathrm{est}})$, where the estimator function $\hat{B^q}$ is the inverse function of $P$, corresponding to the maximum likelihood estimator~\cite{SuppMat}.
The combination of the initial state $\ket{\psi_{\bm{s}}^{\mathrm{SWD}}}$, the measurement $P$ and the estimator $\hat{B}^q$ constitute our implementation of the SWD protocol.  
To quantify the accuracy of the estimates we use the normalized root mean squared error defined as $\mathrm{\overline{RMSE}} = \sqrt{\frac{n}{m}\sum_{i=1}^M (B^q_{\mathrm{cal}}-B^q_{\mathrm{est},i})^2}$, where $m$ is the number of independent estimates, each containing $n$ shots. 
The $\mathrm{\overline{RMSE}}$ combines both systematic and statistical errors, and the factor $\sqrt{n}$ cancels the expected shot-noise scaling.
Using the initial state and measurement of the SWD protocol it is possible to achieve the minimum $\mathrm{\overline{RMSE}}$ over all states and measurements in the $3^2 D_{5/2}$ manifold that are insensitive to the noise fields, as set by the Heisenberg limit~\cite{SuppMat}. Achieving that minimum value requires the optimal estimator, which, in the limit of infinite shots, is the maximum likelihood estimator.

Our separable protocol is achieved by removing the MS gate, which results in the preparation of the separable state $\ket{\psi_{\bm{s}}^{\mathrm{sep}}} = \bigotimes_i (\ket{s_i}+\ket{-s_i})/ \sqrt{2}$, with $\bm{s} = (1, -2, 1)$.
When exposed to overwhelming fluctuations in $B^c$ and $B^l$ during sensing --- which eliminate all off-diagonal coherence terms that are sensitive to them --- 
the separable state $\ket{\psi_{(1, -2, 1)}^{\mathrm{sep}}}$ is partially projected onto the DFS spanned by $\{\ket{1, -2, 1}, \ket{-1, 2, -1}\}$. 
The expected parity has the same form as Equation \ref{equation:parity} with a reduced amplitude of $A = 0.25$.
The initial state and measurement of the separable protocol can achieve the minimum $\mathrm{\overline{RMSE}}$ over all noise-protected separable states restricted to occupying two levels of the $3^2 D_{5/2}$ manifold and projective measurements that don't utilize entanglement~\cite{SuppMat}. 
The separable protocol corresponds to the established technique of correlation spectroscopy~\cite{hainzer_correlation_2024, HumeLifetimeLimited, nicholElementaryQuantumNetwork2022}.
For characterization purposes, state tomography is used to reconstruct the generated states~\cite{SuppMat}. The reconstructed density matrices $\rho^{\textrm{SWD}}$ and $\rho^{\mathrm{sep}}$ are ideally given by $\ket{\psi_{(1, -2, 1)}^{\mathrm{SWD}}}\bra{\psi_{(1, -2, 1)}^{\mathrm{SWD}}}$ and $\ket{\psi_{(1, -2, 1)}^{\mathrm{sep}}}\bra{\psi_{(1, -2, 1)}^{\mathrm{sep}}}$, respectively.

\begin{figure}[t]
	\begin{center}
       \includegraphics[width=1\columnwidth]{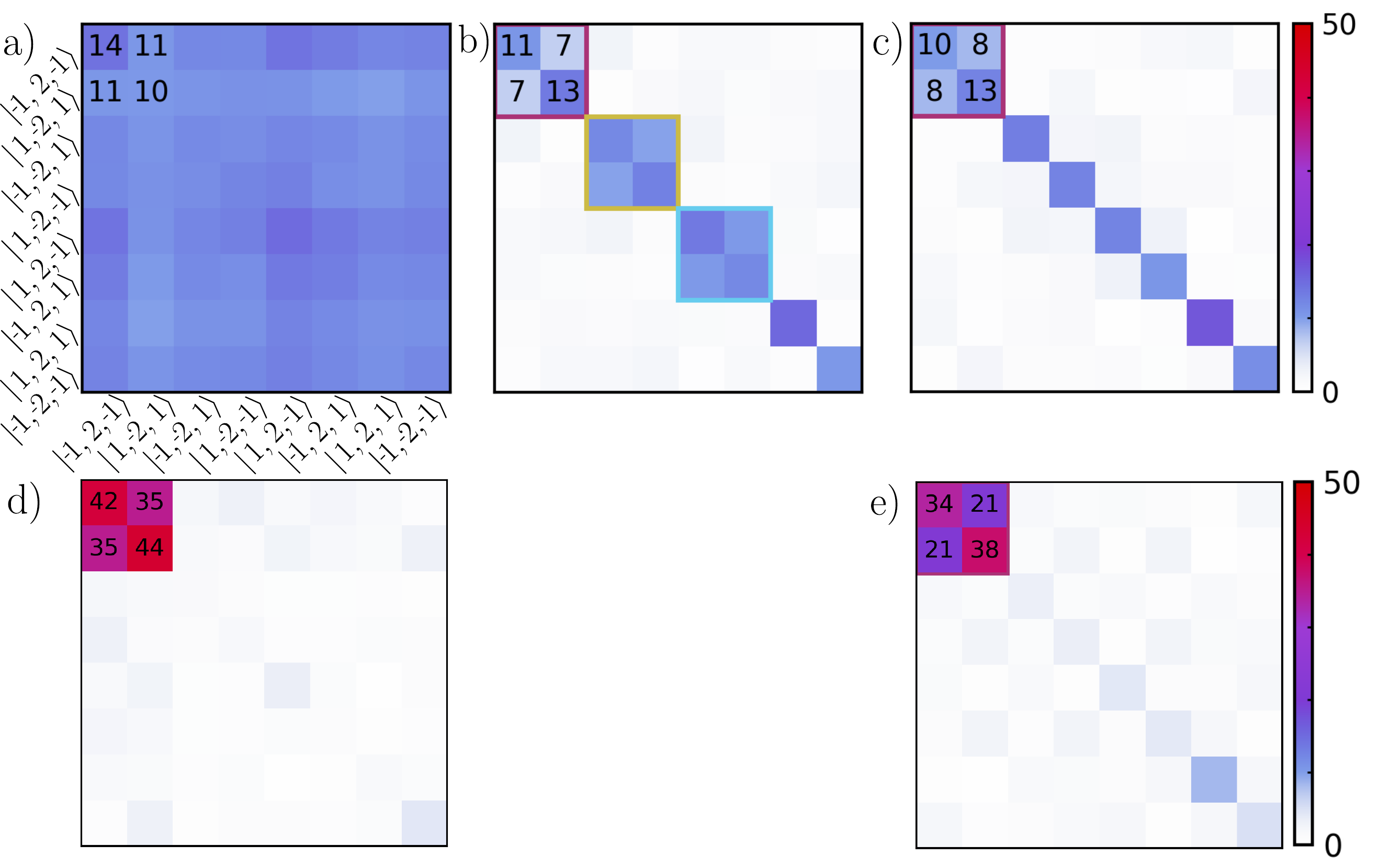}
		\caption{
    	Experimentally-reconstructed density matrices of three-sensor states. Numbers in cells are matrix elements multiplied by 100. 
      \textbf{(a)} Initial separable state $\rho^{\mathrm{sep}}$. 
       \textbf{(b)} $\rho^{\mathrm{sep}}$ after \SI{80}{\milli\second} of noise from CC coils. Red, yellow and blue boxes show DFSs with respect to spatially-constant noise. 
       \textbf{(c)} $\rho^{\mathrm{sep}}$ after \SI{80}{\milli\second} of noise from both the CC and CG coils. Red box shows DFS with respect to both spatially constant and gradient noise.
       \textbf{(d)} $\rho^{\mathrm{SWD}}$.
       \textbf{(e)} $\rho^{\mathrm{SWD}}$ after \SI{80}{\milli\second} noise from both CC and CG coils.
       }\label{fig:noise}
		\vspace{-5 mm}
	\end{center}
\end{figure}

We begin by investigating the spatial properties of the fluctuating fields produced by the two coil pairs across the ion string. 
Tomographic-reconstruction of $\rho^{\mathrm{sep}}$ is seen to populate all coherence terms between eigenstates (Figure \ref{fig:noise}a), making this a useful state to probe and verify the action of the noise fields. 
Next, we subject the state $\rho^{\mathrm{sep}}$ to \SI{80}{\milli\second} of noise generated by the CC coils. The reconstructed state shows that the initial state has been partially projected into an incoherent mixture of three two-dimensional subspaces (Figure \ref{fig:noise}b); coherences that lie outside those subspaces have essentially been eliminated, while those within them remain. The three remaining subspaces are the ones that are expected to be decoherence-free with respect to spatially-constant noise~\cite{SuppMat}, verifying that the action of the CC coils is to produce spatially-constant noise across the ion string. 
Next, we subject the state $\rho^{\mathrm{sep}}$ to \SI{80}{\milli\second} of noise generated by both the CC and GC coils.
The reconstructed state shows that the initial state has now been partially projected into a single subspace, spanned by $\{\ket{1, -2, 1}, \ket{-1, 2,-1}\}$ (Figure \ref{fig:noise}c). That subspace is the one that is expected to be decoherence-free with respect to both spatially constant and gradient noise.  
The fields generated from both coil pairs over \SI{80}{\milli\second} therefore cause overwhelming noise with both spatially constant and gradient functions over the ion string. 

We perform tomographic reconstruction of the SWD state before (Figure \ref{fig:noise}d) and after exposure to the \SI{80}{\milli\second}-long noise channel from both coil pairs (Figure \ref{fig:noise}e). 
The fidelities of the initial and final state with a GHZ state~\cite{SuppMat} are 0.780(14) and 0.564(12) respectively. Since both values are larger than 0.5, the presence of genuine GHZ-type multipartite entanglement is proven~\cite{Guhne2003}. 
The potential of the generated states to serve as sensors of a spatially-quadratic field is determined by the amplitude $A_{\rho}$ of the $\ket{1, -2, 1}\bra{-1, 2, -1}$ coherence term. Specifically, the amplitude of the parity signal in Equation \ref{equation:parity} is ideally given by $A=2A_{\rho}$. After exposure to our noise fields, the observed separable state achieves only $A=0.16(2)$ (Figure \ref{fig:noise}c), while the observed entangled state (Figure \ref{fig:noise}e) achieves $A=0.42(4)$.  
The ideal values for the separable and entangled states are $A=0.25$ and $A=1$, respectively. 
Imperfections in the prepared initial states are predominantly due to laser phase noise~\cite{SuppMat}. Additional imperfections in states reconstructed after \SI{80}{\milli\second} are predominantly due to spontaneous decay of the $3^2\textrm{D}_{5/2}$ manifold, which has a lifetime of $\SI{1.045}{\second}$~\cite{SuppMat}. 
\begin{figure}[t]
	\begin{center}
		\includegraphics[width=1\columnwidth]{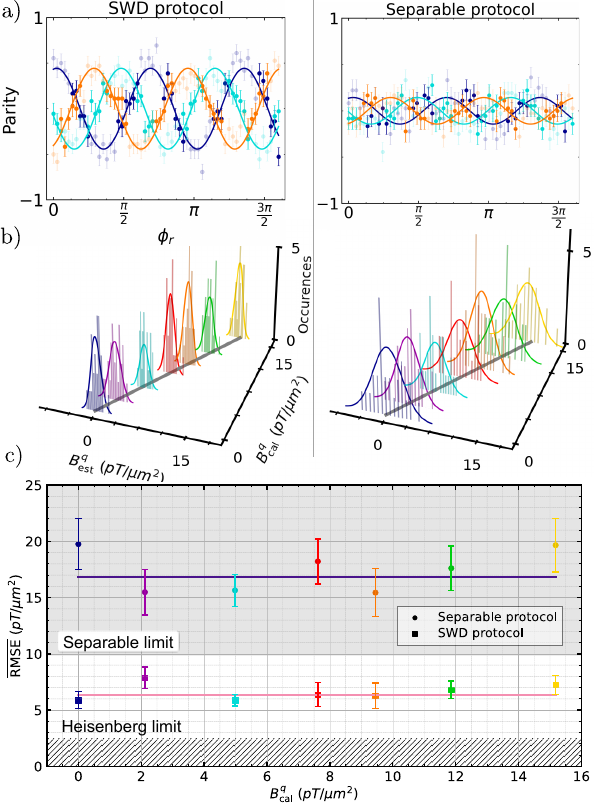}
		\caption{Sensing results. Data coloring reflects applied signal strength $B^q_{\mathrm{cal}}$. \textbf{(a)} Example parity estimates.  Colored shapes show data. Lines show fits of $P(B^q, \phi_r)$ to data. Bold markers are used for estimating $B^q_{\mathrm{est}}$ in (b). 
        \textbf{(b)} Histograms showing that quadratic field strength estimates $B^q_{\mathrm{est}}$ are centred around the calibrated values $B^q_{\mathrm{cal}}$. Grey lines show $B^q_{\mathrm{est}} = B^q_{\mathrm{cal}}$. Coloured curves show Gaussian fits. 
        \textbf{(c)} Markers show $\mathrm{\overline{RMSE}}$ from data in each histogram in (b). Solid lines show corresponding weighted averages. 
        Grey area is achievable by two-level separable estimation protocols. Striped area is beyond the Heisenberg-limit and inaccessible. Errorbars are one standard deviation.
        }
		\label{fig:results2}
		\vspace{-5mm}
	\end{center}
\end{figure}

Next we perform the full sensing protocol, aiming to sense calibrated quadratic field strengths (${B^q_{\mathrm{cal}}}$) generated by the laser, while the sensors are exposed to the noise fields generated by both coils for \SI{80}{\milli\second}.
For each applied quadratic field strength, measurements of parity are made for 60 values of the analysis phase $\phi_r$ that are equally spaced between $\{0,1.6\pi\}$\SI{}{\rad}.
The protocol is carried out both when preparing $\rho^{\textrm{SWD}}$ and $\rho^{\mathrm{sep}}$.
The obtained parity data exhibit the expected sinusoidal oscillations, with phases shifted by different applied quadratic field strengths (exemplified in Figure~\ref{fig:results2}a).
Each parity oscillation is separately fit to Equation \ref{equation:parity}. The  phase offset parameter $\phi_0$, caused by a background quadratic field~\cite{SuppMat}, is determined from the fit to the case $B_{\mathrm{cal}}^q = 0$ (Figure~\ref{fig:results2}a, blue solid line). The total phase of each oscillation is given by $\Phi(B^q) = \omega B^q+3\phi_r + \phi_0$.
$A$ is calculated from the average fitted amplitude, yielding $A = 0.45(2)$ and $A = 0.146(9)$ for the SWD and separable protocols, respectively. 
Those fitted amplitudes are statistically consistent with the ones predicted from tomography.

We turn our attention to extracting estimates of the applied quadratic field strengths $B^q_{\mathrm{est}}$ from the data, and then calculating the $\mathrm{\overline{RMSE}}$ to quantify the accuracy of those estimates. 
Parity estimates are selected around the maximum slope of the parity function, which minimizes the $\mathrm{\overline{RMSE}}$~\cite{SuppMat}. 
Specifically, parity estimates are chosen in the range $\mathrm{mod}_{\pi}\big(\Phi(B^q_{\mathrm{cal}})\big)\in\frac{\pi}{2} \pm 0.73$ (Figure~\ref{fig:results2}a, bold markers). 
Quadratic field strength estimates are computed for each selected parity estimate. The estimates are obtained using the inverse parity function $P^{-1}$ (inverse of Equation \ref{equation:parity}), using the relevant average fitted amplitude and the relevant fitted phase offset parameter ($\phi_0$). 
Histograms comparing the estimated and calibrated quadratic field strengths show that the SWD state yields consistently lower variance than the separable state (Figure~\ref{fig:results2}b). 
The mean values of $B^q_{\mathrm{est}}$ obtained from the two protocols are statistically consistent with each other, but not with those of $B^q_{\mathrm{cal}}$ (Figure~\ref{fig:results2}b, grey line). We attribute this discrepancy to a 6\% change in the intensity of the Stark shift laser between calibration and sensing stages. This systematic error is masked by the larger statistical fluctuations in the RMSE values discussed later.

Figure~\ref{fig:results2}c compares the measured $\mathrm{\overline{RMSE}}$ values --- obtained from the histograms (Figure~\ref{fig:results2}b) --- with those for ideal implementations. First, one sees that the experimental SWD protocol significantly outperforms the experimental separable protocol. The average $\mathrm{\overline{RMSE}}$
measured for the SWD protocol is a factor of $2.6(1)$ smaller than the average measured for the separable protocol, corresponding to a $\SI{4.1(2)}{\decibel}$ improvement. 
Second, the performance of the measured SWD protocol even surpasses an ideal implementation of the  separable protocol. The average $\mathrm{\overline{RMSE}}$ measured for the SWD protocol is a factor 1.49(6) lower than an ideal implementation of the separable protocol, corresponding to a \SI{1.7(2)}{\decibel} improvement. We conclude that the implemented SWD protocol outperforms the best that could possibly be achieved with a separable protocol that utilizes two levels of each sensor.  
The deviation between the measured and ideal performances for both protocols are statistically consistent with the imperfect amplitudes (A).

Finally, we present new theoretical results. 
Consider the task of sensing the highest order of an $M$ order Taylor expansion, whilst being insensitive to fluctuations in all lower orders, using the minimum number of sensors, $N_{\mathrm{min}}$. 
That task requires a minimum of $N_{\mathrm{min}}=M+1$ sensors all at different locations. We find that the SWD protocol achieves a lower $\mathrm{\overline{RMSE}}$ compared to the best two-level separable protocol by an amount that grows exponentially in $M$~\cite{SuppMat}. 
Moreover, we prove that this exponential advantage extends to arbitrary spatial field distributions~\cite{SuppMat}. 

We experimentally demonstrated a protocol which exploits entanglement to optimally sense a spatially-distributed signal, whilst being immune to noise with different spatial dependencies.
Exploiting that protocol in future requires positioning sensors at more locations and the ability to establish multipartite entanglement between them. 
Over distances of hundreds of microns these capabilities are currently under development in various experimental platforms (e.g.,~\cite{PhysRevX.13.041052, Bruzewicz2019, Bluvstein2022, superscompsreview}), and over distances of meters to tens of kilometers there has been significant success establishing entanglement between many of those platforms using traveling photons~\cite{vanLeent2022, Pompili2021, Moehring2007, PhysRevLett.124.110501, Ritter2012, Delteil2016, PhysRevLett.119.010503, PhysRevLett.125.260502, Liu2024, Knaut2024}, including an ion in the trap used in the present work and another \SI{230}{\meter} away~\cite{PhysRevLett.130.050803}.
Converting photons to telecom wavelengths~\cite{PRXQuantum.5.020308} and using quantum repeaters~\cite{PhysRevLett.130.213601} could establish entanglement between ions separated by hundreds of kilometers or more, allowing for the SWD protocol to detect large-scale spatial field variations.
Challenges include significantly increasing the rates at which remote ion-qubits can be entangled beyond the state of the art \cite{PhysRevLett.124.110501} and developing robust ion-qubit memories~\cite{PhysRevLett.130.090803}. 
Over global scales, we speculate that the protocol could be used to investigate gravitational waves emitted by specific sources, whilst hiding from others. 

The data available in Ref.~\cite{zenodo} includes the measurement outcomes necessary to reconstruct the main results of this letter (Figure 3).

\begin{acknowledgments}

We thank Denis Vasilyev for valuable discussions.  
This work was funded in part by; the Austrian Science Fund (FWF) 
[Grant DOIs: 
10.55776/P34055, 10.55776/P36009, 10.55776/P36010 
and 10.55776/COE1]; 
the European Union under the DIGITAL-2021-QCI-01 Digital European Program under Project number No 101091642 and project name `QCI-CAT', and the European Union’s Horizon Europe research and innovation programme under grant agreement No. 101102140 and project name ‘QIA-Phase 1'; the Österreichische Nationalstiftung für Forschung, Technologie und Entwicklung (AQUnet project).  
We acknowledge funding for B.P.L. by the CIFAR Quantum Information Science Program of Canada. The opinions expressed in this document reflect only the author’s view and reflects in no way the European Commission’s opinions. The European Commission is not responsible for any use that may be made of the information it contains. For open access purposes, the author has applied a CC BY public copyright license to any author-accepted manuscript version arising from this submission. 
J. B. took data. J. B., A. H., and B. P. L. analyzed and interpreted data. J. B. and A. H. performed theoretical modeling. J. B., M. C., A. W., Z. K., and V. K. contributed to the experimental setup. A. H. derived theoretical results, with supervision from W. D. The manuscript was written by J. B., A. H. and B. P. L., with all authors providing detailed comments. The project was conceived and supervised by B. P. L with support from W. D. 

\end{acknowledgments}

\begin{thebibliography}{83}%
\makeatletter
\providecommand \@ifxundefined [1]{%
 \@ifx{#1\undefined}
}%
\providecommand \@ifnum [1]{%
 \ifnum #1\expandafter \@firstoftwo
 \else \expandafter \@secondoftwo
 \fi
}%
\providecommand \@ifx [1]{%
 \ifx #1\expandafter \@firstoftwo
 \else \expandafter \@secondoftwo
 \fi
}%
\providecommand \natexlab [1]{#1}%
\providecommand \enquote  [1]{``#1''}%
\providecommand \bibnamefont  [1]{#1}%
\providecommand \bibfnamefont [1]{#1}%
\providecommand \citenamefont [1]{#1}%
\providecommand \href@noop [0]{\@secondoftwo}%
\providecommand \href [0]{\begingroup \@sanitize@url \@href}%
\providecommand \@href[1]{\@@startlink{#1}\@@href}%
\providecommand \@@href[1]{\endgroup#1\@@endlink}%
\providecommand \@sanitize@url [0]{\catcode `\\12\catcode `\$12\catcode `\&12\catcode `\#12\catcode `\^12\catcode `\_12\catcode `\%12\relax}%
\providecommand \@@startlink[1]{}%
\providecommand \@@endlink[0]{}%
\providecommand \url  [0]{\begingroup\@sanitize@url \@url }%
\providecommand \@url [1]{\endgroup\@href {#1}{\urlprefix }}%
\providecommand \urlprefix  [0]{URL }%
\providecommand \Eprint [0]{\href }%
\providecommand \doibase [0]{https://doi.org/}%
\providecommand \selectlanguage [0]{\@gobble}%
\providecommand \bibinfo  [0]{\@secondoftwo}%
\providecommand \bibfield  [0]{\@secondoftwo}%
\providecommand \translation [1]{[#1]}%
\providecommand \BibitemOpen [0]{}%
\providecommand \bibitemStop [0]{}%
\providecommand \bibitemNoStop [0]{.\EOS\space}%
\providecommand \EOS [0]{\spacefactor3000\relax}%
\providecommand \BibitemShut  [1]{\csname bibitem#1\endcsname}%
\let\auto@bib@innerbib\@empty
\bibitem [{\citenamefont {Caves}(1981)}]{PhysRevD.23.1693}%
  \BibitemOpen
  \bibfield  {author} {\bibinfo {author} {\bibfnamefont {C.~M.}\ \bibnamefont {Caves}},\ }\bibfield  {title} {\bibinfo {title} {Quantum-mechanical noise in an interferometer},\ }\href {https://doi.org/10.1103/PhysRevD.23.1693} {\bibfield  {journal} {\bibinfo  {journal} {Phys. Rev. D}\ }\textbf {\bibinfo {volume} {23}},\ \bibinfo {pages} {1693} (\bibinfo {year} {1981})}\BibitemShut {NoStop}%
\bibitem [{\citenamefont {Braunstein}\ and\ \citenamefont {Caves}(1994)}]{PhysRevLett.72.3439}%
  \BibitemOpen
  \bibfield  {author} {\bibinfo {author} {\bibfnamefont {S.~L.}\ \bibnamefont {Braunstein}}\ and\ \bibinfo {author} {\bibfnamefont {C.~M.}\ \bibnamefont {Caves}},\ }\bibfield  {title} {\bibinfo {title} {Statistical distance and the geometry of quantum states},\ }\href {https://doi.org/10.1103/PhysRevLett.72.3439} {\bibfield  {journal} {\bibinfo  {journal} {Phys. Rev. Lett.}\ }\textbf {\bibinfo {volume} {72}},\ \bibinfo {pages} {3439} (\bibinfo {year} {1994})}\BibitemShut {NoStop}%
\bibitem [{\citenamefont {Kitagawa}\ and\ \citenamefont {Ueda}(1993)}]{PhysRevA.47.5138}%
  \BibitemOpen
  \bibfield  {author} {\bibinfo {author} {\bibfnamefont {M.}~\bibnamefont {Kitagawa}}\ and\ \bibinfo {author} {\bibfnamefont {M.}~\bibnamefont {Ueda}},\ }\bibfield  {title} {\bibinfo {title} {Squeezed spin states},\ }\href {https://doi.org/10.1103/PhysRevA.47.5138} {\bibfield  {journal} {\bibinfo  {journal} {Phys. Rev. A}\ }\textbf {\bibinfo {volume} {47}},\ \bibinfo {pages} {5138} (\bibinfo {year} {1993})}\BibitemShut {NoStop}%
\bibitem [{\citenamefont {Bollinger}\ \emph {et~al.}(1996)\citenamefont {Bollinger}, \citenamefont {Itano}, \citenamefont {Wineland},\ and\ \citenamefont {Heinzen}}]{PhysRevA.54.R4649}%
  \BibitemOpen
  \bibfield  {author} {\bibinfo {author} {\bibfnamefont {J.~J.~.}\ \bibnamefont {Bollinger}}, \bibinfo {author} {\bibfnamefont {W.~M.}\ \bibnamefont {Itano}}, \bibinfo {author} {\bibfnamefont {D.~J.}\ \bibnamefont {Wineland}},\ and\ \bibinfo {author} {\bibfnamefont {D.~J.}\ \bibnamefont {Heinzen}},\ }\bibfield  {title} {\bibinfo {title} {Optimal frequency measurements with maximally correlated states},\ }\href {https://doi.org/10.1103/PhysRevA.54.R4649} {\bibfield  {journal} {\bibinfo  {journal} {Phys. Rev. A}\ }\textbf {\bibinfo {volume} {54}},\ \bibinfo {pages} {R4649} (\bibinfo {year} {1996})}\BibitemShut {NoStop}%
\bibitem [{\citenamefont {Giovannetti}\ \emph {et~al.}(2011)\citenamefont {Giovannetti}, \citenamefont {Lloyd},\ and\ \citenamefont {Maccone}}]{giovannettiAdvancesQuantumMetrology2011}%
  \BibitemOpen
  \bibfield  {author} {\bibinfo {author} {\bibfnamefont {V.}~\bibnamefont {Giovannetti}}, \bibinfo {author} {\bibfnamefont {S.}~\bibnamefont {Lloyd}},\ and\ \bibinfo {author} {\bibfnamefont {L.}~\bibnamefont {Maccone}},\ }\bibfield  {title} {\bibinfo {title} {Advances in quantum metrology},\ }\href {https://doi.org/10.1038/nphoton.2011.35} {\bibfield  {journal} {\bibinfo  {journal} {Nature Photonics}\ }\textbf {\bibinfo {volume} {5}},\ \bibinfo {pages} {222} (\bibinfo {year} {2011})}\BibitemShut {NoStop}%
\bibitem [{\citenamefont {Pezz\`e}\ \emph {et~al.}(2018)\citenamefont {Pezz\`e}, \citenamefont {Smerzi}, \citenamefont {Oberthaler}, \citenamefont {Schmied},\ and\ \citenamefont {Treutlein}}]{RevModPhys.90.035005}%
  \BibitemOpen
  \bibfield  {author} {\bibinfo {author} {\bibfnamefont {L.}~\bibnamefont {Pezz\`e}}, \bibinfo {author} {\bibfnamefont {A.}~\bibnamefont {Smerzi}}, \bibinfo {author} {\bibfnamefont {M.~K.}\ \bibnamefont {Oberthaler}}, \bibinfo {author} {\bibfnamefont {R.}~\bibnamefont {Schmied}},\ and\ \bibinfo {author} {\bibfnamefont {P.}~\bibnamefont {Treutlein}},\ }\bibfield  {title} {\bibinfo {title} {Quantum metrology with nonclassical states of atomic ensembles},\ }\href {https://doi.org/10.1103/RevModPhys.90.035005} {\bibfield  {journal} {\bibinfo  {journal} {Rev. Mod. Phys.}\ }\textbf {\bibinfo {volume} {90}},\ \bibinfo {pages} {035005} (\bibinfo {year} {2018})}\BibitemShut {NoStop}%
\bibitem [{\citenamefont {Demkowicz-Dobrzański}\ \emph {et~al.}(2015)\citenamefont {Demkowicz-Dobrzański}, \citenamefont {Jarzyna},\ and\ \citenamefont {Kołodyński}}]{DEMKOWICZDOBRZANSKI2015345}%
  \BibitemOpen
  \bibfield  {author} {\bibinfo {author} {\bibfnamefont {R.}~\bibnamefont {Demkowicz-Dobrzański}}, \bibinfo {author} {\bibfnamefont {M.}~\bibnamefont {Jarzyna}},\ and\ \bibinfo {author} {\bibfnamefont {J.}~\bibnamefont {Kołodyński}},\ }\bibfield  {title} {\bibinfo {title} {Chapter four - quantum limits in optical interferometry}\ }(\bibinfo  {publisher} {Elsevier},\ \bibinfo {year} {2015})\ pp.\ \bibinfo {pages} {345--435}\BibitemShut {NoStop}%
\bibitem [{\citenamefont {Roos}\ \emph {et~al.}(2006)\citenamefont {Roos}, \citenamefont {Chwalla}, \citenamefont {Kim}, \citenamefont {Riebe},\ and\ \citenamefont {Blatt}}]{Roos2006}%
  \BibitemOpen
  \bibfield  {author} {\bibinfo {author} {\bibfnamefont {C.~F.}\ \bibnamefont {Roos}}, \bibinfo {author} {\bibfnamefont {M.}~\bibnamefont {Chwalla}}, \bibinfo {author} {\bibfnamefont {K.}~\bibnamefont {Kim}}, \bibinfo {author} {\bibfnamefont {M.}~\bibnamefont {Riebe}},\ and\ \bibinfo {author} {\bibfnamefont {R.}~\bibnamefont {Blatt}},\ }\bibfield  {title} {\bibinfo {title} {`designer atoms' for quantum metrology},\ }\href {https://doi.org/10.1038/nature05101} {\bibfield  {journal} {\bibinfo  {journal} {Nature}\ }\textbf {\bibinfo {volume} {443}},\ \bibinfo {pages} {316} (\bibinfo {year} {2006})}\BibitemShut {NoStop}%
\bibitem [{\citenamefont {Gilmore}\ \emph {et~al.}(2021)\citenamefont {Gilmore}, \citenamefont {Affolter}, \citenamefont {Lewis-Swan}, \citenamefont {Barberena}, \citenamefont {Jordan}, \citenamefont {Rey},\ and\ \citenamefont {Bollinger}}]{Gilmore2021}%
  \BibitemOpen
  \bibfield  {author} {\bibinfo {author} {\bibfnamefont {K.~A.}\ \bibnamefont {Gilmore}}, \bibinfo {author} {\bibfnamefont {M.}~\bibnamefont {Affolter}}, \bibinfo {author} {\bibfnamefont {R.~J.}\ \bibnamefont {Lewis-Swan}}, \bibinfo {author} {\bibfnamefont {D.}~\bibnamefont {Barberena}}, \bibinfo {author} {\bibfnamefont {E.}~\bibnamefont {Jordan}}, \bibinfo {author} {\bibfnamefont {A.~M.}\ \bibnamefont {Rey}},\ and\ \bibinfo {author} {\bibfnamefont {J.~J.}\ \bibnamefont {Bollinger}},\ }\bibfield  {title} {\bibinfo {title} {Quantum-enhanced sensing of displacements and electric fields with two-dimensional trapped-ion crystals},\ }\href {https://doi.org/10.1126/science.abi5226} {\bibfield  {journal} {\bibinfo  {journal} {Science}\ }\textbf {\bibinfo {volume} {373}},\ \bibinfo {pages} {673} (\bibinfo {year} {2021})}\BibitemShut {NoStop}%
\bibitem [{\citenamefont {Taylor}\ \emph {et~al.}(2008)\citenamefont {Taylor}, \citenamefont {Cappellaro}, \citenamefont {Childress}, \citenamefont {Jiang}, \citenamefont {Budker}, \citenamefont {Hemmer}, \citenamefont {Yacoby}, \citenamefont {Walsworth},\ and\ \citenamefont {Lukin}}]{Taylor2008}%
  \BibitemOpen
  \bibfield  {author} {\bibinfo {author} {\bibfnamefont {J.~M.}\ \bibnamefont {Taylor}}, \bibinfo {author} {\bibfnamefont {P.}~\bibnamefont {Cappellaro}}, \bibinfo {author} {\bibfnamefont {L.}~\bibnamefont {Childress}}, \bibinfo {author} {\bibfnamefont {L.}~\bibnamefont {Jiang}}, \bibinfo {author} {\bibfnamefont {D.}~\bibnamefont {Budker}}, \bibinfo {author} {\bibfnamefont {P.~R.}\ \bibnamefont {Hemmer}}, \bibinfo {author} {\bibfnamefont {A.}~\bibnamefont {Yacoby}}, \bibinfo {author} {\bibfnamefont {R.}~\bibnamefont {Walsworth}},\ and\ \bibinfo {author} {\bibfnamefont {M.~D.}\ \bibnamefont {Lukin}},\ }\bibfield  {title} {\bibinfo {title} {High-sensitivity diamond magnetometer with nanoscale resolution},\ }\href {https://doi.org/10.1038/nphys1075} {\bibfield  {journal} {\bibinfo  {journal} {Nature Physics}\ }\textbf {\bibinfo {volume} {4}},\ \bibinfo {pages} {810} (\bibinfo {year} {2008})}\BibitemShut {NoStop}%
\bibitem [{\citenamefont {{LIGO Scientific Collaboration}}(2013)}]{Aasi2013}%
  \BibitemOpen
  \bibfield  {author} {\bibinfo {author} {\bibnamefont {{LIGO Scientific Collaboration}}},\ }\bibfield  {title} {\bibinfo {title} {Enhanced sensitivity of the ligo gravitational wave detector by using squeezed states of light},\ }\href {https://doi.org/10.1038/nphoton.2013.177} {\bibfield  {journal} {\bibinfo  {journal} {Nature Photonics}\ }\textbf {\bibinfo {volume} {7}},\ \bibinfo {pages} {613} (\bibinfo {year} {2013})}\BibitemShut {NoStop}%
\bibitem [{\citenamefont {Salducci}\ \emph {et~al.}(2024)\citenamefont {Salducci}, \citenamefont {Bidel}, \citenamefont {Cadoret}, \citenamefont {Darmon}, \citenamefont {Zahzam}, \citenamefont {Bonnin}, \citenamefont {Schwartz}, \citenamefont {Blanchard},\ and\ \citenamefont {Bresson}}]{ref1}%
  \BibitemOpen
  \bibfield  {author} {\bibinfo {author} {\bibfnamefont {C.}~\bibnamefont {Salducci}}, \bibinfo {author} {\bibfnamefont {Y.}~\bibnamefont {Bidel}}, \bibinfo {author} {\bibfnamefont {M.}~\bibnamefont {Cadoret}}, \bibinfo {author} {\bibfnamefont {S.}~\bibnamefont {Darmon}}, \bibinfo {author} {\bibfnamefont {N.}~\bibnamefont {Zahzam}}, \bibinfo {author} {\bibfnamefont {A.}~\bibnamefont {Bonnin}}, \bibinfo {author} {\bibfnamefont {S.}~\bibnamefont {Schwartz}}, \bibinfo {author} {\bibfnamefont {C.}~\bibnamefont {Blanchard}},\ and\ \bibinfo {author} {\bibfnamefont {A.}~\bibnamefont {Bresson}},\ }\bibfield  {title} {\bibinfo {title} {Quantum sensing of acceleration and rotation by interfering magnetically launched atoms},\ }\href {https://doi.org/10.1126/sciadv.adq4498} {\bibfield  {journal} {\bibinfo  {journal} {Science Advances}\ }\textbf {\bibinfo {volume} {10}},\ \bibinfo {pages} {eadq4498} (\bibinfo {year} {2024})}\BibitemShut {NoStop}%
\bibitem [{\citenamefont {Hempel}\ \emph {et~al.}(2013)\citenamefont {Hempel}, \citenamefont {Lanyon}, \citenamefont {Jurcevic}, \citenamefont {Gerritsma}, \citenamefont {Blatt},\ and\ \citenamefont {Roos}}]{Hempel2013}%
  \BibitemOpen
  \bibfield  {author} {\bibinfo {author} {\bibfnamefont {C.}~\bibnamefont {Hempel}}, \bibinfo {author} {\bibfnamefont {B.~P.}\ \bibnamefont {Lanyon}}, \bibinfo {author} {\bibfnamefont {P.}~\bibnamefont {Jurcevic}}, \bibinfo {author} {\bibfnamefont {R.}~\bibnamefont {Gerritsma}}, \bibinfo {author} {\bibfnamefont {R.}~\bibnamefont {Blatt}},\ and\ \bibinfo {author} {\bibfnamefont {C.~F.}\ \bibnamefont {Roos}},\ }\bibfield  {title} {\bibinfo {title} {Entanglement-enhanced detection of single-photon scattering events},\ }\href {https://doi.org/10.1038/nphoton.2013.172} {\bibfield  {journal} {\bibinfo  {journal} {Nature Photonics}\ }\textbf {\bibinfo {volume} {7}},\ \bibinfo {pages} {630} (\bibinfo {year} {2013})}\BibitemShut {NoStop}%
\bibitem [{\citenamefont {Wolf}\ \emph {et~al.}(2019)\citenamefont {Wolf}, \citenamefont {Shi}, \citenamefont {Heip}, \citenamefont {Gessner}, \citenamefont {Pezz{\`e}}, \citenamefont {Smerzi}, \citenamefont {Schulte}, \citenamefont {Hammerer},\ and\ \citenamefont {Schmidt}}]{Wolf2019}%
  \BibitemOpen
  \bibfield  {author} {\bibinfo {author} {\bibfnamefont {F.}~\bibnamefont {Wolf}}, \bibinfo {author} {\bibfnamefont {C.}~\bibnamefont {Shi}}, \bibinfo {author} {\bibfnamefont {J.~C.}\ \bibnamefont {Heip}}, \bibinfo {author} {\bibfnamefont {M.}~\bibnamefont {Gessner}}, \bibinfo {author} {\bibfnamefont {L.}~\bibnamefont {Pezz{\`e}}}, \bibinfo {author} {\bibfnamefont {A.}~\bibnamefont {Smerzi}}, \bibinfo {author} {\bibfnamefont {M.}~\bibnamefont {Schulte}}, \bibinfo {author} {\bibfnamefont {K.}~\bibnamefont {Hammerer}},\ and\ \bibinfo {author} {\bibfnamefont {P.~O.}\ \bibnamefont {Schmidt}},\ }\bibfield  {title} {\bibinfo {title} {Motional fock states for quantum-enhanced amplitude and phase measurements with trapped ions},\ }\href {https://doi.org/10.1038/s41467-019-10576-4} {\bibfield  {journal} {\bibinfo  {journal} {Nature Communications}\ }\textbf {\bibinfo {volume} {10}},\ \bibinfo {pages} {2929} (\bibinfo {year} {2019})}\BibitemShut {NoStop}%
\bibitem [{\citenamefont {Proctor}\ \emph {et~al.}(2018)\citenamefont {Proctor}, \citenamefont {Knott},\ and\ \citenamefont {Dunningham}}]{proctorMultiparameterEstimationNetworked2018}%
  \BibitemOpen
  \bibfield  {author} {\bibinfo {author} {\bibfnamefont {T.~J.}\ \bibnamefont {Proctor}}, \bibinfo {author} {\bibfnamefont {P.~A.}\ \bibnamefont {Knott}},\ and\ \bibinfo {author} {\bibfnamefont {J.~A.}\ \bibnamefont {Dunningham}},\ }\bibfield  {title} {\bibinfo {title} {Multiparameter {{Estimation}} in {{Networked Quantum Sensors}}},\ }\href {https://doi.org/10.1103/PhysRevLett.120.080501} {\bibfield  {journal} {\bibinfo  {journal} {Physical Review Letters}\ }\textbf {\bibinfo {volume} {120}},\ \bibinfo {pages} {080501} (\bibinfo {year} {2018})}\BibitemShut {NoStop}%
\bibitem [{\citenamefont {Eldredge}\ \emph {et~al.}(2018)\citenamefont {Eldredge}, \citenamefont {{Foss-Feig}}, \citenamefont {Gross}, \citenamefont {Rolston},\ and\ \citenamefont {Gorshkov}}]{eldredgeOptimalSecureMeasurement2018}%
  \BibitemOpen
  \bibfield  {author} {\bibinfo {author} {\bibfnamefont {Z.}~\bibnamefont {Eldredge}}, \bibinfo {author} {\bibfnamefont {M.}~\bibnamefont {{Foss-Feig}}}, \bibinfo {author} {\bibfnamefont {J.~A.}\ \bibnamefont {Gross}}, \bibinfo {author} {\bibfnamefont {S.~L.}\ \bibnamefont {Rolston}},\ and\ \bibinfo {author} {\bibfnamefont {A.~V.}\ \bibnamefont {Gorshkov}},\ }\bibfield  {title} {\bibinfo {title} {Optimal and secure measurement protocols for quantum sensor networks},\ }\href {https://doi.org/10.1103/PhysRevA.97.042337} {\bibfield  {journal} {\bibinfo  {journal} {Physical Review A}\ }\textbf {\bibinfo {volume} {97}},\ \bibinfo {pages} {042337} (\bibinfo {year} {2018})}\BibitemShut {NoStop}%
\bibitem [{\citenamefont {Qian}\ \emph {et~al.}(2019)\citenamefont {Qian}, \citenamefont {Eldredge}, \citenamefont {Ge}, \citenamefont {Pagano}, \citenamefont {Monroe}, \citenamefont {Porto},\ and\ \citenamefont {Gorshkov}}]{qianHeisenbergscalingMeasurementProtocol2019}%
  \BibitemOpen
  \bibfield  {author} {\bibinfo {author} {\bibfnamefont {K.}~\bibnamefont {Qian}}, \bibinfo {author} {\bibfnamefont {Z.}~\bibnamefont {Eldredge}}, \bibinfo {author} {\bibfnamefont {W.}~\bibnamefont {Ge}}, \bibinfo {author} {\bibfnamefont {G.}~\bibnamefont {Pagano}}, \bibinfo {author} {\bibfnamefont {C.}~\bibnamefont {Monroe}}, \bibinfo {author} {\bibfnamefont {J.~V.}\ \bibnamefont {Porto}},\ and\ \bibinfo {author} {\bibfnamefont {A.~V.}\ \bibnamefont {Gorshkov}},\ }\bibfield  {title} {\bibinfo {title} {Heisenberg-scaling measurement protocol for analytic functions with quantum sensor networks},\ }\href {https://doi.org/10.1103/PhysRevA.100.042304} {\bibfield  {journal} {\bibinfo  {journal} {Physical Review A}\ }\textbf {\bibinfo {volume} {100}},\ \bibinfo {pages} {042304} (\bibinfo {year} {2019})}\BibitemShut {NoStop}%
\bibitem [{\citenamefont {Shettell}\ and\ \citenamefont {Markham}(2020)}]{shettellGraphStatesResource2020}%
  \BibitemOpen
  \bibfield  {author} {\bibinfo {author} {\bibfnamefont {N.}~\bibnamefont {Shettell}}\ and\ \bibinfo {author} {\bibfnamefont {D.}~\bibnamefont {Markham}},\ }\bibfield  {title} {\bibinfo {title} {Graph {{States}} as a {{Resource}} for {{Quantum Metrology}}},\ }\href {https://doi.org/10.1103/PhysRevLett.124.110502} {\bibfield  {journal} {\bibinfo  {journal} {Physical Review Letters}\ }\textbf {\bibinfo {volume} {124}},\ \bibinfo {pages} {110502} (\bibinfo {year} {2020})}\BibitemShut {NoStop}%
\bibitem [{\citenamefont {Rubio}\ \emph {et~al.}(2020)\citenamefont {Rubio}, \citenamefont {Knott}, \citenamefont {Proctor},\ and\ \citenamefont {Dunningham}}]{rubioQuantumSensingNetworks2020}%
  \BibitemOpen
  \bibfield  {author} {\bibinfo {author} {\bibfnamefont {J.}~\bibnamefont {Rubio}}, \bibinfo {author} {\bibfnamefont {P.~A.}\ \bibnamefont {Knott}}, \bibinfo {author} {\bibfnamefont {T.~J.}\ \bibnamefont {Proctor}},\ and\ \bibinfo {author} {\bibfnamefont {J.~A.}\ \bibnamefont {Dunningham}},\ }\bibfield  {title} {\bibinfo {title} {Quantum sensing networks for the estimation of linear functions},\ }\href {https://doi.org/10.1088/1751-8121/ab9d46} {\bibfield  {journal} {\bibinfo  {journal} {Journal of Physics A: Mathematical and Theoretical}\ }\textbf {\bibinfo {volume} {53}},\ \bibinfo {pages} {344001} (\bibinfo {year} {2020})}\BibitemShut {NoStop}%
\bibitem [{\citenamefont {Bringewatt}\ \emph {et~al.}(2021)\citenamefont {Bringewatt}, \citenamefont {Boettcher}, \citenamefont {Niroula}, \citenamefont {Bienias},\ and\ \citenamefont {Gorshkov}}]{bringewattProtocolsEstimatingMultiple2021}%
  \BibitemOpen
  \bibfield  {author} {\bibinfo {author} {\bibfnamefont {J.}~\bibnamefont {Bringewatt}}, \bibinfo {author} {\bibfnamefont {I.}~\bibnamefont {Boettcher}}, \bibinfo {author} {\bibfnamefont {P.}~\bibnamefont {Niroula}}, \bibinfo {author} {\bibfnamefont {P.}~\bibnamefont {Bienias}},\ and\ \bibinfo {author} {\bibfnamefont {A.~V.}\ \bibnamefont {Gorshkov}},\ }\bibfield  {title} {\bibinfo {title} {Protocols for estimating multiple functions with quantum sensor networks: {{Geometry}} and performance},\ }\href {https://doi.org/10.1103/PhysRevResearch.3.033011} {\bibfield  {journal} {\bibinfo  {journal} {Physical Review Research}\ }\textbf {\bibinfo {volume} {3}},\ \bibinfo {pages} {033011} (\bibinfo {year} {2021})}\BibitemShut {NoStop}%
\bibitem [{\citenamefont {Shettell}\ \emph {et~al.}(2022)\citenamefont {Shettell}, \citenamefont {Hassani},\ and\ \citenamefont {Markham}}]{shettellPrivateNetworkParameter2022}%
  \BibitemOpen
  \bibfield  {author} {\bibinfo {author} {\bibfnamefont {N.}~\bibnamefont {Shettell}}, \bibinfo {author} {\bibfnamefont {M.}~\bibnamefont {Hassani}},\ and\ \bibinfo {author} {\bibfnamefont {D.}~\bibnamefont {Markham}},\ }\href {https://doi.org/10.48550/arXiv.2207.14450} {\bibinfo {title} {Private network parameter estimation with quantum sensors}} (\bibinfo {year} {2022}),\ \Eprint {https://arxiv.org/abs/2207.14450} {arXiv:2207.14450} \BibitemShut {NoStop}%
\bibitem [{\citenamefont {Bugalho}\ \emph {et~al.}(2024)\citenamefont {Bugalho}, \citenamefont {Hassani}, \citenamefont {Omar},\ and\ \citenamefont {Markham}}]{bugalhoPrivateRobustStates2024}%
  \BibitemOpen
  \bibfield  {author} {\bibinfo {author} {\bibfnamefont {L.}~\bibnamefont {Bugalho}}, \bibinfo {author} {\bibfnamefont {M.}~\bibnamefont {Hassani}}, \bibinfo {author} {\bibfnamefont {Y.}~\bibnamefont {Omar}},\ and\ \bibinfo {author} {\bibfnamefont {D.}~\bibnamefont {Markham}},\ }\href {https://doi.org/10.48550/arXiv.2407.21701} {\bibinfo {title} {Private and {{Robust States}} for {{Distributed Quantum Sensing}}}} (\bibinfo {year} {2024}),\ \Eprint {https://arxiv.org/abs/2407.21701} {arXiv:2407.21701} \BibitemShut {NoStop}%
\bibitem [{\citenamefont {Hassani}\ \emph {et~al.}(2024)\citenamefont {Hassani}, \citenamefont {Scheiner}, \citenamefont {Paris},\ and\ \citenamefont {Markham}}]{hassaniPrivacyNetworksQuantum2024}%
  \BibitemOpen
  \bibfield  {author} {\bibinfo {author} {\bibfnamefont {M.}~\bibnamefont {Hassani}}, \bibinfo {author} {\bibfnamefont {S.}~\bibnamefont {Scheiner}}, \bibinfo {author} {\bibfnamefont {M.~G.~A.}\ \bibnamefont {Paris}},\ and\ \bibinfo {author} {\bibfnamefont {D.}~\bibnamefont {Markham}},\ }\href {https://doi.org/10.48550/arXiv.2408.01711} {\bibinfo {title} {Privacy in networks of quantum sensors}} (\bibinfo {year} {2024}),\ \Eprint {https://arxiv.org/abs/2408.01711} {arXiv:2408.01711} \BibitemShut {NoStop}%
\bibitem [{\citenamefont {Liu}\ \emph {et~al.}(2021)\citenamefont {Liu}, \citenamefont {Zhang}, \citenamefont {Li}, \citenamefont {Zhang}, \citenamefont {Yin}, \citenamefont {Fei}, \citenamefont {Li}, \citenamefont {Liu}, \citenamefont {Xu}, \citenamefont {Chen},\ and\ \citenamefont {Pan}}]{Liu2021}%
  \BibitemOpen
  \bibfield  {author} {\bibinfo {author} {\bibfnamefont {L.-Z.}\ \bibnamefont {Liu}}, \bibinfo {author} {\bibfnamefont {Y.-Z.}\ \bibnamefont {Zhang}}, \bibinfo {author} {\bibfnamefont {Z.-D.}\ \bibnamefont {Li}}, \bibinfo {author} {\bibfnamefont {R.}~\bibnamefont {Zhang}}, \bibinfo {author} {\bibfnamefont {X.-F.}\ \bibnamefont {Yin}}, \bibinfo {author} {\bibfnamefont {Y.-Y.}\ \bibnamefont {Fei}}, \bibinfo {author} {\bibfnamefont {L.}~\bibnamefont {Li}}, \bibinfo {author} {\bibfnamefont {N.-L.}\ \bibnamefont {Liu}}, \bibinfo {author} {\bibfnamefont {F.}~\bibnamefont {Xu}}, \bibinfo {author} {\bibfnamefont {Y.-A.}\ \bibnamefont {Chen}},\ and\ \bibinfo {author} {\bibfnamefont {J.-W.}\ \bibnamefont {Pan}},\ }\bibfield  {title} {\bibinfo {title} {Distributed quantum phase estimation with entangled photons},\ }\href {https://doi.org/10.1038/s41566-020-00718-2} {\bibfield  {journal} {\bibinfo  {journal} {Nature Photonics}\ }\textbf {\bibinfo {volume} {15}},\ \bibinfo {pages} {137} (\bibinfo {year} {2021})}\BibitemShut {NoStop}%
\bibitem [{\citenamefont {Kim}\ \emph {et~al.}(2024)\citenamefont {Kim}, \citenamefont {Hong}, \citenamefont {Kim}, \citenamefont {Kim}, \citenamefont {Lee}, \citenamefont {Pooser}, \citenamefont {Oh}, \citenamefont {Lee}, \citenamefont {Lee},\ and\ \citenamefont {Lim}}]{Kim2024}%
  \BibitemOpen
  \bibfield  {author} {\bibinfo {author} {\bibfnamefont {D.-H.}\ \bibnamefont {Kim}}, \bibinfo {author} {\bibfnamefont {S.}~\bibnamefont {Hong}}, \bibinfo {author} {\bibfnamefont {Y.-S.}\ \bibnamefont {Kim}}, \bibinfo {author} {\bibfnamefont {Y.}~\bibnamefont {Kim}}, \bibinfo {author} {\bibfnamefont {S.-W.}\ \bibnamefont {Lee}}, \bibinfo {author} {\bibfnamefont {R.~C.}\ \bibnamefont {Pooser}}, \bibinfo {author} {\bibfnamefont {K.}~\bibnamefont {Oh}}, \bibinfo {author} {\bibfnamefont {S.-Y.}\ \bibnamefont {Lee}}, \bibinfo {author} {\bibfnamefont {C.}~\bibnamefont {Lee}},\ and\ \bibinfo {author} {\bibfnamefont {H.-T.}\ \bibnamefont {Lim}},\ }\bibfield  {title} {\bibinfo {title} {Distributed quantum sensing of multiple phases with fewer photons},\ }\href {https://doi.org/10.1038/s41467-023-44204-z} {\bibfield  {journal} {\bibinfo  {journal} {Nature Communications}\ }\textbf {\bibinfo {volume} {15}},\ \bibinfo {pages} {266} (\bibinfo {year} {2024})}\BibitemShut {NoStop}%
\bibitem [{\citenamefont {Zhao}\ \emph {et~al.}(2021)\citenamefont {Zhao}, \citenamefont {Zhang}, \citenamefont {Liu}, \citenamefont {Guan}, \citenamefont {Zhang}, \citenamefont {Li}, \citenamefont {Bai}, \citenamefont {Li}, \citenamefont {Liu}, \citenamefont {You}, \citenamefont {Zhang}, \citenamefont {Fan}, \citenamefont {Xu}, \citenamefont {Zhang},\ and\ \citenamefont {Pan}}]{PhysRevX.11.031009}%
  \BibitemOpen
  \bibfield  {author} {\bibinfo {author} {\bibfnamefont {S.-R.}\ \bibnamefont {Zhao}}, \bibinfo {author} {\bibfnamefont {Y.-Z.}\ \bibnamefont {Zhang}}, \bibinfo {author} {\bibfnamefont {W.-Z.}\ \bibnamefont {Liu}}, \bibinfo {author} {\bibfnamefont {J.-Y.}\ \bibnamefont {Guan}}, \bibinfo {author} {\bibfnamefont {W.}~\bibnamefont {Zhang}}, \bibinfo {author} {\bibfnamefont {C.-L.}\ \bibnamefont {Li}}, \bibinfo {author} {\bibfnamefont {B.}~\bibnamefont {Bai}}, \bibinfo {author} {\bibfnamefont {M.-H.}\ \bibnamefont {Li}}, \bibinfo {author} {\bibfnamefont {Y.}~\bibnamefont {Liu}}, \bibinfo {author} {\bibfnamefont {L.}~\bibnamefont {You}}, \bibinfo {author} {\bibfnamefont {J.}~\bibnamefont {Zhang}}, \bibinfo {author} {\bibfnamefont {J.}~\bibnamefont {Fan}}, \bibinfo {author} {\bibfnamefont {F.}~\bibnamefont {Xu}}, \bibinfo {author} {\bibfnamefont {Q.}~\bibnamefont {Zhang}},\ and\ \bibinfo {author} {\bibfnamefont {J.-W.}\ \bibnamefont {Pan}},\ }\bibfield  {title} {\bibinfo {title} {Field demonstration of distributed quantum sensing without post-selection},\ }\href {https://doi.org/10.1103/PhysRevX.11.031009} {\bibfield  {journal} {\bibinfo  {journal} {Phys. Rev. X}\ }\textbf {\bibinfo {volume} {11}},\ \bibinfo {pages} {031009} (\bibinfo {year} {2021})}\BibitemShut {NoStop}%
\bibitem [{\citenamefont {Nichol}\ \emph {et~al.}(2022)\citenamefont {Nichol}, \citenamefont {Srinivas}, \citenamefont {Nadlinger}, \citenamefont {Drmota}, \citenamefont {Main}, \citenamefont {Araneda}, \citenamefont {Ballance},\ and\ \citenamefont {Lucas}}]{nicholElementaryQuantumNetwork2022}%
  \BibitemOpen
  \bibfield  {author} {\bibinfo {author} {\bibfnamefont {B.~C.}\ \bibnamefont {Nichol}}, \bibinfo {author} {\bibfnamefont {R.}~\bibnamefont {Srinivas}}, \bibinfo {author} {\bibfnamefont {D.~P.}\ \bibnamefont {Nadlinger}}, \bibinfo {author} {\bibfnamefont {P.}~\bibnamefont {Drmota}}, \bibinfo {author} {\bibfnamefont {D.}~\bibnamefont {Main}}, \bibinfo {author} {\bibfnamefont {G.}~\bibnamefont {Araneda}}, \bibinfo {author} {\bibfnamefont {C.~J.}\ \bibnamefont {Ballance}},\ and\ \bibinfo {author} {\bibfnamefont {D.~M.}\ \bibnamefont {Lucas}},\ }\bibfield  {title} {\bibinfo {title} {An elementary quantum network of entangled optical atomic clocks},\ }\href {https://doi.org/10.1038/s41586-022-05088-z} {\bibfield  {journal} {\bibinfo  {journal} {Nature}\ }\textbf {\bibinfo {volume} {609}},\ \bibinfo {pages} {689} (\bibinfo {year} {2022})}\BibitemShut {NoStop}%
\bibitem [{\citenamefont {Hainzer}\ \emph {et~al.}()\citenamefont {Hainzer}, \citenamefont {Kiesenhofer}, \citenamefont {Ollikainen}, \citenamefont {Bock}, \citenamefont {Kranzl}, \citenamefont {Joshi}, \citenamefont {Yoeli}, \citenamefont {Blatt}, \citenamefont {Gefen},\ and\ \citenamefont {Roos}}]{hainzer_correlation_2024}%
  \BibitemOpen
  \bibfield  {author} {\bibinfo {author} {\bibfnamefont {H.}~\bibnamefont {Hainzer}}, \bibinfo {author} {\bibfnamefont {D.}~\bibnamefont {Kiesenhofer}}, \bibinfo {author} {\bibfnamefont {T.}~\bibnamefont {Ollikainen}}, \bibinfo {author} {\bibfnamefont {M.}~\bibnamefont {Bock}}, \bibinfo {author} {\bibfnamefont {F.}~\bibnamefont {Kranzl}}, \bibinfo {author} {\bibfnamefont {M.~K.}\ \bibnamefont {Joshi}}, \bibinfo {author} {\bibfnamefont {G.}~\bibnamefont {Yoeli}}, \bibinfo {author} {\bibfnamefont {R.}~\bibnamefont {Blatt}}, \bibinfo {author} {\bibfnamefont {T.}~\bibnamefont {Gefen}},\ and\ \bibinfo {author} {\bibfnamefont {C.~F.}\ \bibnamefont {Roos}},\ }\bibfield  {title} {\bibinfo {title} {Correlation spectroscopy with multiqubit-enhanced phase estimation},\ }\href {https://doi.org/10.1103/PhysRevX.14.011033} {\bibfield  {journal} {\bibinfo  {journal} {Phys. Rev. X}\ }\textbf {\bibinfo {volume} {14}},\ \bibinfo {pages} {011033}}\BibitemShut {NoStop}%
\bibitem [{\citenamefont {Malia}\ \emph {et~al.}(2022)\citenamefont {Malia}, \citenamefont {Wu}, \citenamefont {{Mart{\'i}nez-Rinc{\'o}n}},\ and\ \citenamefont {Kasevich}}]{maliaDistributedQuantumSensing2022}%
  \BibitemOpen
  \bibfield  {author} {\bibinfo {author} {\bibfnamefont {B.~K.}\ \bibnamefont {Malia}}, \bibinfo {author} {\bibfnamefont {Y.}~\bibnamefont {Wu}}, \bibinfo {author} {\bibfnamefont {J.}~\bibnamefont {{Mart{\'i}nez-Rinc{\'o}n}}},\ and\ \bibinfo {author} {\bibfnamefont {M.~A.}\ \bibnamefont {Kasevich}},\ }\bibfield  {title} {\bibinfo {title} {Distributed quantum sensing with mode-entangled spin-squeezed atomic states},\ }\href {https://doi.org/10.1038/s41586-022-05363-z} {\bibfield  {journal} {\bibinfo  {journal} {Nature}\ }\textbf {\bibinfo {volume} {612}},\ \bibinfo {pages} {661} (\bibinfo {year} {2022})}\BibitemShut {NoStop}%
\bibitem [{\citenamefont {{Demkowicz-Dobrza{\'n}ski}}\ \emph {et~al.}(2017)\citenamefont {{Demkowicz-Dobrza{\'n}ski}}, \citenamefont {Czajkowski},\ and\ \citenamefont {Sekatski}}]{demkowicz-dobrzanskiAdaptiveQuantumMetrology2017a}%
  \BibitemOpen
  \bibfield  {author} {\bibinfo {author} {\bibfnamefont {R.}~\bibnamefont {{Demkowicz-Dobrza{\'n}ski}}}, \bibinfo {author} {\bibfnamefont {J.}~\bibnamefont {Czajkowski}},\ and\ \bibinfo {author} {\bibfnamefont {P.}~\bibnamefont {Sekatski}},\ }\bibfield  {title} {\bibinfo {title} {Adaptive {{Quantum Metrology}} under {{General Markovian Noise}}},\ }\href {https://doi.org/10.1103/PhysRevX.7.041009} {\bibfield  {journal} {\bibinfo  {journal} {Physical Review X}\ }\textbf {\bibinfo {volume} {7}},\ \bibinfo {pages} {041009} (\bibinfo {year} {2017})}\BibitemShut {NoStop}%
\bibitem [{\citenamefont {Zhou}\ \emph {et~al.}(2018)\citenamefont {Zhou}, \citenamefont {Zhang}, \citenamefont {Preskill},\ and\ \citenamefont {Jiang}}]{zhouAchievingHeisenbergLimit2018a}%
  \BibitemOpen
  \bibfield  {author} {\bibinfo {author} {\bibfnamefont {S.}~\bibnamefont {Zhou}}, \bibinfo {author} {\bibfnamefont {M.}~\bibnamefont {Zhang}}, \bibinfo {author} {\bibfnamefont {J.}~\bibnamefont {Preskill}},\ and\ \bibinfo {author} {\bibfnamefont {L.}~\bibnamefont {Jiang}},\ }\bibfield  {title} {\bibinfo {title} {Achieving the {{Heisenberg}} limit in quantum metrology using quantum error correction},\ }\href {https://doi.org/10.1038/s41467-017-02510-3} {\bibfield  {journal} {\bibinfo  {journal} {Nature Communications}\ }\textbf {\bibinfo {volume} {9}},\ \bibinfo {pages} {78} (\bibinfo {year} {2018})}\BibitemShut {NoStop}%
\bibitem [{\citenamefont {Monz}\ \emph {et~al.}(2011)\citenamefont {Monz}, \citenamefont {Schindler}, \citenamefont {Barreiro}, \citenamefont {Chwalla}, \citenamefont {Nigg}, \citenamefont {Coish}, \citenamefont {Harlander}, \citenamefont {H\"ansel}, \citenamefont {Hennrich},\ and\ \citenamefont {Blatt}}]{PhysRevLett.106.130506}%
  \BibitemOpen
  \bibfield  {author} {\bibinfo {author} {\bibfnamefont {T.}~\bibnamefont {Monz}}, \bibinfo {author} {\bibfnamefont {P.}~\bibnamefont {Schindler}}, \bibinfo {author} {\bibfnamefont {J.~T.}\ \bibnamefont {Barreiro}}, \bibinfo {author} {\bibfnamefont {M.}~\bibnamefont {Chwalla}}, \bibinfo {author} {\bibfnamefont {D.}~\bibnamefont {Nigg}}, \bibinfo {author} {\bibfnamefont {W.~A.}\ \bibnamefont {Coish}}, \bibinfo {author} {\bibfnamefont {M.}~\bibnamefont {Harlander}}, \bibinfo {author} {\bibfnamefont {W.}~\bibnamefont {H\"ansel}}, \bibinfo {author} {\bibfnamefont {M.}~\bibnamefont {Hennrich}},\ and\ \bibinfo {author} {\bibfnamefont {R.}~\bibnamefont {Blatt}},\ }\bibfield  {title} {\bibinfo {title} {14-qubit entanglement: Creation and coherence},\ }\href {https://doi.org/10.1103/PhysRevLett.106.130506} {\bibfield  {journal} {\bibinfo  {journal} {Phys. Rev. Lett.}\ }\textbf {\bibinfo {volume} {106}},\ \bibinfo {pages} {130506} (\bibinfo {year} {2011})}\BibitemShut {NoStop}%
\bibitem [{\citenamefont {Omran}\ \emph {et~al.}(2019)\citenamefont {Omran}, \citenamefont {Levine}, \citenamefont {Keesling}, \citenamefont {Semeghini}, \citenamefont {Wang}, \citenamefont {Ebadi}, \citenamefont {Bernien}, \citenamefont {Zibrov}, \citenamefont {Pichler}, \citenamefont {Choi}, \citenamefont {Cui}, \citenamefont {Rossignolo}, \citenamefont {Rembold}, \citenamefont {Montangero}, \citenamefont {Calarco}, \citenamefont {Endres}, \citenamefont {Greiner}, \citenamefont {Vuleti{\'{c}}},\ and\ \citenamefont {Lukin}}]{Omran2019}%
  \BibitemOpen
  \bibfield  {author} {\bibinfo {author} {\bibfnamefont {A.}~\bibnamefont {Omran}}, \bibinfo {author} {\bibfnamefont {H.}~\bibnamefont {Levine}}, \bibinfo {author} {\bibfnamefont {A.}~\bibnamefont {Keesling}}, \bibinfo {author} {\bibfnamefont {G.}~\bibnamefont {Semeghini}}, \bibinfo {author} {\bibfnamefont {T.~T.}\ \bibnamefont {Wang}}, \bibinfo {author} {\bibfnamefont {S.}~\bibnamefont {Ebadi}}, \bibinfo {author} {\bibfnamefont {H.}~\bibnamefont {Bernien}}, \bibinfo {author} {\bibfnamefont {A.~S.}\ \bibnamefont {Zibrov}}, \bibinfo {author} {\bibfnamefont {H.}~\bibnamefont {Pichler}}, \bibinfo {author} {\bibfnamefont {S.}~\bibnamefont {Choi}}, \bibinfo {author} {\bibfnamefont {J.}~\bibnamefont {Cui}}, \bibinfo {author} {\bibfnamefont {M.}~\bibnamefont {Rossignolo}}, \bibinfo {author} {\bibfnamefont {P.}~\bibnamefont {Rembold}}, \bibinfo {author} {\bibfnamefont {S.}~\bibnamefont {Montangero}}, \bibinfo {author} {\bibfnamefont {T.}~\bibnamefont {Calarco}}, \bibinfo {author} {\bibfnamefont {M.}~\bibnamefont {Endres}}, \bibinfo {author} {\bibfnamefont {M.}~\bibnamefont {Greiner}}, \bibinfo {author} {\bibfnamefont {V.}~\bibnamefont {Vuleti{\'{c}}}},\ and\ \bibinfo {author} {\bibfnamefont {M.~D.}\ \bibnamefont {Lukin}},\ }\bibfield  {title} {\bibinfo {title} {Generation and manipulation of schr{\"o}dinger cat states in rydberg atom arrays},\ }\href {https://doi.org/10.1126/science.aax9743} {\bibfield  {journal} {\bibinfo  {journal} {Science}\ }\textbf {\bibinfo {volume} {365}},\ \bibinfo {pages} {570} (\bibinfo {year} {2019})}\BibitemShut {NoStop}%
\bibitem [{\citenamefont {Bao}\ \emph {et~al.}(2024)\citenamefont {Bao}, \citenamefont {Xu}, \citenamefont {Song}, \citenamefont {Wang}, \citenamefont {Xiang}, \citenamefont {Zhu}, \citenamefont {Chen}, \citenamefont {Jin}, \citenamefont {Zhu}, \citenamefont {Gao} \emph {et~al.}}]{Bao2024}%
  \BibitemOpen
  \bibfield  {author} {\bibinfo {author} {\bibfnamefont {Z.}~\bibnamefont {Bao}}, \bibinfo {author} {\bibfnamefont {S.}~\bibnamefont {Xu}}, \bibinfo {author} {\bibfnamefont {Z.}~\bibnamefont {Song}}, \bibinfo {author} {\bibfnamefont {K.}~\bibnamefont {Wang}}, \bibinfo {author} {\bibfnamefont {L.}~\bibnamefont {Xiang}}, \bibinfo {author} {\bibfnamefont {Z.}~\bibnamefont {Zhu}}, \bibinfo {author} {\bibfnamefont {J.}~\bibnamefont {Chen}}, \bibinfo {author} {\bibfnamefont {F.}~\bibnamefont {Jin}}, \bibinfo {author} {\bibfnamefont {X.}~\bibnamefont {Zhu}}, \bibinfo {author} {\bibfnamefont {Y.}~\bibnamefont {Gao}}, \emph {et~al.},\ }\bibfield  {title} {\bibinfo {title} {Creating and controlling global greenberger-horne-zeilinger entanglement on quantum processors},\ }\href {https://doi.org/10.1038/s41467-024-53140-5} {\bibfield  {journal} {\bibinfo  {journal} {Nature Communications}\ }\textbf {\bibinfo {volume} {15}},\ \bibinfo {pages} {8823} (\bibinfo {year} {2024})}\BibitemShut {NoStop}%
\bibitem [{\citenamefont {Kessler}\ \emph {et~al.}(2014)\citenamefont {Kessler}, \citenamefont {Lovchinsky}, \citenamefont {Sushkov},\ and\ \citenamefont {Lukin}}]{kesslerQuantumErrorCorrection2014}%
  \BibitemOpen
  \bibfield  {author} {\bibinfo {author} {\bibfnamefont {E.~M.}\ \bibnamefont {Kessler}}, \bibinfo {author} {\bibfnamefont {I.}~\bibnamefont {Lovchinsky}}, \bibinfo {author} {\bibfnamefont {A.~O.}\ \bibnamefont {Sushkov}},\ and\ \bibinfo {author} {\bibfnamefont {M.~D.}\ \bibnamefont {Lukin}},\ }\bibfield  {title} {\bibinfo {title} {Quantum {{Error Correction}} for {{Metrology}}},\ }\href {https://doi.org/10.1103/PhysRevLett.112.150802} {\bibfield  {journal} {\bibinfo  {journal} {Physical Review Letters}\ }\textbf {\bibinfo {volume} {112}},\ \bibinfo {pages} {150802} (\bibinfo {year} {2014})}\BibitemShut {NoStop}%
\bibitem [{\citenamefont {Arrad}\ \emph {et~al.}(2014)\citenamefont {Arrad}, \citenamefont {Vinkler}, \citenamefont {Aharonov},\ and\ \citenamefont {Retzker}}]{arradIncreasingSensingResolution2014}%
  \BibitemOpen
  \bibfield  {author} {\bibinfo {author} {\bibfnamefont {G.}~\bibnamefont {Arrad}}, \bibinfo {author} {\bibfnamefont {Y.}~\bibnamefont {Vinkler}}, \bibinfo {author} {\bibfnamefont {D.}~\bibnamefont {Aharonov}},\ and\ \bibinfo {author} {\bibfnamefont {A.}~\bibnamefont {Retzker}},\ }\bibfield  {title} {\bibinfo {title} {Increasing {{Sensing Resolution}} with {{Error Correction}}},\ }\href {https://doi.org/10.1103/PhysRevLett.112.150801} {\bibfield  {journal} {\bibinfo  {journal} {Physical Review Letters}\ }\textbf {\bibinfo {volume} {112}},\ \bibinfo {pages} {150801} (\bibinfo {year} {2014})}\BibitemShut {NoStop}%
\bibitem [{\citenamefont {Sekatski}\ \emph {et~al.}(2017)\citenamefont {Sekatski}, \citenamefont {Skotiniotis}, \citenamefont {Ko{\l}ody{\'n}ski},\ and\ \citenamefont {D{\"u}r}}]{sekatskiQuantumMetrologyFull2017}%
  \BibitemOpen
  \bibfield  {author} {\bibinfo {author} {\bibfnamefont {P.}~\bibnamefont {Sekatski}}, \bibinfo {author} {\bibfnamefont {M.}~\bibnamefont {Skotiniotis}}, \bibinfo {author} {\bibfnamefont {J.}~\bibnamefont {Ko{\l}ody{\'n}ski}},\ and\ \bibinfo {author} {\bibfnamefont {W.}~\bibnamefont {D{\"u}r}},\ }\bibfield  {title} {\bibinfo {title} {Quantum metrology with full and fast quantum control},\ }\href {https://doi.org/10.22331/q-2017-09-06-27} {\bibfield  {journal} {\bibinfo  {journal} {Quantum}\ }\textbf {\bibinfo {volume} {1}},\ \bibinfo {pages} {27} (\bibinfo {year} {2017})}\BibitemShut {NoStop}%
\bibitem [{\citenamefont {Faist}\ \emph {et~al.}(2023)\citenamefont {Faist}, \citenamefont {Woods}, \citenamefont {Albert}, \citenamefont {Renes}, \citenamefont {Eisert},\ and\ \citenamefont {Preskill}}]{faistTimeEnergyUncertaintyRelation2023}%
  \BibitemOpen
  \bibfield  {author} {\bibinfo {author} {\bibfnamefont {P.}~\bibnamefont {Faist}}, \bibinfo {author} {\bibfnamefont {M.~P.}\ \bibnamefont {Woods}}, \bibinfo {author} {\bibfnamefont {V.~V.}\ \bibnamefont {Albert}}, \bibinfo {author} {\bibfnamefont {J.~M.}\ \bibnamefont {Renes}}, \bibinfo {author} {\bibfnamefont {J.}~\bibnamefont {Eisert}},\ and\ \bibinfo {author} {\bibfnamefont {J.}~\bibnamefont {Preskill}},\ }\bibfield  {title} {\bibinfo {title} {Time-{{Energy Uncertainty Relation}} for {{Noisy Quantum Metrology}}},\ }\href {https://doi.org/10.1103/PRXQuantum.4.040336} {\bibfield  {journal} {\bibinfo  {journal} {PRX Quantum}\ }\textbf {\bibinfo {volume} {4}},\ \bibinfo {pages} {040336} (\bibinfo {year} {2023})}\BibitemShut {NoStop}%
\bibitem [{\citenamefont {Sekatski}\ \emph {et~al.}(2020)\citenamefont {Sekatski}, \citenamefont {W{\"o}lk},\ and\ \citenamefont {D{\"u}r}}]{sekatskiOptimalDistributedSensing2020}%
  \BibitemOpen
  \bibfield  {author} {\bibinfo {author} {\bibfnamefont {P.}~\bibnamefont {Sekatski}}, \bibinfo {author} {\bibfnamefont {S.}~\bibnamefont {W{\"o}lk}},\ and\ \bibinfo {author} {\bibfnamefont {W.}~\bibnamefont {D{\"u}r}},\ }\bibfield  {title} {\bibinfo {title} {Optimal distributed sensing in noisy environments},\ }\href {https://doi.org/10.1103/PhysRevResearch.2.023052} {\bibfield  {journal} {\bibinfo  {journal} {Physical Review Research}\ }\textbf {\bibinfo {volume} {2}},\ \bibinfo {pages} {023052} (\bibinfo {year} {2020})}\BibitemShut {NoStop}%
\bibitem [{\citenamefont {Helstrom}(1969)}]{helstrom1969quantum}%
  \BibitemOpen
  \bibfield  {author} {\bibinfo {author} {\bibfnamefont {C.~W.}\ \bibnamefont {Helstrom}},\ }\bibfield  {title} {\bibinfo {title} {Quantum detection and estimation theory},\ }\href@noop {} {\bibfield  {journal} {\bibinfo  {journal} {Journal of Statistical Physics}\ }\textbf {\bibinfo {volume} {1}},\ \bibinfo {pages} {231} (\bibinfo {year} {1969})}\BibitemShut {NoStop}%
\bibitem [{\citenamefont {Zanardi}\ and\ \citenamefont {Rasetti}(1997)}]{PhysRevLett.79.3306}%
  \BibitemOpen
  \bibfield  {author} {\bibinfo {author} {\bibfnamefont {P.}~\bibnamefont {Zanardi}}\ and\ \bibinfo {author} {\bibfnamefont {M.}~\bibnamefont {Rasetti}},\ }\bibfield  {title} {\bibinfo {title} {Noiseless quantum codes},\ }\href {https://doi.org/10.1103/PhysRevLett.79.3306} {\bibfield  {journal} {\bibinfo  {journal} {Phys. Rev. Lett.}\ }\textbf {\bibinfo {volume} {79}},\ \bibinfo {pages} {3306} (\bibinfo {year} {1997})}\BibitemShut {NoStop}%
\bibitem [{\citenamefont {Hamann}\ \emph {et~al.}(2024)\citenamefont {Hamann}, \citenamefont {Sekatski},\ and\ \citenamefont {D{\"u}r}}]{hamannOptimalDistributedMultiparameter2024}%
  \BibitemOpen
  \bibfield  {author} {\bibinfo {author} {\bibfnamefont {A.}~\bibnamefont {Hamann}}, \bibinfo {author} {\bibfnamefont {P.}~\bibnamefont {Sekatski}},\ and\ \bibinfo {author} {\bibfnamefont {W.}~\bibnamefont {D{\"u}r}},\ }\bibfield  {title} {\bibinfo {title} {Optimal distributed multi-parameter estimation in noisy environments},\ }\href {https://doi.org/10.1088/2058-9565/ad37d5} {\bibfield  {journal} {\bibinfo  {journal} {Quantum Science and Technology}\ }\textbf {\bibinfo {volume} {9}},\ \bibinfo {pages} {035005} (\bibinfo {year} {2024})}\BibitemShut {NoStop}%
\bibitem [{\citenamefont {W{\"o}lk}\ \emph {et~al.}(2020)\citenamefont {W{\"o}lk}, \citenamefont {Sekatski},\ and\ \citenamefont {D{\"u}r}}]{wolkNoisyDistributedSensing2020}%
  \BibitemOpen
  \bibfield  {author} {\bibinfo {author} {\bibfnamefont {S.}~\bibnamefont {W{\"o}lk}}, \bibinfo {author} {\bibfnamefont {P.}~\bibnamefont {Sekatski}},\ and\ \bibinfo {author} {\bibfnamefont {W.}~\bibnamefont {D{\"u}r}},\ }\bibfield  {title} {\bibinfo {title} {Noisy distributed sensing in the {{Bayesian}} regime},\ }\href {https://doi.org/10.1088/2058-9565/ab9ba5} {\bibfield  {journal} {\bibinfo  {journal} {Quantum Science and Technology}\ }\textbf {\bibinfo {volume} {5}},\ \bibinfo {pages} {045003} (\bibinfo {year} {2020})}\BibitemShut {NoStop}%
\bibitem [{\citenamefont {Hamann}\ \emph {et~al.}(2022)\citenamefont {Hamann}, \citenamefont {Sekatski},\ and\ \citenamefont {D{\"u}r}}]{hamannApproximateDecoherenceFree2022}%
  \BibitemOpen
  \bibfield  {author} {\bibinfo {author} {\bibfnamefont {A.}~\bibnamefont {Hamann}}, \bibinfo {author} {\bibfnamefont {P.}~\bibnamefont {Sekatski}},\ and\ \bibinfo {author} {\bibfnamefont {W.}~\bibnamefont {D{\"u}r}},\ }\bibfield  {title} {\bibinfo {title} {Approximate decoherence free subspaces for distributed sensing},\ }\href {https://doi.org/10.1088/2058-9565/ac44de} {\bibfield  {journal} {\bibinfo  {journal} {Quantum Science and Technology}\ }\textbf {\bibinfo {volume} {7}},\ \bibinfo {pages} {025003} (\bibinfo {year} {2022})}\BibitemShut {NoStop}%
\bibitem [{\citenamefont {Landini}\ \emph {et~al.}(2014)\citenamefont {Landini}, \citenamefont {Fattori}, \citenamefont {Pezz{\`e}},\ and\ \citenamefont {Smerzi}}]{landiniPhasenoiseProtectionQuantumenhanced2014}%
  \BibitemOpen
  \bibfield  {author} {\bibinfo {author} {\bibfnamefont {M.}~\bibnamefont {Landini}}, \bibinfo {author} {\bibfnamefont {M.}~\bibnamefont {Fattori}}, \bibinfo {author} {\bibfnamefont {L.}~\bibnamefont {Pezz{\`e}}},\ and\ \bibinfo {author} {\bibfnamefont {A.}~\bibnamefont {Smerzi}},\ }\bibfield  {title} {\bibinfo {title} {Phase-noise protection in quantum-enhanced differential interferometry},\ }\href {https://doi.org/10.1088/1367-2630/16/11/113074} {\bibfield  {journal} {\bibinfo  {journal} {New Journal of Physics}\ }\textbf {\bibinfo {volume} {16}},\ \bibinfo {pages} {113074} (\bibinfo {year} {2014})}\BibitemShut {NoStop}%
\bibitem [{Sup()}]{SuppMat}%
  \BibitemOpen
  \href@noop {} {\bibinfo {title} {See supplemental material including refs.~\cite{Rao1992,CramerHarald1946Mmos, fisherMathematicalFoundationsTheoretical1922,helstrom1969quantum,parisQuantumEstimationQuantum2009,giovannettiAdvancesQuantumMetrology2011,sekatskiOptimalDistributedSensing2020,MarcoThesis,PhysRevLett.106.130506,James1998,RoosThesis,ringbauerUniversalQuditQuantum2022,Milena,Chessa2021,Wilde_2013,PhysRevA.62.022311, Kirchmair_2009} at [url will be inserted by publisher] for details on the estimation protocols, experimental methods, and supporting experimental and theoretical results.}}\BibitemShut {Stop}%
\bibitem [{\citenamefont {Greenberger}\ \emph {et~al.}(1989)\citenamefont {Greenberger}, \citenamefont {Horne},\ and\ \citenamefont {Zeilinger}}]{Greenberger_1989}%
  \BibitemOpen
  \bibfield  {author} {\bibinfo {author} {\bibfnamefont {D.~M.}\ \bibnamefont {Greenberger}}, \bibinfo {author} {\bibfnamefont {M.~A.}\ \bibnamefont {Horne}},\ and\ \bibinfo {author} {\bibfnamefont {A.}~\bibnamefont {Zeilinger}},\ }\bibinfo {title} {Going beyond bell’s theorem},\ in\ \href {https://doi.org/10.1007/978-94-017-0849-4_10} {\emph {\bibinfo {booktitle} {Bell’s Theorem, Quantum Theory and Conceptions of the Universe}}}\ (\bibinfo  {publisher} {Springer Netherlands},\ \bibinfo {year} {1989})\ p.\ \bibinfo {pages} {69–72}\BibitemShut {NoStop}%
\bibitem [{\citenamefont {M\o{}lmer}\ and\ \citenamefont {S\o{}rensen}(1999)}]{PhysRevLett.82.1835}%
  \BibitemOpen
  \bibfield  {author} {\bibinfo {author} {\bibfnamefont {K.}~\bibnamefont {M\o{}lmer}}\ and\ \bibinfo {author} {\bibfnamefont {A.}~\bibnamefont {S\o{}rensen}},\ }\bibfield  {title} {\bibinfo {title} {Multiparticle entanglement of hot trapped ions},\ }\href {https://doi.org/10.1103/PhysRevLett.82.1835} {\bibfield  {journal} {\bibinfo  {journal} {Phys. Rev. Lett.}\ }\textbf {\bibinfo {volume} {82}},\ \bibinfo {pages} {1835} (\bibinfo {year} {1999})}\BibitemShut {NoStop}%
\bibitem [{\citenamefont {Tan}\ \emph {et~al.}(2013)\citenamefont {Tan}, \citenamefont {Gaebler}, \citenamefont {Bowler}, \citenamefont {Lin}, \citenamefont {Jost}, \citenamefont {Leibfried},\ and\ \citenamefont {Wineland}}]{PhysRevLett.110.263002}%
  \BibitemOpen
  \bibfield  {author} {\bibinfo {author} {\bibfnamefont {T.~R.}\ \bibnamefont {Tan}}, \bibinfo {author} {\bibfnamefont {J.~P.}\ \bibnamefont {Gaebler}}, \bibinfo {author} {\bibfnamefont {R.}~\bibnamefont {Bowler}}, \bibinfo {author} {\bibfnamefont {Y.}~\bibnamefont {Lin}}, \bibinfo {author} {\bibfnamefont {J.~D.}\ \bibnamefont {Jost}}, \bibinfo {author} {\bibfnamefont {D.}~\bibnamefont {Leibfried}},\ and\ \bibinfo {author} {\bibfnamefont {D.~J.}\ \bibnamefont {Wineland}},\ }\bibfield  {title} {\bibinfo {title} {Demonstration of a dressed-state phase gate for trapped ions},\ }\href {https://doi.org/10.1103/PhysRevLett.110.263002} {\bibfield  {journal} {\bibinfo  {journal} {Phys. Rev. Lett.}\ }\textbf {\bibinfo {volume} {110}},\ \bibinfo {pages} {263002} (\bibinfo {year} {2013})}\BibitemShut {NoStop}%
\bibitem [{\citenamefont {Clements}\ \emph {et~al.}(2020)\citenamefont {Clements}, \citenamefont {Kim}, \citenamefont {Cui}, \citenamefont {Hankin}, \citenamefont {Brewer}, \citenamefont {Valencia}, \citenamefont {Chen}, \citenamefont {Chou}, \citenamefont {Leibrandt},\ and\ \citenamefont {Hume}}]{HumeLifetimeLimited}%
  \BibitemOpen
  \bibfield  {author} {\bibinfo {author} {\bibfnamefont {E.~R.}\ \bibnamefont {Clements}}, \bibinfo {author} {\bibfnamefont {M.~E.}\ \bibnamefont {Kim}}, \bibinfo {author} {\bibfnamefont {K.}~\bibnamefont {Cui}}, \bibinfo {author} {\bibfnamefont {A.~M.}\ \bibnamefont {Hankin}}, \bibinfo {author} {\bibfnamefont {S.~M.}\ \bibnamefont {Brewer}}, \bibinfo {author} {\bibfnamefont {J.}~\bibnamefont {Valencia}}, \bibinfo {author} {\bibfnamefont {J.-S.}\ \bibnamefont {Chen}}, \bibinfo {author} {\bibfnamefont {C.-W.}\ \bibnamefont {Chou}}, \bibinfo {author} {\bibfnamefont {D.~R.}\ \bibnamefont {Leibrandt}},\ and\ \bibinfo {author} {\bibfnamefont {D.~B.}\ \bibnamefont {Hume}},\ }\bibfield  {title} {\bibinfo {title} {Lifetime-limited interrogation of two independent $^{27}{\mathrm{al}}^{+}$ clocks using correlation spectroscopy},\ }\href {https://doi.org/10.1103/PhysRevLett.125.243602} {\bibfield  {journal} {\bibinfo  {journal} {Phys. Rev. Lett.}\ }\textbf {\bibinfo {volume} {125}},\ \bibinfo {pages} {243602} (\bibinfo {year} {2020})}\BibitemShut {NoStop}%
\bibitem [{\citenamefont {G{\"u}hne}\ and\ \citenamefont {Hyllus}(2003)}]{Guhne2003}%
  \BibitemOpen
  \bibfield  {author} {\bibinfo {author} {\bibfnamefont {O.}~\bibnamefont {G{\"u}hne}}\ and\ \bibinfo {author} {\bibfnamefont {P.}~\bibnamefont {Hyllus}},\ }\bibfield  {title} {\bibinfo {title} {Investigating three qubit entanglement with local measurements},\ }\href {https://doi.org/10.1023/A:1025422606845} {\bibfield  {journal} {\bibinfo  {journal} {International Journal of Theoretical Physics}\ }\textbf {\bibinfo {volume} {42}},\ \bibinfo {pages} {1001} (\bibinfo {year} {2003})}\BibitemShut {NoStop}%
\bibitem [{\citenamefont {Moses}\ \emph {et~al.}(2023)\citenamefont {Moses}, \citenamefont {Baldwin}, \citenamefont {Allman}, \citenamefont {Ancona}, \citenamefont {Ascarrunz}, \citenamefont {Barnes}, \citenamefont {Bartolotta}, \citenamefont {Bjork}, \citenamefont {Blanchard}, \citenamefont {Bohn} \emph {et~al.}}]{PhysRevX.13.041052}%
  \BibitemOpen
  \bibfield  {author} {\bibinfo {author} {\bibfnamefont {S.~A.}\ \bibnamefont {Moses}}, \bibinfo {author} {\bibfnamefont {C.~H.}\ \bibnamefont {Baldwin}}, \bibinfo {author} {\bibfnamefont {M.~S.}\ \bibnamefont {Allman}}, \bibinfo {author} {\bibfnamefont {R.}~\bibnamefont {Ancona}}, \bibinfo {author} {\bibfnamefont {L.}~\bibnamefont {Ascarrunz}}, \bibinfo {author} {\bibfnamefont {C.}~\bibnamefont {Barnes}}, \bibinfo {author} {\bibfnamefont {J.}~\bibnamefont {Bartolotta}}, \bibinfo {author} {\bibfnamefont {B.}~\bibnamefont {Bjork}}, \bibinfo {author} {\bibfnamefont {P.}~\bibnamefont {Blanchard}}, \bibinfo {author} {\bibfnamefont {M.}~\bibnamefont {Bohn}}, \emph {et~al.},\ }\bibfield  {title} {\bibinfo {title} {A race-track trapped-ion quantum processor},\ }\href {https://doi.org/10.1103/PhysRevX.13.041052} {\bibfield  {journal} {\bibinfo  {journal} {Phys. Rev. X}\ }\textbf {\bibinfo {volume} {13}},\ \bibinfo {pages} {041052} (\bibinfo {year} {2023})}\BibitemShut {NoStop}%
\bibitem [{\citenamefont {Bruzewicz}\ \emph {et~al.}(2019)\citenamefont {Bruzewicz}, \citenamefont {Chiaverini}, \citenamefont {McConnell},\ and\ \citenamefont {Sage}}]{Bruzewicz2019}%
  \BibitemOpen
  \bibfield  {author} {\bibinfo {author} {\bibfnamefont {C.~D.}\ \bibnamefont {Bruzewicz}}, \bibinfo {author} {\bibfnamefont {J.}~\bibnamefont {Chiaverini}}, \bibinfo {author} {\bibfnamefont {R.}~\bibnamefont {McConnell}},\ and\ \bibinfo {author} {\bibfnamefont {J.~M.}\ \bibnamefont {Sage}},\ }\bibfield  {title} {\bibinfo {title} {Trapped-ion quantum computing: Progress and challenges},\ }\href {https://doi.org/10.1063/1.5088164} {\bibfield  {journal} {\bibinfo  {journal} {Applied Physics Reviews}\ }\textbf {\bibinfo {volume} {6}},\ \bibinfo {pages} {021314} (\bibinfo {year} {2019})}\BibitemShut {NoStop}%
\bibitem [{\citenamefont {Bluvstein}\ \emph {et~al.}(2022)\citenamefont {Bluvstein}, \citenamefont {Levine}, \citenamefont {Semeghini}, \citenamefont {Wang}, \citenamefont {Ebadi}, \citenamefont {Kalinowski}, \citenamefont {Keesling}, \citenamefont {Maskara}, \citenamefont {Pichler}, \citenamefont {Greiner}, \citenamefont {Vuleti{\'{c}}},\ and\ \citenamefont {Lukin}}]{Bluvstein2022}%
  \BibitemOpen
  \bibfield  {author} {\bibinfo {author} {\bibfnamefont {D.}~\bibnamefont {Bluvstein}}, \bibinfo {author} {\bibfnamefont {H.}~\bibnamefont {Levine}}, \bibinfo {author} {\bibfnamefont {G.}~\bibnamefont {Semeghini}}, \bibinfo {author} {\bibfnamefont {T.~T.}\ \bibnamefont {Wang}}, \bibinfo {author} {\bibfnamefont {S.}~\bibnamefont {Ebadi}}, \bibinfo {author} {\bibfnamefont {M.}~\bibnamefont {Kalinowski}}, \bibinfo {author} {\bibfnamefont {A.}~\bibnamefont {Keesling}}, \bibinfo {author} {\bibfnamefont {N.}~\bibnamefont {Maskara}}, \bibinfo {author} {\bibfnamefont {H.}~\bibnamefont {Pichler}}, \bibinfo {author} {\bibfnamefont {M.}~\bibnamefont {Greiner}}, \bibinfo {author} {\bibfnamefont {V.}~\bibnamefont {Vuleti{\'{c}}}},\ and\ \bibinfo {author} {\bibfnamefont {M.~D.}\ \bibnamefont {Lukin}},\ }\bibfield  {title} {\bibinfo {title} {A quantum processor based on coherent transport of entangled atom arrays},\ }\href {https://doi.org/10.1038/s41586-022-04592-6} {\bibfield  {journal} {\bibinfo  {journal} {Nature}\ }\textbf {\bibinfo {volume} {604}},\ \bibinfo {pages} {451} (\bibinfo {year} {2022})}\BibitemShut {NoStop}%
\bibitem [{\citenamefont {Kjaergaard}\ \emph {et~al.}(2020)\citenamefont {Kjaergaard}, \citenamefont {Schwartz}, \citenamefont {Braumüller}, \citenamefont {Krantz}, \citenamefont {Wang}, \citenamefont {Gustavsson},\ and\ \citenamefont {Oliver}}]{superscompsreview}%
  \BibitemOpen
  \bibfield  {author} {\bibinfo {author} {\bibfnamefont {M.}~\bibnamefont {Kjaergaard}}, \bibinfo {author} {\bibfnamefont {M.~E.}\ \bibnamefont {Schwartz}}, \bibinfo {author} {\bibfnamefont {J.}~\bibnamefont {Braumüller}}, \bibinfo {author} {\bibfnamefont {P.}~\bibnamefont {Krantz}}, \bibinfo {author} {\bibfnamefont {J.~I.-J.}\ \bibnamefont {Wang}}, \bibinfo {author} {\bibfnamefont {S.}~\bibnamefont {Gustavsson}},\ and\ \bibinfo {author} {\bibfnamefont {W.~D.}\ \bibnamefont {Oliver}},\ }\bibfield  {title} {\bibinfo {title} {Superconducting qubits: Current state of play},\ }\href {https://doi.org/https://doi.org/10.1146/annurev-conmatphys-031119-050605} {\bibfield  {journal} {\bibinfo  {journal} {Annual Review of Condensed Matter Physics}\ }\textbf {\bibinfo {volume} {11}},\ \bibinfo {pages} {369} (\bibinfo {year} {2020})}\BibitemShut {NoStop}%
\bibitem [{\citenamefont {van Leent}\ \emph {et~al.}(2022)\citenamefont {van Leent}, \citenamefont {Bock}, \citenamefont {Fertig}, \citenamefont {Garthoff}, \citenamefont {Eppelt}, \citenamefont {Zhou}, \citenamefont {Malik}, \citenamefont {Seubert}, \citenamefont {Bauer}, \citenamefont {Rosenfeld}, \citenamefont {Zhang}, \citenamefont {Becher},\ and\ \citenamefont {Weinfurter}}]{vanLeent2022}%
  \BibitemOpen
  \bibfield  {author} {\bibinfo {author} {\bibfnamefont {T.}~\bibnamefont {van Leent}}, \bibinfo {author} {\bibfnamefont {M.}~\bibnamefont {Bock}}, \bibinfo {author} {\bibfnamefont {F.}~\bibnamefont {Fertig}}, \bibinfo {author} {\bibfnamefont {R.}~\bibnamefont {Garthoff}}, \bibinfo {author} {\bibfnamefont {S.}~\bibnamefont {Eppelt}}, \bibinfo {author} {\bibfnamefont {Y.}~\bibnamefont {Zhou}}, \bibinfo {author} {\bibfnamefont {P.}~\bibnamefont {Malik}}, \bibinfo {author} {\bibfnamefont {M.}~\bibnamefont {Seubert}}, \bibinfo {author} {\bibfnamefont {T.}~\bibnamefont {Bauer}}, \bibinfo {author} {\bibfnamefont {W.}~\bibnamefont {Rosenfeld}}, \bibinfo {author} {\bibfnamefont {W.}~\bibnamefont {Zhang}}, \bibinfo {author} {\bibfnamefont {C.}~\bibnamefont {Becher}},\ and\ \bibinfo {author} {\bibfnamefont {H.}~\bibnamefont {Weinfurter}},\ }\bibfield  {title} {\bibinfo {title} {Entangling single atoms over 33{\thinspace}km telecom fibre},\ }\href {https://doi.org/10.1038/s41586-022-04764-4} {\bibfield  {journal} {\bibinfo  {journal} {Nature}\ }\textbf {\bibinfo {volume} {607}},\ \bibinfo {pages} {69} (\bibinfo {year} {2022})}\BibitemShut {NoStop}%
\bibitem [{\citenamefont {Pompili}\ \emph {et~al.}(2021)\citenamefont {Pompili}, \citenamefont {Hermans}, \citenamefont {Baier}, \citenamefont {Beukers}, \citenamefont {Humphreys}, \citenamefont {Schouten}, \citenamefont {Vermeulen}, \citenamefont {Tiggelman}, \citenamefont {dos Santos~Martins}, \citenamefont {Dirkse}, \citenamefont {Wehner},\ and\ \citenamefont {Hanson}}]{Pompili2021}%
  \BibitemOpen
  \bibfield  {author} {\bibinfo {author} {\bibfnamefont {M.}~\bibnamefont {Pompili}}, \bibinfo {author} {\bibfnamefont {S.~L.~N.}\ \bibnamefont {Hermans}}, \bibinfo {author} {\bibfnamefont {S.}~\bibnamefont {Baier}}, \bibinfo {author} {\bibfnamefont {H.~K.~C.}\ \bibnamefont {Beukers}}, \bibinfo {author} {\bibfnamefont {P.~C.}\ \bibnamefont {Humphreys}}, \bibinfo {author} {\bibfnamefont {R.~N.}\ \bibnamefont {Schouten}}, \bibinfo {author} {\bibfnamefont {R.~F.~L.}\ \bibnamefont {Vermeulen}}, \bibinfo {author} {\bibfnamefont {M.~J.}\ \bibnamefont {Tiggelman}}, \bibinfo {author} {\bibfnamefont {L.}~\bibnamefont {dos Santos~Martins}}, \bibinfo {author} {\bibfnamefont {B.}~\bibnamefont {Dirkse}}, \bibinfo {author} {\bibfnamefont {S.}~\bibnamefont {Wehner}},\ and\ \bibinfo {author} {\bibfnamefont {R.}~\bibnamefont {Hanson}},\ }\bibfield  {title} {\bibinfo {title} {Realization of a multinode quantum network of remote solid-state qubits},\ }\href {https://doi.org/10.1126/science.abg1919} {\bibfield  {journal} {\bibinfo  {journal} {Science}\ }\textbf {\bibinfo {volume} {372}},\ \bibinfo {pages} {259} (\bibinfo {year} {2021})}\BibitemShut {NoStop}%
\bibitem [{\citenamefont {Moehring}\ \emph {et~al.}(2007)\citenamefont {Moehring}, \citenamefont {Maunz}, \citenamefont {Olmschenk}, \citenamefont {Younge}, \citenamefont {Matsukevich}, \citenamefont {Duan},\ and\ \citenamefont {Monroe}}]{Moehring2007}%
  \BibitemOpen
  \bibfield  {author} {\bibinfo {author} {\bibfnamefont {D.~L.}\ \bibnamefont {Moehring}}, \bibinfo {author} {\bibfnamefont {P.}~\bibnamefont {Maunz}}, \bibinfo {author} {\bibfnamefont {S.}~\bibnamefont {Olmschenk}}, \bibinfo {author} {\bibfnamefont {K.~C.}\ \bibnamefont {Younge}}, \bibinfo {author} {\bibfnamefont {D.~N.}\ \bibnamefont {Matsukevich}}, \bibinfo {author} {\bibfnamefont {L.-M.}\ \bibnamefont {Duan}},\ and\ \bibinfo {author} {\bibfnamefont {C.}~\bibnamefont {Monroe}},\ }\bibfield  {title} {\bibinfo {title} {Entanglement of single-atom quantum bits at a distance},\ }\href {https://doi.org/10.1038/nature06118} {\bibfield  {journal} {\bibinfo  {journal} {Nature}\ }\textbf {\bibinfo {volume} {449}},\ \bibinfo {pages} {68} (\bibinfo {year} {2007})}\BibitemShut {NoStop}%
\bibitem [{\citenamefont {Stephenson}\ \emph {et~al.}(2020)\citenamefont {Stephenson}, \citenamefont {Nadlinger}, \citenamefont {Nichol}, \citenamefont {An}, \citenamefont {Drmota}, \citenamefont {Ballance}, \citenamefont {Thirumalai}, \citenamefont {Goodwin}, \citenamefont {Lucas},\ and\ \citenamefont {Ballance}}]{PhysRevLett.124.110501}%
  \BibitemOpen
  \bibfield  {author} {\bibinfo {author} {\bibfnamefont {L.~J.}\ \bibnamefont {Stephenson}}, \bibinfo {author} {\bibfnamefont {D.~P.}\ \bibnamefont {Nadlinger}}, \bibinfo {author} {\bibfnamefont {B.~C.}\ \bibnamefont {Nichol}}, \bibinfo {author} {\bibfnamefont {S.}~\bibnamefont {An}}, \bibinfo {author} {\bibfnamefont {P.}~\bibnamefont {Drmota}}, \bibinfo {author} {\bibfnamefont {T.~G.}\ \bibnamefont {Ballance}}, \bibinfo {author} {\bibfnamefont {K.}~\bibnamefont {Thirumalai}}, \bibinfo {author} {\bibfnamefont {J.~F.}\ \bibnamefont {Goodwin}}, \bibinfo {author} {\bibfnamefont {D.~M.}\ \bibnamefont {Lucas}},\ and\ \bibinfo {author} {\bibfnamefont {C.~J.}\ \bibnamefont {Ballance}},\ }\bibfield  {title} {\bibinfo {title} {High-rate, high-fidelity entanglement of qubits across an elementary quantum network},\ }\href {https://doi.org/10.1103/PhysRevLett.124.110501} {\bibfield  {journal} {\bibinfo  {journal} {Phys. Rev. Lett.}\ }\textbf {\bibinfo {volume} {124}},\ \bibinfo {pages} {110501} (\bibinfo {year} {2020})}\BibitemShut {NoStop}%
\bibitem [{\citenamefont {Ritter}\ \emph {et~al.}(2012)\citenamefont {Ritter}, \citenamefont {N{\"o}lleke}, \citenamefont {Hahn}, \citenamefont {Reiserer}, \citenamefont {Neuzner}, \citenamefont {Uphoff}, \citenamefont {M{\"u}cke}, \citenamefont {Figueroa}, \citenamefont {Bochmann},\ and\ \citenamefont {Rempe}}]{Ritter2012}%
  \BibitemOpen
  \bibfield  {author} {\bibinfo {author} {\bibfnamefont {S.}~\bibnamefont {Ritter}}, \bibinfo {author} {\bibfnamefont {C.}~\bibnamefont {N{\"o}lleke}}, \bibinfo {author} {\bibfnamefont {C.}~\bibnamefont {Hahn}}, \bibinfo {author} {\bibfnamefont {A.}~\bibnamefont {Reiserer}}, \bibinfo {author} {\bibfnamefont {A.}~\bibnamefont {Neuzner}}, \bibinfo {author} {\bibfnamefont {M.}~\bibnamefont {Uphoff}}, \bibinfo {author} {\bibfnamefont {M.}~\bibnamefont {M{\"u}cke}}, \bibinfo {author} {\bibfnamefont {E.}~\bibnamefont {Figueroa}}, \bibinfo {author} {\bibfnamefont {J.}~\bibnamefont {Bochmann}},\ and\ \bibinfo {author} {\bibfnamefont {G.}~\bibnamefont {Rempe}},\ }\bibfield  {title} {\bibinfo {title} {An elementary quantum network of single atoms in optical cavities},\ }\href {https://doi.org/10.1038/nature11023} {\bibfield  {journal} {\bibinfo  {journal} {Nature}\ }\textbf {\bibinfo {volume} {484}},\ \bibinfo {pages} {195} (\bibinfo {year} {2012})}\BibitemShut {NoStop}%
\bibitem [{\citenamefont {Delteil}\ \emph {et~al.}(2016)\citenamefont {Delteil}, \citenamefont {Sun}, \citenamefont {Gao}, \citenamefont {Togan}, \citenamefont {Faelt},\ and\ \citenamefont {Imamo{\u{g}}lu}}]{Delteil2016}%
  \BibitemOpen
  \bibfield  {author} {\bibinfo {author} {\bibfnamefont {A.}~\bibnamefont {Delteil}}, \bibinfo {author} {\bibfnamefont {Z.}~\bibnamefont {Sun}}, \bibinfo {author} {\bibfnamefont {W.-b.}\ \bibnamefont {Gao}}, \bibinfo {author} {\bibfnamefont {E.}~\bibnamefont {Togan}}, \bibinfo {author} {\bibfnamefont {S.}~\bibnamefont {Faelt}},\ and\ \bibinfo {author} {\bibfnamefont {A.}~\bibnamefont {Imamo{\u{g}}lu}},\ }\bibfield  {title} {\bibinfo {title} {Generation of heralded entanglement between distant hole spins},\ }\href {https://doi.org/10.1038/nphys3605} {\bibfield  {journal} {\bibinfo  {journal} {Nature Physics}\ }\textbf {\bibinfo {volume} {12}},\ \bibinfo {pages} {218} (\bibinfo {year} {2016})}\BibitemShut {NoStop}%
\bibitem [{\citenamefont {Stockill}\ \emph {et~al.}(2017)\citenamefont {Stockill}, \citenamefont {Stanley}, \citenamefont {Huthmacher}, \citenamefont {Clarke}, \citenamefont {Hugues}, \citenamefont {Miller}, \citenamefont {Matthiesen}, \citenamefont {Le~Gall},\ and\ \citenamefont {Atat\"ure}}]{PhysRevLett.119.010503}%
  \BibitemOpen
  \bibfield  {author} {\bibinfo {author} {\bibfnamefont {R.}~\bibnamefont {Stockill}}, \bibinfo {author} {\bibfnamefont {M.~J.}\ \bibnamefont {Stanley}}, \bibinfo {author} {\bibfnamefont {L.}~\bibnamefont {Huthmacher}}, \bibinfo {author} {\bibfnamefont {E.}~\bibnamefont {Clarke}}, \bibinfo {author} {\bibfnamefont {M.}~\bibnamefont {Hugues}}, \bibinfo {author} {\bibfnamefont {A.~J.}\ \bibnamefont {Miller}}, \bibinfo {author} {\bibfnamefont {C.}~\bibnamefont {Matthiesen}}, \bibinfo {author} {\bibfnamefont {C.}~\bibnamefont {Le~Gall}},\ and\ \bibinfo {author} {\bibfnamefont {M.}~\bibnamefont {Atat\"ure}},\ }\bibfield  {title} {\bibinfo {title} {Phase-tuned entangled state generation between distant spin qubits},\ }\href {https://doi.org/10.1103/PhysRevLett.119.010503} {\bibfield  {journal} {\bibinfo  {journal} {Phys. Rev. Lett.}\ }\textbf {\bibinfo {volume} {119}},\ \bibinfo {pages} {010503} (\bibinfo {year} {2017})}\BibitemShut {NoStop}%
\bibitem [{\citenamefont {Magnard}\ \emph {et~al.}(2020)\citenamefont {Magnard}, \citenamefont {Storz}, \citenamefont {Kurpiers}, \citenamefont {Sch\"ar}, \citenamefont {Marxer}, \citenamefont {L\"utolf}, \citenamefont {Walter}, \citenamefont {Besse}, \citenamefont {Gabureac}, \citenamefont {Reuer}, \citenamefont {Akin}, \citenamefont {Royer}, \citenamefont {Blais},\ and\ \citenamefont {Wallraff}}]{PhysRevLett.125.260502}%
  \BibitemOpen
  \bibfield  {author} {\bibinfo {author} {\bibfnamefont {P.}~\bibnamefont {Magnard}}, \bibinfo {author} {\bibfnamefont {S.}~\bibnamefont {Storz}}, \bibinfo {author} {\bibfnamefont {P.}~\bibnamefont {Kurpiers}}, \bibinfo {author} {\bibfnamefont {J.}~\bibnamefont {Sch\"ar}}, \bibinfo {author} {\bibfnamefont {F.}~\bibnamefont {Marxer}}, \bibinfo {author} {\bibfnamefont {J.}~\bibnamefont {L\"utolf}}, \bibinfo {author} {\bibfnamefont {T.}~\bibnamefont {Walter}}, \bibinfo {author} {\bibfnamefont {J.-C.}\ \bibnamefont {Besse}}, \bibinfo {author} {\bibfnamefont {M.}~\bibnamefont {Gabureac}}, \bibinfo {author} {\bibfnamefont {K.}~\bibnamefont {Reuer}}, \bibinfo {author} {\bibfnamefont {A.}~\bibnamefont {Akin}}, \bibinfo {author} {\bibfnamefont {B.}~\bibnamefont {Royer}}, \bibinfo {author} {\bibfnamefont {A.}~\bibnamefont {Blais}},\ and\ \bibinfo {author} {\bibfnamefont {A.}~\bibnamefont {Wallraff}},\ }\bibfield  {title} {\bibinfo {title} {Microwave quantum link between superconducting circuits housed in spatially separated cryogenic systems},\ }\href {https://doi.org/10.1103/PhysRevLett.125.260502} {\bibfield  {journal} {\bibinfo  {journal} {Phys. Rev. Lett.}\ }\textbf {\bibinfo {volume} {125}},\ \bibinfo {pages} {260502} (\bibinfo {year} {2020})}\BibitemShut {NoStop}%
\bibitem [{\citenamefont {Liu}\ \emph {et~al.}(2024)\citenamefont {Liu}, \citenamefont {Luo}, \citenamefont {Yu}, \citenamefont {Wang}, \citenamefont {Wang}, \citenamefont {Hu}, \citenamefont {Li}, \citenamefont {Zheng}, \citenamefont {Yao}, \citenamefont {Yan} \emph {et~al.}}]{Liu2024}%
  \BibitemOpen
  \bibfield  {author} {\bibinfo {author} {\bibfnamefont {J.-L.}\ \bibnamefont {Liu}}, \bibinfo {author} {\bibfnamefont {X.-Y.}\ \bibnamefont {Luo}}, \bibinfo {author} {\bibfnamefont {Y.}~\bibnamefont {Yu}}, \bibinfo {author} {\bibfnamefont {C.-Y.}\ \bibnamefont {Wang}}, \bibinfo {author} {\bibfnamefont {B.}~\bibnamefont {Wang}}, \bibinfo {author} {\bibfnamefont {Y.}~\bibnamefont {Hu}}, \bibinfo {author} {\bibfnamefont {J.}~\bibnamefont {Li}}, \bibinfo {author} {\bibfnamefont {M.-Y.}\ \bibnamefont {Zheng}}, \bibinfo {author} {\bibfnamefont {B.}~\bibnamefont {Yao}}, \bibinfo {author} {\bibfnamefont {Z.}~\bibnamefont {Yan}}, \emph {et~al.},\ }\bibfield  {title} {\bibinfo {title} {Creation of memory--memory entanglement in a metropolitan quantum network},\ }\href {https://doi.org/10.1038/s41586-024-07308-0} {\bibfield  {journal} {\bibinfo  {journal} {Nature}\ }\textbf {\bibinfo {volume} {629}},\ \bibinfo {pages} {579} (\bibinfo {year} {2024})}\BibitemShut {NoStop}%
\bibitem [{\citenamefont {Knaut}\ \emph {et~al.}(2024)\citenamefont {Knaut}, \citenamefont {Suleymanzade}, \citenamefont {Wei}, \citenamefont {Assumpcao}, \citenamefont {Stas}, \citenamefont {Huan}, \citenamefont {Machielse}, \citenamefont {Knall}, \citenamefont {Sutula}, \citenamefont {Baranes}, \citenamefont {Lukin} \emph {et~al.}}]{Knaut2024}%
  \BibitemOpen
  \bibfield  {author} {\bibinfo {author} {\bibfnamefont {C.~M.}\ \bibnamefont {Knaut}}, \bibinfo {author} {\bibfnamefont {A.}~\bibnamefont {Suleymanzade}}, \bibinfo {author} {\bibfnamefont {Y.-C.}\ \bibnamefont {Wei}}, \bibinfo {author} {\bibfnamefont {D.~R.}\ \bibnamefont {Assumpcao}}, \bibinfo {author} {\bibfnamefont {P.-J.}\ \bibnamefont {Stas}}, \bibinfo {author} {\bibfnamefont {Y.~Q.}\ \bibnamefont {Huan}}, \bibinfo {author} {\bibfnamefont {B.}~\bibnamefont {Machielse}}, \bibinfo {author} {\bibfnamefont {E.~N.}\ \bibnamefont {Knall}}, \bibinfo {author} {\bibfnamefont {M.}~\bibnamefont {Sutula}}, \bibinfo {author} {\bibfnamefont {G.}~\bibnamefont {Baranes}}, \bibinfo {author} {\bibfnamefont {M.~D.}\ \bibnamefont {Lukin}}, \emph {et~al.},\ }\bibfield  {title} {\bibinfo {title} {Entanglement of nanophotonic quantum memory nodes in a telecom network},\ }\href {https://doi.org/10.1038/s41586-024-07252-z} {\bibfield  {journal} {\bibinfo  {journal} {Nature}\ }\textbf {\bibinfo {volume} {629}},\ \bibinfo {pages} {573} (\bibinfo {year} {2024})}\BibitemShut {NoStop}%
\bibitem [{\citenamefont {Krutyanskiy}\ \emph {et~al.}(2023{\natexlab{a}})\citenamefont {Krutyanskiy}, \citenamefont {Galli}, \citenamefont {Krcmarsky}, \citenamefont {Baier}, \citenamefont {Fioretto}, \citenamefont {Pu}, \citenamefont {Mazloom}, \citenamefont {Sekatski}, \citenamefont {Canteri}, \citenamefont {Teller}, \citenamefont {Schupp}, \citenamefont {Bate}, \citenamefont {Meraner}, \citenamefont {Sangouard}, \citenamefont {Lanyon},\ and\ \citenamefont {Northup}}]{PhysRevLett.130.050803}%
  \BibitemOpen
  \bibfield  {author} {\bibinfo {author} {\bibfnamefont {V.}~\bibnamefont {Krutyanskiy}}, \bibinfo {author} {\bibfnamefont {M.}~\bibnamefont {Galli}}, \bibinfo {author} {\bibfnamefont {V.}~\bibnamefont {Krcmarsky}}, \bibinfo {author} {\bibfnamefont {S.}~\bibnamefont {Baier}}, \bibinfo {author} {\bibfnamefont {D.~A.}\ \bibnamefont {Fioretto}}, \bibinfo {author} {\bibfnamefont {Y.}~\bibnamefont {Pu}}, \bibinfo {author} {\bibfnamefont {A.}~\bibnamefont {Mazloom}}, \bibinfo {author} {\bibfnamefont {P.}~\bibnamefont {Sekatski}}, \bibinfo {author} {\bibfnamefont {M.}~\bibnamefont {Canteri}}, \bibinfo {author} {\bibfnamefont {M.}~\bibnamefont {Teller}}, \bibinfo {author} {\bibfnamefont {J.}~\bibnamefont {Schupp}}, \bibinfo {author} {\bibfnamefont {J.}~\bibnamefont {Bate}}, \bibinfo {author} {\bibfnamefont {M.}~\bibnamefont {Meraner}}, \bibinfo {author} {\bibfnamefont {N.}~\bibnamefont {Sangouard}}, \bibinfo {author} {\bibfnamefont {B.~P.}\ \bibnamefont {Lanyon}},\ and\ \bibinfo {author} {\bibfnamefont {T.~E.}\ \bibnamefont {Northup}},\ }\bibfield  {title} {\bibinfo {title} {Entanglement of trapped-ion qubits separated by 230 meters},\ }\href {https://doi.org/10.1103/PhysRevLett.130.050803} {\bibfield  {journal} {\bibinfo  {journal} {Phys. Rev. Lett.}\ }\textbf {\bibinfo {volume} {130}},\ \bibinfo {pages} {050803} (\bibinfo {year} {2023}{\natexlab{a}})}\BibitemShut {NoStop}%
\bibitem [{\citenamefont {Krutyanskiy}\ \emph {et~al.}(2024)\citenamefont {Krutyanskiy}, \citenamefont {Canteri}, \citenamefont {Meraner}, \citenamefont {Krcmarsky},\ and\ \citenamefont {Lanyon}}]{PRXQuantum.5.020308}%
  \BibitemOpen
  \bibfield  {author} {\bibinfo {author} {\bibfnamefont {V.}~\bibnamefont {Krutyanskiy}}, \bibinfo {author} {\bibfnamefont {M.}~\bibnamefont {Canteri}}, \bibinfo {author} {\bibfnamefont {M.}~\bibnamefont {Meraner}}, \bibinfo {author} {\bibfnamefont {V.}~\bibnamefont {Krcmarsky}},\ and\ \bibinfo {author} {\bibfnamefont {B.}~\bibnamefont {Lanyon}},\ }\bibfield  {title} {\bibinfo {title} {Multimode ion-photon entanglement over 101 kilometers},\ }\href {https://doi.org/10.1103/PRXQuantum.5.020308} {\bibfield  {journal} {\bibinfo  {journal} {PRX Quantum}\ }\textbf {\bibinfo {volume} {5}},\ \bibinfo {pages} {020308} (\bibinfo {year} {2024})}\BibitemShut {NoStop}%
\bibitem [{\citenamefont {Krutyanskiy}\ \emph {et~al.}(2023{\natexlab{b}})\citenamefont {Krutyanskiy}, \citenamefont {Canteri}, \citenamefont {Meraner}, \citenamefont {Bate}, \citenamefont {Krcmarsky}, \citenamefont {Schupp}, \citenamefont {Sangouard},\ and\ \citenamefont {Lanyon}}]{PhysRevLett.130.213601}%
  \BibitemOpen
  \bibfield  {author} {\bibinfo {author} {\bibfnamefont {V.}~\bibnamefont {Krutyanskiy}}, \bibinfo {author} {\bibfnamefont {M.}~\bibnamefont {Canteri}}, \bibinfo {author} {\bibfnamefont {M.}~\bibnamefont {Meraner}}, \bibinfo {author} {\bibfnamefont {J.}~\bibnamefont {Bate}}, \bibinfo {author} {\bibfnamefont {V.}~\bibnamefont {Krcmarsky}}, \bibinfo {author} {\bibfnamefont {J.}~\bibnamefont {Schupp}}, \bibinfo {author} {\bibfnamefont {N.}~\bibnamefont {Sangouard}},\ and\ \bibinfo {author} {\bibfnamefont {B.~P.}\ \bibnamefont {Lanyon}},\ }\bibfield  {title} {\bibinfo {title} {Telecom-wavelength quantum repeater node based on a trapped-ion processor},\ }\href {https://doi.org/10.1103/PhysRevLett.130.213601} {\bibfield  {journal} {\bibinfo  {journal} {Phys. Rev. Lett.}\ }\textbf {\bibinfo {volume} {130}},\ \bibinfo {pages} {213601} (\bibinfo {year} {2023}{\natexlab{b}})}\BibitemShut {NoStop}%
\bibitem [{\citenamefont {Drmota}\ \emph {et~al.}(2023)\citenamefont {Drmota}, \citenamefont {Main}, \citenamefont {Nadlinger}, \citenamefont {Nichol}, \citenamefont {Weber}, \citenamefont {Ainley}, \citenamefont {Agrawal}, \citenamefont {Srinivas}, \citenamefont {Araneda}, \citenamefont {Ballance},\ and\ \citenamefont {Lucas}}]{PhysRevLett.130.090803}%
  \BibitemOpen
  \bibfield  {author} {\bibinfo {author} {\bibfnamefont {P.}~\bibnamefont {Drmota}}, \bibinfo {author} {\bibfnamefont {D.}~\bibnamefont {Main}}, \bibinfo {author} {\bibfnamefont {D.~P.}\ \bibnamefont {Nadlinger}}, \bibinfo {author} {\bibfnamefont {B.~C.}\ \bibnamefont {Nichol}}, \bibinfo {author} {\bibfnamefont {M.~A.}\ \bibnamefont {Weber}}, \bibinfo {author} {\bibfnamefont {E.~M.}\ \bibnamefont {Ainley}}, \bibinfo {author} {\bibfnamefont {A.}~\bibnamefont {Agrawal}}, \bibinfo {author} {\bibfnamefont {R.}~\bibnamefont {Srinivas}}, \bibinfo {author} {\bibfnamefont {G.}~\bibnamefont {Araneda}}, \bibinfo {author} {\bibfnamefont {C.~J.}\ \bibnamefont {Ballance}},\ and\ \bibinfo {author} {\bibfnamefont {D.~M.}\ \bibnamefont {Lucas}},\ }\bibfield  {title} {\bibinfo {title} {Robust quantum memory in a trapped-ion quantum network node},\ }\href {https://doi.org/10.1103/PhysRevLett.130.090803} {\bibfield  {journal} {\bibinfo  {journal} {Phys. Rev. Lett.}\ }\textbf {\bibinfo {volume} {130}},\ \bibinfo {pages} {090803} (\bibinfo {year} {2023})}\BibitemShut {NoStop}%
\bibitem [{zen()}]{zenodo}%
  \BibitemOpen
  \href@noop {} {\bibinfo {title} {The authors will supply the zenodo repository identification number and title on acceptance for publication.}}\BibitemShut {Stop}%
\bibitem [{\citenamefont {Rao}(1992)}]{Rao1992}%
  \BibitemOpen
  \bibfield  {author} {\bibinfo {author} {\bibfnamefont {C.~R.}\ \bibnamefont {Rao}},\ }\bibfield  {title} {\bibinfo {title} {Information and the accuracy attainable in the estimation of statistical parameters},\ }in\ \href {https://doi.org/10.1007/978-1-4612-0919-5_16} {\emph {\bibinfo {booktitle} {Breakthroughs in Statistics: {{Foundations}} and Basic Theory}}},\ \bibinfo {editor} {edited by\ \bibinfo {editor} {\bibfnamefont {S.}~\bibnamefont {Kotz}}\ and\ \bibinfo {editor} {\bibfnamefont {N.~L.}\ \bibnamefont {Johnson}}}\ (\bibinfo  {publisher} {Springer New York},\ \bibinfo {address} {New York, NY},\ \bibinfo {year} {1992})\ pp.\ \bibinfo {pages} {235--247}\BibitemShut {NoStop}%
\bibitem [{\citenamefont {Cramér}(1946)}]{CramerHarald1946Mmos}%
  \BibitemOpen
  \bibfield  {author} {\bibinfo {author} {\bibfnamefont {H.}~\bibnamefont {Cramér}},\ }\href@noop {} {\emph {\bibinfo {title} {Mathematical methods of statistics}}},\ \bibinfo {edition} {1st}\ ed.,\ Princeton mathematical series\ (\bibinfo {year} {1946})\BibitemShut {NoStop}%
\bibitem [{\citenamefont {Fisher}\ and\ \citenamefont {Russell}(1922)}]{fisherMathematicalFoundationsTheoretical1922}%
  \BibitemOpen
  \bibfield  {author} {\bibinfo {author} {\bibfnamefont {R.~A.}\ \bibnamefont {Fisher}}\ and\ \bibinfo {author} {\bibfnamefont {E.~J.}\ \bibnamefont {Russell}},\ }\bibfield  {title} {\bibinfo {title} {On the mathematical foundations of theoretical statistics},\ }\href {https://doi.org/10.1098/rsta.1922.0009} {\bibfield  {journal} {\bibinfo  {journal} {Philosophical Transactions of the Royal Society of London. Series A, Containing Papers of a Mathematical or Physical Character}\ }\textbf {\bibinfo {volume} {222}},\ \bibinfo {pages} {309} (\bibinfo {year} {1922})}\BibitemShut {NoStop}%
\bibitem [{\citenamefont {Paris}(2009)}]{parisQuantumEstimationQuantum2009}%
  \BibitemOpen
  \bibfield  {author} {\bibinfo {author} {\bibfnamefont {M.~G.~A.}\ \bibnamefont {Paris}},\ }\bibfield  {title} {\bibinfo {title} {Quantum estimation for quantum technology},\ }\href {https://doi.org/10.1142/S0219749909004839} {\bibfield  {journal} {\bibinfo  {journal} {International Journal of Quantum Information}\ }\textbf {\bibinfo {volume} {07}},\ \bibinfo {pages} {125} (\bibinfo {year} {2009})}\BibitemShut {NoStop}%
\bibitem [{\citenamefont {Canteri}(2020)}]{MarcoThesis}%
  \BibitemOpen
  \bibfield  {author} {\bibinfo {author} {\bibfnamefont {M.}~\bibnamefont {Canteri}},\ }\emph {\bibinfo {title} {Single-atom-focused laser for photon generation and qubit control}},\ \href {https://quantumoptics.at/images/publications/diploma/master_Marco_Canteri.pdf} {Master's thesis},\ \bibinfo  {school} {University of Innsbruck} (\bibinfo {year} {2020})\BibitemShut {NoStop}%
\bibitem [{\citenamefont {James}(1998)}]{James1998}%
  \BibitemOpen
  \bibfield  {author} {\bibinfo {author} {\bibfnamefont {D.~F.~V.}\ \bibnamefont {James}},\ }\bibfield  {title} {\bibinfo {title} {Quantum dynamics of cold trapped ions with application to quantum computation},\ }\href {https://doi.org/10.1007/s003400050373} {\bibfield  {journal} {\bibinfo  {journal} {Applied Physics B}\ }\textbf {\bibinfo {volume} {66}},\ \bibinfo {pages} {181} (\bibinfo {year} {1998})}\BibitemShut {NoStop}%
\bibitem [{\citenamefont {Roos}(2000)}]{RoosThesis}%
  \BibitemOpen
  \bibfield  {author} {\bibinfo {author} {\bibfnamefont {C.}~\bibnamefont {Roos}},\ }\emph {\bibinfo {title} {Controlling the quantum state of trapped ions}},\ \href {https://www.quantumoptics.at/images/publications/dissertation/roos-diss.pdf} {Ph.D. thesis},\ \bibinfo  {school} {University of Innsbruck} (\bibinfo {year} {2000})\BibitemShut {NoStop}%
\bibitem [{\citenamefont {Ringbauer}\ \emph {et~al.}(2022)\citenamefont {Ringbauer}, \citenamefont {Meth}, \citenamefont {Postler}, \citenamefont {Stricker}, \citenamefont {Blatt}, \citenamefont {Schindler},\ and\ \citenamefont {Monz}}]{ringbauerUniversalQuditQuantum2022}%
  \BibitemOpen
  \bibfield  {author} {\bibinfo {author} {\bibfnamefont {M.}~\bibnamefont {Ringbauer}}, \bibinfo {author} {\bibfnamefont {M.}~\bibnamefont {Meth}}, \bibinfo {author} {\bibfnamefont {L.}~\bibnamefont {Postler}}, \bibinfo {author} {\bibfnamefont {R.}~\bibnamefont {Stricker}}, \bibinfo {author} {\bibfnamefont {R.}~\bibnamefont {Blatt}}, \bibinfo {author} {\bibfnamefont {P.}~\bibnamefont {Schindler}},\ and\ \bibinfo {author} {\bibfnamefont {T.}~\bibnamefont {Monz}},\ }\bibfield  {title} {\bibinfo {title} {A universal qudit quantum processor with trapped ions},\ }\href {https://doi.org/10.1038/s41567-022-01658-0} {\bibfield  {journal} {\bibinfo  {journal} {Nature Physics}\ }\textbf {\bibinfo {volume} {18}},\ \bibinfo {pages} {1053} (\bibinfo {year} {2022})}\BibitemShut {NoStop}%
\bibitem [{\citenamefont {Bertsch}()}]{Milena}%
  \BibitemOpen
  \bibfield  {author} {\bibinfo {author} {\bibfnamefont {M.~G.}\ \bibnamefont {Bertsch}},\ }\emph {\bibinfo {title} {Optical clocks with trapped ions}},\ \href {https://www.quantumoptics.at/images/publications/dissertation/Thesis_Milena_Guevara_Bertsch.pdf} {Ph.D. thesis},\ \bibinfo  {school} {University of Innsbruck}\BibitemShut {NoStop}%
\bibitem [{\citenamefont {Chessa}\ and\ \citenamefont {Giovannetti}(2021)}]{Chessa2021}%
  \BibitemOpen
  \bibfield  {author} {\bibinfo {author} {\bibfnamefont {S.}~\bibnamefont {Chessa}}\ and\ \bibinfo {author} {\bibfnamefont {V.}~\bibnamefont {Giovannetti}},\ }\bibfield  {title} {\bibinfo {title} {Quantum capacity analysis of multi-level amplitude damping channels},\ }\href {https://doi.org/10.1038/s42005-021-00524-4} {\bibfield  {journal} {\bibinfo  {journal} {Communications Physics}\ }\textbf {\bibinfo {volume} {4}},\ \bibinfo {pages} {22} (\bibinfo {year} {2021})}\BibitemShut {NoStop}%
\bibitem [{\citenamefont {Wilde}(2013)}]{Wilde_2013}%
  \BibitemOpen
  \bibfield  {author} {\bibinfo {author} {\bibfnamefont {M.~M.}\ \bibnamefont {Wilde}},\ }\href@noop {} {\emph {\bibinfo {title} {Quantum Information Theory}}}\ (\bibinfo  {publisher} {Cambridge University Press},\ \bibinfo {year} {2013})\ p.\ \bibinfo {pages} {138}\BibitemShut {NoStop}%
\bibitem [{\citenamefont {S\o{}rensen}\ and\ \citenamefont {M\o{}lmer}(2000)}]{PhysRevA.62.022311}%
  \BibitemOpen
  \bibfield  {author} {\bibinfo {author} {\bibfnamefont {A.}~\bibnamefont {S\o{}rensen}}\ and\ \bibinfo {author} {\bibfnamefont {K.}~\bibnamefont {M\o{}lmer}},\ }\bibfield  {title} {\bibinfo {title} {Entanglement and quantum computation with ions in thermal motion},\ }\href {https://doi.org/10.1103/PhysRevA.62.022311} {\bibfield  {journal} {\bibinfo  {journal} {Phys. Rev. A}\ }\textbf {\bibinfo {volume} {62}},\ \bibinfo {pages} {022311} (\bibinfo {year} {2000})}\BibitemShut {NoStop}%
\bibitem [{\citenamefont {Kirchmair}\ \emph {et~al.}(2009)\citenamefont {Kirchmair}, \citenamefont {Benhelm}, \citenamefont {Zähringer}, \citenamefont {Gerritsma}, \citenamefont {Roos},\ and\ \citenamefont {Blatt}}]{Kirchmair_2009}%
  \BibitemOpen
  \bibfield  {author} {\bibinfo {author} {\bibfnamefont {G.}~\bibnamefont {Kirchmair}}, \bibinfo {author} {\bibfnamefont {J.}~\bibnamefont {Benhelm}}, \bibinfo {author} {\bibfnamefont {F.}~\bibnamefont {Zähringer}}, \bibinfo {author} {\bibfnamefont {R.}~\bibnamefont {Gerritsma}}, \bibinfo {author} {\bibfnamefont {C.~F.}\ \bibnamefont {Roos}},\ and\ \bibinfo {author} {\bibfnamefont {R.}~\bibnamefont {Blatt}},\ }\bibfield  {title} {\bibinfo {title} {Deterministic entanglement of ions in thermal states of motion},\ }\href {https://doi.org/10.1088/1367-2630/11/2/023002} {\bibfield  {journal} {\bibinfo  {journal} {New Journal of Physics}\ }\textbf {\bibinfo {volume} {11}},\ \bibinfo {pages} {023002} (\bibinfo {year} {2009})}\BibitemShut {NoStop}%
\end{thebibliography}%


\begin{thebibliography}{20}%
\makeatletter
\providecommand \@ifxundefined [1]{%
 \@ifx{#1\undefined}
}%
\providecommand \@ifnum [1]{%
 \ifnum #1\expandafter \@firstoftwo
 \else \expandafter \@secondoftwo
 \fi
}%
\providecommand \@ifx [1]{%
 \ifx #1\expandafter \@firstoftwo
 \else \expandafter \@secondoftwo
 \fi
}%
\providecommand \natexlab [1]{#1}%
\providecommand \enquote  [1]{``#1''}%
\providecommand \bibnamefont  [1]{#1}%
\providecommand \bibfnamefont [1]{#1}%
\providecommand \citenamefont [1]{#1}%
\providecommand \href@noop [0]{\@secondoftwo}%
\providecommand \href [0]{\begingroup \@sanitize@url \@href}%
\providecommand \@href[1]{\@@startlink{#1}\@@href}%
\providecommand \@@href[1]{\endgroup#1\@@endlink}%
\providecommand \@sanitize@url [0]{\catcode `\\12\catcode `\$12\catcode `\&12\catcode `\#12\catcode `\^12\catcode `\_12\catcode `\%12\relax}%
\providecommand \@@startlink[1]{}%
\providecommand \@@endlink[0]{}%
\providecommand \url  [0]{\begingroup\@sanitize@url \@url }%
\providecommand \@url [1]{\endgroup\@href {#1}{\urlprefix }}%
\providecommand \urlprefix  [0]{URL }%
\providecommand \Eprint [0]{\href }%
\providecommand \doibase [0]{https://doi.org/}%
\providecommand \selectlanguage [0]{\@gobble}%
\providecommand \bibinfo  [0]{\@secondoftwo}%
\providecommand \bibfield  [0]{\@secondoftwo}%
\providecommand \translation [1]{[#1]}%
\providecommand \BibitemOpen [0]{}%
\providecommand \bibitemStop [0]{}%
\providecommand \bibitemNoStop [0]{.\EOS\space}%
\providecommand \EOS [0]{\spacefactor3000\relax}%
\providecommand \BibitemShut  [1]{\csname bibitem#1\endcsname}%
\let\auto@bib@innerbib\@empty
\bibitem [{\citenamefont {Canteri}(2020)}]{MarcoThesis}%
  \BibitemOpen
  \bibfield  {author} {\bibinfo {author} {\bibfnamefont {M.}~\bibnamefont {Canteri}},\ }\emph {\bibinfo {title} {Single-atom-focused laser for photon generation and qubit control}},\ \href {https://quantumoptics.at/images/publications/diploma/master_Marco_Canteri.pdf} {Master's thesis},\ \bibinfo  {school} {University of Innsbruck} (\bibinfo {year} {2020})\BibitemShut {NoStop}%
\bibitem [{\citenamefont {S\o{}rensen}\ and\ \citenamefont {M\o{}lmer}(2000)}]{PhysRevA.62.022311}%
  \BibitemOpen
  \bibfield  {author} {\bibinfo {author} {\bibfnamefont {A.}~\bibnamefont {S\o{}rensen}}\ and\ \bibinfo {author} {\bibfnamefont {K.}~\bibnamefont {M\o{}lmer}},\ }\bibfield  {title} {\bibinfo {title} {Entanglement and quantum computation with ions in thermal motion},\ }\href {https://doi.org/10.1103/PhysRevA.62.022311} {\bibfield  {journal} {\bibinfo  {journal} {Phys. Rev. A}\ }\textbf {\bibinfo {volume} {62}},\ \bibinfo {pages} {022311} (\bibinfo {year} {2000})}\BibitemShut {NoStop}%
\bibitem [{\citenamefont {Kirchmair}\ \emph {et~al.}(2009)\citenamefont {Kirchmair}, \citenamefont {Benhelm}, \citenamefont {Zähringer}, \citenamefont {Gerritsma}, \citenamefont {Roos},\ and\ \citenamefont {Blatt}}]{Kirchmair_2009}%
  \BibitemOpen
  \bibfield  {author} {\bibinfo {author} {\bibfnamefont {G.}~\bibnamefont {Kirchmair}}, \bibinfo {author} {\bibfnamefont {J.}~\bibnamefont {Benhelm}}, \bibinfo {author} {\bibfnamefont {F.}~\bibnamefont {Zähringer}}, \bibinfo {author} {\bibfnamefont {R.}~\bibnamefont {Gerritsma}}, \bibinfo {author} {\bibfnamefont {C.~F.}\ \bibnamefont {Roos}},\ and\ \bibinfo {author} {\bibfnamefont {R.}~\bibnamefont {Blatt}},\ }\bibfield  {title} {\bibinfo {title} {Deterministic entanglement of ions in thermal states of motion},\ }\href {https://doi.org/10.1088/1367-2630/11/2/023002} {\bibfield  {journal} {\bibinfo  {journal} {New Journal of Physics}\ }\textbf {\bibinfo {volume} {11}},\ \bibinfo {pages} {023002} (\bibinfo {year} {2009})}\BibitemShut {NoStop}%
\bibitem [{\citenamefont {Monz}\ \emph {et~al.}(2011)\citenamefont {Monz}, \citenamefont {Schindler}, \citenamefont {Barreiro}, \citenamefont {Chwalla}, \citenamefont {Nigg}, \citenamefont {Coish}, \citenamefont {Harlander}, \citenamefont {H\"ansel}, \citenamefont {Hennrich},\ and\ \citenamefont {Blatt}}]{PhysRevLett.106.130506}%
  \BibitemOpen
  \bibfield  {author} {\bibinfo {author} {\bibfnamefont {T.}~\bibnamefont {Monz}}, \bibinfo {author} {\bibfnamefont {P.}~\bibnamefont {Schindler}}, \bibinfo {author} {\bibfnamefont {J.~T.}\ \bibnamefont {Barreiro}}, \bibinfo {author} {\bibfnamefont {M.}~\bibnamefont {Chwalla}}, \bibinfo {author} {\bibfnamefont {D.}~\bibnamefont {Nigg}}, \bibinfo {author} {\bibfnamefont {W.~A.}\ \bibnamefont {Coish}}, \bibinfo {author} {\bibfnamefont {M.}~\bibnamefont {Harlander}}, \bibinfo {author} {\bibfnamefont {W.}~\bibnamefont {H\"ansel}}, \bibinfo {author} {\bibfnamefont {M.}~\bibnamefont {Hennrich}},\ and\ \bibinfo {author} {\bibfnamefont {R.}~\bibnamefont {Blatt}},\ }\bibfield  {title} {\bibinfo {title} {14-qubit entanglement: Creation and coherence},\ }\href {https://doi.org/10.1103/PhysRevLett.106.130506} {\bibfield  {journal} {\bibinfo  {journal} {Phys. Rev. Lett.}\ }\textbf {\bibinfo {volume} {106}},\ \bibinfo {pages} {130506} (\bibinfo {year} {2011})}\BibitemShut {NoStop}%
\bibitem [{\citenamefont {Wilde}(2013)}]{Wilde_2013}%
  \BibitemOpen
  \bibfield  {author} {\bibinfo {author} {\bibfnamefont {M.~M.}\ \bibnamefont {Wilde}},\ }\href@noop {} {\emph {\bibinfo {title} {Quantum Information Theory}}}\ (\bibinfo  {publisher} {Cambridge University Press},\ \bibinfo {year} {2013})\ p.\ \bibinfo {pages} {138}\BibitemShut {NoStop}%
\bibitem [{\citenamefont {Stroinski}(2024)}]{TabeaThesis}%
  \BibitemOpen
  \bibfield  {author} {\bibinfo {author} {\bibfnamefont {T.}~\bibnamefont {Stroinski}},\ }\emph {\bibinfo {title} {Implementation and Application of a Controlled-NOT Gate in a Trapped-Ion Quantum Network Node}},\ \href {https://quantumoptics.at/images/publications/diploma/master_Stroinski.pdf} {Master's thesis},\ \bibinfo  {school} {University of Innsbruck} (\bibinfo {year} {2024})\BibitemShut {NoStop}%
\bibitem [{\citenamefont {James}(1998)}]{James1998}%
  \BibitemOpen
  \bibfield  {author} {\bibinfo {author} {\bibfnamefont {D.~F.~V.}\ \bibnamefont {James}},\ }\bibfield  {title} {\bibinfo {title} {Quantum dynamics of cold trapped ions with application to quantum computation},\ }\href {https://doi.org/10.1007/s003400050373} {\bibfield  {journal} {\bibinfo  {journal} {Applied Physics B}\ }\textbf {\bibinfo {volume} {66}},\ \bibinfo {pages} {181} (\bibinfo {year} {1998})}\BibitemShut {NoStop}%
\bibitem [{\citenamefont {Chessa}\ and\ \citenamefont {Giovannetti}(2021)}]{Chessa2021}%
  \BibitemOpen
  \bibfield  {author} {\bibinfo {author} {\bibfnamefont {S.}~\bibnamefont {Chessa}}\ and\ \bibinfo {author} {\bibfnamefont {V.}~\bibnamefont {Giovannetti}},\ }\bibfield  {title} {\bibinfo {title} {Quantum capacity analysis of multi-level amplitude damping channels},\ }\href {https://doi.org/10.1038/s42005-021-00524-4} {\bibfield  {journal} {\bibinfo  {journal} {Communications Physics}\ }\textbf {\bibinfo {volume} {4}},\ \bibinfo {pages} {22} (\bibinfo {year} {2021})}\BibitemShut {NoStop}%
\bibitem [{\citenamefont {Roos}(2000)}]{RoosThesis}%
  \BibitemOpen
  \bibfield  {author} {\bibinfo {author} {\bibfnamefont {C.}~\bibnamefont {Roos}},\ }\emph {\bibinfo {title} {Controlling the quantum state of trapped ions}},\ \href {https://www.quantumoptics.at/images/publications/dissertation/roos-diss.pdf} {Ph.D. thesis},\ \bibinfo  {school} {University of Innsbruck} (\bibinfo {year} {2000})\BibitemShut {NoStop}%
\bibitem [{\citenamefont {Bertsch}()}]{Milena}%
  \BibitemOpen
  \bibfield  {author} {\bibinfo {author} {\bibfnamefont {M.~G.}\ \bibnamefont {Bertsch}},\ }\emph {\bibinfo {title} {Optical clocks with trapped ions}},\ \href {https://www.quantumoptics.at/images/publications/dissertation/Thesis_Milena_Guevara_Bertsch.pdf} {Ph.D. thesis},\ \bibinfo  {school} {University of Innsbruck}\BibitemShut {NoStop}%
\bibitem [{\citenamefont {Rao}(1992)}]{Rao1992}%
  \BibitemOpen
  \bibfield  {author} {\bibinfo {author} {\bibfnamefont {C.~R.}\ \bibnamefont {Rao}},\ }\bibfield  {title} {\bibinfo {title} {Information and the accuracy attainable in the estimation of statistical parameters},\ }in\ \href {https://doi.org/10.1007/978-1-4612-0919-5_16} {\emph {\bibinfo {booktitle} {Breakthroughs in Statistics: {{Foundations}} and Basic Theory}}},\ \bibinfo {editor} {edited by\ \bibinfo {editor} {\bibfnamefont {S.}~\bibnamefont {Kotz}}\ and\ \bibinfo {editor} {\bibfnamefont {N.~L.}\ \bibnamefont {Johnson}}}\ (\bibinfo  {publisher} {Springer New York},\ \bibinfo {address} {New York, NY},\ \bibinfo {year} {1992})\ pp.\ \bibinfo {pages} {235--247}\BibitemShut {NoStop}%
\bibitem [{\citenamefont {Cramér}(1946)}]{CramerHarald1946Mmos}%
  \BibitemOpen
  \bibfield  {author} {\bibinfo {author} {\bibfnamefont {H.}~\bibnamefont {Cramér}},\ }\href@noop {} {\emph {\bibinfo {title} {Mathematical methods of statistics}}},\ \bibinfo {edition} {1st}\ ed.,\ Princeton mathematical series\ (\bibinfo {year} {1946})\BibitemShut {NoStop}%
\bibitem [{\citenamefont {Fisher}\ and\ \citenamefont {Russell}(1922)}]{fisherMathematicalFoundationsTheoretical1922}%
  \BibitemOpen
  \bibfield  {author} {\bibinfo {author} {\bibfnamefont {R.~A.}\ \bibnamefont {Fisher}}\ and\ \bibinfo {author} {\bibfnamefont {E.~J.}\ \bibnamefont {Russell}},\ }\bibfield  {title} {\bibinfo {title} {On the mathematical foundations of theoretical statistics},\ }\href {https://doi.org/10.1098/rsta.1922.0009} {\bibfield  {journal} {\bibinfo  {journal} {Philosophical Transactions of the Royal Society of London. Series A, Containing Papers of a Mathematical or Physical Character}\ }\textbf {\bibinfo {volume} {222}},\ \bibinfo {pages} {309} (\bibinfo {year} {1922})}\BibitemShut {NoStop}%
\bibitem [{\citenamefont {Helstrom}(1969)}]{helstrom1969quantum}%
  \BibitemOpen
  \bibfield  {author} {\bibinfo {author} {\bibfnamefont {C.~W.}\ \bibnamefont {Helstrom}},\ }\bibfield  {title} {\bibinfo {title} {Quantum detection and estimation theory},\ }\href@noop {} {\bibfield  {journal} {\bibinfo  {journal} {Journal of Statistical Physics}\ }\textbf {\bibinfo {volume} {1}},\ \bibinfo {pages} {231} (\bibinfo {year} {1969})}\BibitemShut {NoStop}%
\bibitem [{\citenamefont {Paris}(2009)}]{parisQuantumEstimationQuantum2009}%
  \BibitemOpen
  \bibfield  {author} {\bibinfo {author} {\bibfnamefont {M.~G.~A.}\ \bibnamefont {Paris}},\ }\bibfield  {title} {\bibinfo {title} {Quantum estimation for quantum technology},\ }\href {https://doi.org/10.1142/S0219749909004839} {\bibfield  {journal} {\bibinfo  {journal} {International Journal of Quantum Information}\ }\textbf {\bibinfo {volume} {07}},\ \bibinfo {pages} {125} (\bibinfo {year} {2009})}\BibitemShut {NoStop}%
\bibitem [{\citenamefont {Giovannetti}\ \emph {et~al.}(2011)\citenamefont {Giovannetti}, \citenamefont {Lloyd},\ and\ \citenamefont {Maccone}}]{giovannettiAdvancesQuantumMetrology2011}%
  \BibitemOpen
  \bibfield  {author} {\bibinfo {author} {\bibfnamefont {V.}~\bibnamefont {Giovannetti}}, \bibinfo {author} {\bibfnamefont {S.}~\bibnamefont {Lloyd}},\ and\ \bibinfo {author} {\bibfnamefont {L.}~\bibnamefont {Maccone}},\ }\bibfield  {title} {\bibinfo {title} {Advances in quantum metrology},\ }\href {https://doi.org/10.1038/nphoton.2011.35} {\bibfield  {journal} {\bibinfo  {journal} {Nature Photonics}\ }\textbf {\bibinfo {volume} {5}},\ \bibinfo {pages} {222} (\bibinfo {year} {2011})}\BibitemShut {NoStop}%
\bibitem [{\citenamefont {Sekatski}\ \emph {et~al.}(2020)\citenamefont {Sekatski}, \citenamefont {W{\"o}lk},\ and\ \citenamefont {D{\"u}r}}]{sekatskiOptimalDistributedSensing2020}%
  \BibitemOpen
  \bibfield  {author} {\bibinfo {author} {\bibfnamefont {P.}~\bibnamefont {Sekatski}}, \bibinfo {author} {\bibfnamefont {S.}~\bibnamefont {W{\"o}lk}},\ and\ \bibinfo {author} {\bibfnamefont {W.}~\bibnamefont {D{\"u}r}},\ }\bibfield  {title} {\bibinfo {title} {Optimal distributed sensing in noisy environments},\ }\href {https://doi.org/10.1103/PhysRevResearch.2.023052} {\bibfield  {journal} {\bibinfo  {journal} {Physical Review Research}\ }\textbf {\bibinfo {volume} {2}},\ \bibinfo {pages} {023052} (\bibinfo {year} {2020})}\BibitemShut {NoStop}%
\bibitem [{\citenamefont {Hamann}\ \emph {et~al.}(2022)\citenamefont {Hamann}, \citenamefont {Sekatski},\ and\ \citenamefont {D{\"u}r}}]{hamannApproximateDecoherenceFree2022}%
  \BibitemOpen
  \bibfield  {author} {\bibinfo {author} {\bibfnamefont {A.}~\bibnamefont {Hamann}}, \bibinfo {author} {\bibfnamefont {P.}~\bibnamefont {Sekatski}},\ and\ \bibinfo {author} {\bibfnamefont {W.}~\bibnamefont {D{\"u}r}},\ }\bibfield  {title} {\bibinfo {title} {Approximate decoherence free subspaces for distributed sensing},\ }\href {https://doi.org/10.1088/2058-9565/ac44de} {\bibfield  {journal} {\bibinfo  {journal} {Quantum Science and Technology}\ }\textbf {\bibinfo {volume} {7}},\ \bibinfo {pages} {025003} (\bibinfo {year} {2022})}\BibitemShut {NoStop}%
\bibitem [{\citenamefont {Ringbauer}\ \emph {et~al.}(2022)\citenamefont {Ringbauer}, \citenamefont {Meth}, \citenamefont {Postler}, \citenamefont {Stricker}, \citenamefont {Blatt}, \citenamefont {Schindler},\ and\ \citenamefont {Monz}}]{ringbauerUniversalQuditQuantum2022}%
  \BibitemOpen
  \bibfield  {author} {\bibinfo {author} {\bibfnamefont {M.}~\bibnamefont {Ringbauer}}, \bibinfo {author} {\bibfnamefont {M.}~\bibnamefont {Meth}}, \bibinfo {author} {\bibfnamefont {L.}~\bibnamefont {Postler}}, \bibinfo {author} {\bibfnamefont {R.}~\bibnamefont {Stricker}}, \bibinfo {author} {\bibfnamefont {R.}~\bibnamefont {Blatt}}, \bibinfo {author} {\bibfnamefont {P.}~\bibnamefont {Schindler}},\ and\ \bibinfo {author} {\bibfnamefont {T.}~\bibnamefont {Monz}},\ }\bibfield  {title} {\bibinfo {title} {A universal qudit quantum processor with trapped ions},\ }\href {https://doi.org/10.1038/s41567-022-01658-0} {\bibfield  {journal} {\bibinfo  {journal} {Nature Physics}\ }\textbf {\bibinfo {volume} {18}},\ \bibinfo {pages} {1053} (\bibinfo {year} {2022})}\BibitemShut {NoStop}%
\bibitem [{\citenamefont {Horn}\ and\ \citenamefont {Johnson}(1985)}]{matrixAnalysis}%
  \BibitemOpen
  \bibfield  {author} {\bibinfo {author} {\bibfnamefont {R.~A.}\ \bibnamefont {Horn}}\ and\ \bibinfo {author} {\bibfnamefont {C.~R.}\ \bibnamefont {Johnson}},\ }\href@noop {} {\emph {\bibinfo {title} {Matrix Analysis}}}\ (\bibinfo  {publisher} {Cambridge University Press},\ \bibinfo {year} {1985})\BibitemShut {NoStop}%
\end{thebibliography}%

%

\newpage

\newpage

\end{document}


\title{Experimental distributed quantum sensing in a noisy environment: Supplemental material}

\author{J.~Bate\,\orcidlink{0000-0002-3570-5102}}
\affiliation{Universit\"at Innsbruck, Institut f\"ur Experimentalphysik, Technikerstr. 25, 6020 Innsbruck, Austria}
\author{A.~Hamann\,\orcidlink{https://orcid.org/0000-0002-9016-3641}}
\affiliation{Universit\"at Innsbruck, Institut f\"ur Theoretische Physik, Technikerstr. 21a, 6020 Innsbruck, Austria}
\author{M.~Canteri\,\orcidlink{0000-0001-9726-2434}}
\affiliation{Universit\"at Innsbruck, Institut f\"ur Experimentalphysik, Technikerstr. 25, 6020 Innsbruck, Austria}
\author{A.~Winkler\,\orcidlink{0000-0002-9459-027X}}
\affiliation{Universit\"at Innsbruck, Institut f\"ur Experimentalphysik, Technikerstr. 25, 6020 Innsbruck, Austria}
\author{Z.~X.~Koong\,\orcidlink{0000-0001-8066-754X}}
\affiliation{Universit\"at Innsbruck, Institut f\"ur Experimentalphysik, Technikerstr. 25, 6020 Innsbruck, Austria}
\author{V.~Krutyanskiy\,\orcidlink{0000-0003-0620-4648}}
\affiliation{Universit\"at Innsbruck, Institut f\"ur Experimentalphysik, Technikerstr. 25, 6020 Innsbruck, Austria}
\author{W.~Dür\,\orcidlink{0000-0002-0234-7425}}
\affiliation{Universit\"at Innsbruck, Institut f\"ur Theoretische Physik, Technikerstr. 21a, 6020 Innsbruck, Austria}
\author{B.~P.~Lanyon\,\orcidlink{0000-0002-7379-4572}}
\affiliation{Universit\"at Innsbruck, Institut f\"ur Experimentalphysik, Technikerstr. 25, 6020 Innsbruck, Austria}
\email[Correspondence should be sent to ]{ben.lanyon@uibk.ac.at}

\maketitle

\tableofcontents

\section{Experimental methods and protocol}
\subsection{Quantum-logic operations implemented in the experiment}
\label{appendix:quantumoperations}

Coherent rotations operating equally on all ions are implemented by a broadly-focused \SI{729}{nm} laser that couples approximately equally to all ions. This laser performs rotations on optical qubits encoded in superpositions that span the $4^2\textrm{S}_{1/2}$ and $3^2\textrm{D}_{5/2}$ manifolds. This is achieved by setting the frequency to be resonant with the corresponding transition. In the ideal case, this rotation operation is described by the following unitary: 
\begin{equation}\label{eq:rotglob}
    R(\theta, \phi) =  \bigotimes_{j=1}^3 e^{-i\frac{\theta}{2}({\sigma^{j}_x}\cos{\phi} + {\sigma^{j}_y}\sin{\phi})},
\end{equation}
where $\sigma^{j}_{\alpha}$ represents the Pauli operator $\sigma_{\alpha}$ acting on an optical qubit encoded in the $j$-th ion, $\phi$ is controlled by the phase of the laser, and $\theta$ is controlled by the intensity and duration of the applied laser pulse.
Addressed operations on a chosen ion are implemented by a single-ion-focused \SI{393}{\nano\meter} laser with a \SI{1.3(1)}{\um} waist, deflected along the ion-string direction using an acousto-optic deflector~\cite{MarcoThesis}. This \SI{393}{\nano\meter} laser is detuned by \SI{-6.64(2)}{\giga\hertz} from the $4^2\textrm{S}_{1/2}\leftrightarrow 4^2\textrm{P}_{3/2}$ transition, and implements an equal AC stark shift on the states $\ket{S}$ and $\ket{S'}$ with respect to states in the $3^2\textrm{D}_{5/2}$ manifold (as defined in Figure \ref{fig:fullexperimentalsequence}).
In the ideal case, this operation on optically-encoded qubits is described by the following unitary: 
\begin{equation}\label{eq:rotadd}
   R^j_z(\theta) = e^{-i\frac{\theta}{2}\sigma^{j}_z},
\end{equation}
where $\theta$ is controlled by the intensity and duration of the applied \SI{393}{\nano\meter} laser pulse. 
Entanglement between the ions is achieved using a M{\o}lmer-S{\o}rensen logic gate (MS-gate)~\cite{PhysRevA.62.022311, Kirchmair_2009}. 
This gate is implemented by applying a \SI{729}{nm} laser pulse with two frequencies.
Those frequencies are $\nu \pm (f_{\mathrm{COM}} + \delta)$, where $\nu$ is the frequency of the relevant qubit `carrier' transition, $f_{\mathrm{COM}}$ is the frequency of the axial center of mass motional mode and $\delta$ is an overall detuning. We set $\delta = $ \SI{6.9}{\kilo\hertz} and we measure $f_{\mathrm{COM}} = $ \SI{0.9620(4)}{\mega\hertz}. The applied pulse is \SI{143}{\us} long. 
In the ideal case, this operation is defined by the unitary:

\begin{equation}\label{eq:MS}
    R_{\mathrm{MS}} = e^{-i\frac{\pi}{8}(\sigma_x^1+\sigma_x^2+\sigma_x^3)^2}.
\end{equation}

\subsection{Experimental sequence}
\label{appendix:experimentalsequence}

In this section, we show the experimental pulse sequence for the SWD and separable protocols.
We present the experimental sequence on the level of individual quantum-logic operations, and how these quantum-logic operations realize the quantum circuit shown in Figure 1c in the main text. 

\begin{figure}[t]
	\begin{center}
        \includegraphics[width=1\columnwidth]{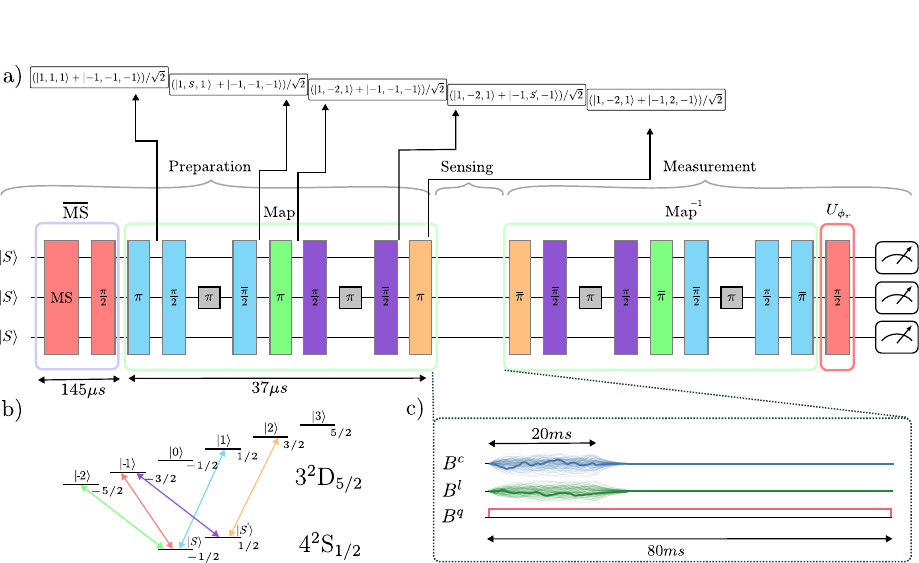}
		\caption{ \textbf{(a)} Full experimental sequence, except for Doppler cooling, sideband cooling and optical pumping required to prepare the ions in $\ket{S}$. Pulses (which are color-filled) are grouped (colored outlines) to relate them to Figure 1c of the main text.
        The pulse labeled MS represents $R_{\mathrm{MS}}$ (Equation \ref{eq:MS}). Colored pulses acting on all ions represent $R(\theta, \phi)$ (Equation \ref{eq:rotglob}), where $\theta$ is labeled inside the box and $\phi = \pi$ when there is an overline and $\phi = 0$ when there is no overline. Grey pulses acting on the middle ion represent $R_z^{2}(\theta)$ (Equation \ref{eq:rotadd}), where $\theta$ is labeled inside the pulse. The action of any single-ion addressed pulse when combined with its immediate neighbours (one pulse before and one after) is to change only the state of the addressed ion. For example, consider the first addressed pulse on the middle ion. When combined with the preceding and following blue pulses, the effect is to move the $\ket{1}$ state of ion 2 to state $\ket{S}$. \textbf{(b)} Relevant level scheme of the $^{40}$Ca$^+$ ion. The pulses in \textbf{(a)} are color-coded to represent the transition on which they act. \textbf{(c)} During the sensing step, the noise fields are switched on for \SI{20}{\ms}, and subsequently switched off. The quadratic field is applied for \SI{80}{\ms}. 
             }
    		\label{fig:fullexperimentalsequence}
		\vspace{-5mm}
	\end{center}
\end{figure}

First, Doppler cooling is performed for \SI{30}{\milli\second} on all ions\footnote{For technical reasons, that duration of Doppler cooling was required for our control system to run the $\sim$\SI{120}{\milli\second}-long experimental sequence.}, where a combination of \SI{397}{nm}, \SI{866}{nm} and \SI{854}{nm} laser light is applied. 
This is directly followed by optical pumping for \SI{20}{\micro\second}, achieved using $\sigma^-$ polarized \SI{397}{nm} light beam with a propagation direction parallel to the principle magnetic field. 
Next, sideband cooling is performed on the axial center of mass (COM) motional mode, where a combination of \SI{729}{nm} and \SI{854}{nm} laser light is applied for \SI{2}{\milli\second}, followed by a second round of optical pumping. After these steps, the ions are each ideally prepared in $\ket{S}$ and the axial COM mode is in its ground state.
Next, the operations $\overline{\mathrm{MS}}$ and Map (Figure 1c, main text) are implemented using the sequences of individual quantum-logic operations shown in Figure \ref{fig:fullexperimentalsequence}, which ideally prepare the ions into the SWD state ($\ket{\psi_{(1, -2, 1)}^{\mathrm{SWD}}}$). This ends the preparation step.
Next comes the sensing step.
The noisy voltage generators are switched on for each coil pair (CC and GC) at the beginning of the sensing step using a solid state switch and off again \SI{20}{\ms} later. 
That \SI{20}{\ms} was chosen as it is the shortest time necessary to eliminate any sensor state coherences lying outside the subspace targeted by the entangled SWD state with the noise voltage generators at full power.
We find that after switching current on or off, the magnetic field measured at the ion does not change instantaneously, but with a profile that we associate with eddy currents in the steel vacuum chamber, generated by the nearby coils. 
We measured that the spatially constant magnetic field measured by the ions changes according to an exponential function ($e^{-t/\tau}$) with a time constant of $\tau = $ \SI{2.27(4)}{ms} for the CC coil pair, and the spatially gradient magnetic field measured by the ions changes according to the same exponential function with a time constant of \SI{3.3(6)}{ms} for the GC coil pair\footnote{The data used to calculate the time constant of the GC coil used 40 shots per point, compared to 80 shots per point for the CC coil. This is the reason that the GC time constant has a larger errorbar.}. 
We attribute those time constants with the inductance and resistance of the system.
A measurement of the spatially constant magnetic field generated by the GC coil pair, which is two orders of magnitude larger than that created by the CC coil pair\footnote{That large spatially constant magnetic field is there because the ions are not positioned exactly equidistantly between the GC coils., and does not affect the interpretation of the results}, finds deviations from that exponential function that are oscillatory, which we attribute to additional capacitances of the system.
For the GC pair, we find that a wait time of \SI{60}{\ms} is needed before the magnetic fields die down enough such that they don't cause an observable effect on the fidelity of the subsequent quantum-logic operations (Section IIC).
A vacuum chamber composed of a non-conductive material like glass would eliminate these effects.
The sensing step therefore lasts a further \SI{60}{\ms} after the \SI{20}{\ms} of applied noise, resulting in a total time of \SI{80}{\ms}. 

The \SI{729}{\nano\meter} laser is applied throughout these \SI{80}{\ms} to generate an effective quadratic field, with a frequency detuned from the $\ket{S}\leftrightarrow\ket{-2}$ transition by $\Delta_{q}=$ \SI{-700}{\kilo\hertz}. 
During the final measurement step, the operations $\mathrm{Map}^{-1}$ and $U_{\phi_r}$ (as shown in Figure \ref{fig:fullexperimentalsequence}, using the notation of Section \ref{appendix:quantumoperations}) are implemented using sequences of individual quantum-logic operations.
Lastly, state detection is achieved by imaging \SI{397}{nm} fluorescence onto a CCD camera for \SI{7}{\ms}.

\subsection{State tomography}
\label{tomo}

Three-qubit state tomography is implemented by replacing $U_{\phi_r}$ in the experimental sequence shown in Figure \ref{fig:fullexperimentalsequence} by single-qubit rotations that effectively rotate the measurement basis. Measurements are made in 27 different bases, corresponding to all permutations of Pauli operators for three qubits. The single-qubit rotations are chosen to minimize the required number of addressed Stark shift gates (Section \ref{appendix:quantumoperations}). Using the notation of Section \ref{appendix:quantumoperations},
some examples are as follows: the measurement basis $X\otimes X\otimes -Y$ is implemented with the single-qubit rotations $R(\pi/2, 3\pi/2)R^3_z(\pi/2)$; the measurement basis $X\otimes Y\otimes Z$ is implemented with the single-qubit rotations $R(\pi/2, \pi/2)R^2_z(\pi/2)R(\pi/2, 3\pi/2)R(\pi/2, 0)R^1_z(\pi/2)R(\pi/2, \pi)$; the measurement basis $X\otimes X\otimes Z$ is implemented with the single-qubit rotations $R(\pi/2, 3\pi/2)R(\pi/2, \pi)R^3_z(\pi/2)R(\pi/2, 0)$. 

For each of the 27 measurement bases, an estimate of the eight outcome probabilities is made from 480 repetitions of the experimental sequence. A reconstructed density matrix is found that maximizes the likelihood of the observed outcome probabilities from all measurement bases.
Quantities of interest such as GHZ-fidelity (Section \ref{appendix:imperfections}) and the amplitude of a parity oscillation ($A$, as defined in the main text) are extracted from that reconstructed density matrix.
Statistical uncertainties in those quantities of interest are found through the method of Monte Carlo sampling. Specifically, a probability distribution of the eight measurement outcomes for the $j$-th measurement basis is simulated as a multinomial distribution, where the probability of observing each of the eight outcomes is set to the experimentally observed value. $N_s = 480$ samples are then taken from the probability distribution, resulting in the $i$-th outcome being observed $N_{ij}$ times. The probability of the $i$-th outcome is estimated to be $p_{ij} = N_{ij}/N_s$. From the resulting estimated outcome probabilities $\{p_{ij}\}$, a density matrix is reconstructed in the same way as described above. This process is repeated 300 times, resulting in a set of 300 reconstructed density matrices. The quantity of interest is then extracted for each of the 300 density matrices, and an errorbar in the quantity of interest is calculated from the standard deviation in the 300 extracted quantities of interest. The errorbar is centered around the value calculated from the raw data. 

The fidelity ($F$) of the reconstructed sensor states with respect to their corresponding target state is calculated in the following way.  
For the reconstructed SWD state ($\rho^{\mathrm{SWD}}$), the fidelity with respect to the target state ($\ket{\psi_{(1, -2, 1)}^{\mathrm{SWD}}}$, as defined in the main text) is calculated by $F = \max_{\phi}\bra{\psi_{(1, -2, 1)}^{\mathrm{SWD}}}(R_z^1(\phi)\otimes R_z^2(\phi)\otimes R_z^3(\phi))^\dag\rho^{\mathrm{SWD}}R_z^1(\phi)\otimes R_z^2(\phi)\otimes R_z^3(\phi)\ket{\psi_{(1, -2, 1)}^{\mathrm{SWD}}}$, where $R_z^1(\phi)\otimes R_z^2(\phi)\otimes R_z^3(\phi)$ represents single-qubit phase shifts that are the same on each qubit ($R_z^j$ is defined in Equation \ref{eq:rotadd}), which is chosen to maximize $F$.
For the reconstructed separable state ($\rho^{\mathrm{sep}}$), the fidelity with respect to the target state ($\ket{\psi_{(1, -2, 1)}^{\mathrm{sep}}}$, as defined in the main text) is calculated by $F = \max_{\phi}\bra{\psi_{(1, -2, 1)}^{\mathrm{sep}}}(R_z^1(\phi)\otimes R_z^2(\phi)\otimes R_z^3(\phi))^\dag\rho^{\mathrm{sep}}R_z^1(\phi)\otimes R_z^2(\phi)\otimes R_z^3(\phi)\ket{\psi_{(1, -2, 1)}^{\mathrm{sep}}}$.

\section{Supporting experimental results}

\subsection{Infidelities in the prepared sensor states}
\label{appendix:imperfections}
\subsubsection{State preparation}

We perform additional experiments with the aim of providing information on sources of infidelity in the entangled initial sensor state $\rho^{\mathrm{SWD}}$.
Recall that the entangled sensor state $\rho^{\mathrm{SWD}}$ is the one after the preparation sequence of operations shown in Figure 1a of the main text. 
We re-prepare the state $\rho^{\mathrm{SWD}}$ and characterize its fidelity ($F$) with the ideal state via the method of \cite{PhysRevLett.106.130506}, where the expression $F = \frac{P + A}{2}$ is used, with $P$ being the sum of the two on-diagonal population terms within the DFS. Using that method, a state fidelity of $F=0.74(1)$ was obtained\footnote{The new value ($F=0.74(1)$) is lower than the one obtained from state tomography in the main text (0.78(1)), which we attribute to an improved calibration of the MS-gate parameters in the latter case.}. Next, we characterise the initial entangled state in the absence of the mapping pulses, which we label as $\hat{\rho}^{\mathrm{SWD}}$, obtaining a state fidelity of 0.85(1). The mapping pulses (both mapping and unmapping) were thus responsible for a drop in entangled state fidelity of 0.11(2). We attribute the errors in the mapping pulses and those in preparing the state $\hat{\rho}^{\mathrm{SWD}}$ predominantly to phase instabilities in our \SI{729}{\nano\meter} laser. The laser exhibits phase noise at both higher frequencies of a hundred kilohertz or more and at lower frequencies in the acoustic regime or lower. Indeed, several years ago, when using a different and more phase-stable laser, we prepared a three-qubit GHZ state in our system (on the $\ket{S}\leftrightarrow \ket{-2}$ transition) with a fidelity of $0.97(1)$. In the following paragraph we estimate and model errors in the mapping pulses. \\

\noindent \textbf{Mapping pulse error model.}

\noindent The mapping pulses and their inverse (Figure \ref{fig:fullexperimentalsequence}), consist of a total of six $\pi$ pulses and four Ramsey-type sequences.  Specifically, the latter each consist of two $\pi$/2 pulses with an `addressed' laser pulse on the middle ion in between. 
We model the infidelity introduced by the entire pulse sequence (mapping followed by its inverse), for a given input state, via a simple three-qubit uncorrelated depolarizing channel \cite{Wilde_2013}. The channel is parameterised by three probabilities $p_i$, corresponding to the probability of a depolarisation error occurring on each qubit individually. To obtain an estimate for the error probabilities $p_i$ we use the results of the calibration experiments carried out when setting up the pulses in the mapping sequence, as now described. 

The $\pi$ pulses on each transition ($\ket{S}\leftrightarrow\ket{j}$ for $j = -2, -1, 1$ as defined in Figure \ref{fig:fullexperimentalsequence}) are calibrated in an experiment (one for each transition) in which five pulses are performed in a row on the initial state $\ket{SSS}$, which should ideally prepare the state $\ket{jjj}$. Once optimised, we extract the probability that each ion is not in its target state, finding that the results are statistically consistent over all transitions and between the ions, and obtaining a single-qubit error probability for the five $\pi$ pulses of $p_{\pi}=0.011(4)$. A similar experiment is carried out when calibrating the addressed Ramsey-type sequences, by applying four of them sequentially to the state $\ket{SSS}$, which should ideally return the state $\ket{SSS}$. Once optimized, we measure the probability that each ion is not in its target state, finding that the results are statistically consistent over all transitions, and obtaining single-qubit error probabilities of $P_{a,1}=0.040(9)$, $P_{a,2}=0.02(2)$ and $P_{a,3}=0.040(9)$ for ions 1, 2 and 3, respectively.

The three-qubit uncorrelated depolarizing channel is given by
\begin{equation*}
    \mathcal{E}(\rho) = \sum_{K_1 \in \mathcal{K}^{\mathrm{1}} }\sum_{K_2 \in \mathcal{K}^{\mathrm{2}} }\sum_{K_3 \in \mathcal{K}^{\mathrm{3}} }(K_1\otimes K_2\otimes K_3)\rho( K_1^{\dag}\otimes K_2^{\dag}\otimes K_3^{\dag}),
\end{equation*}
where the sets $\mathcal{K}^{i}$ contain the following kraus operators
\begin{equation*}
    \begin{pmatrix}
    \sqrt{1 - 3p_i/4} & 0\\
    0& \sqrt{1 - 3p_i/4}\\
    \end{pmatrix},
    \hspace{1cm}
    \begin{pmatrix}
    0 & \sqrt{p_i/4}\\
    \sqrt{p_i/4}& 0\\
    \end{pmatrix},
    \hspace{1cm}
    \begin{pmatrix}
    \sqrt{p_i/4} & 0\\
    0& -\sqrt{p_i/4}\\
    \end{pmatrix},
    \hspace{1cm}
    \begin{pmatrix}
    0 & -i\sqrt{p_i/4}\\
    i\sqrt{p_i/4}& 0\\
    \end{pmatrix},
\end{equation*}
with error probabilities $p_1 = p_3 = 0.051(14)$ and $p_2 = 0.024(20)$. 
Those error probabilities, due to the combination of the errors mentioned in the previous paragraph, are estimated via $p_i=1-(1-p_{\pi})(1-P_{a,i})$. Although this misses the additional error due to one extra $\pi$-pulse, the effect is expected to be negligible compared to the statistical uncertainty in the values $p_i$. 

Applying that depolarizing model of the mapping pulses to the ideal entangled state $\rho = \ket{\psi_{(1, -2, 1)}^{\mathrm{SWD}}}\bra{\psi_{(1, -2, 1)}^{\mathrm{SWD}}}$ yields a fidelity drop of 0.09(2), which is comparable with the observed fidelity drop of 0.11(2). Applying the same model to the ideal separable state $\rho = \ket{\psi_{(1, -2, 1)}^{\mathrm{sep}}}\bra{\psi_{(1, -2, 1)}^{\mathrm{sep}}}$ yields a fidelity drop of 0.06(1), which is consistent with the measured state fidelity of 0.94(1) and a near-perfect implementation of the initial state in the absence of the mapping pulses $\hat{\rho}^{\mathrm{sep}}$\footnote{Since $\hat{\rho}^{\mathrm{sep}}$ requires a single $\pi/2$ pulse, we expect to have achieved a near-perfect implementation of that state.}. 

We attribute errors in the $\pi$ pulses to the aforementioned 729nm laser phase noise, since that is the only laser involved.
A separate investigation reveals that the size of the observed errors in the Ramsey-type sequences are also consistent with that expected from the 729nm laser phase noise~\cite{TabeaThesis}. 
Developing a detailed infidelity budget at the level of a percent is beyond the scope of this work.\\

\subsubsection{Spontaneous emission during the sensing step}
\label{appendix:spontaneousemission}

Figures 2d and 2e of the main text present the tomographically-reconstructed entangled sensor state ($\rho^{\mathrm{SWD}}$) before and after the \SI{80}{\milli\second}-long full noise channel. The obtained GHZ-state fidelities are 0.78(1) and 0.56(1) respectively; a fidelity drop of 0.22(2). In this subsection, we present a model for the expected drop in fidelity due to spontaneous decay of the $3^2\textrm{D}_{5/2}$ manifold. The model predicts that the initially prepared sensor state of Figure 2d  would suffer a fidelity drop of 0.16(1) over \SI{80}{\milli\second}, from which we conclude that the majority of the observed loss of fidelity is due to spontaneous decay. We associate the remaining fidelity drop of 0.06(3) to a drift in the performance of the MS gate in the time (several hours) between taking the data of Figure 2d and 2e. 

Our model of spontaneous decay considers three qutrits, each with a ground state ($\ket{g} = \begin{pmatrix}0&0&1\end{pmatrix}$), and two excited states ($\ket{e_1} = \begin{pmatrix}1&0&0\end{pmatrix}$ and $\ket{e_2} = \begin{pmatrix}0&1&0\end{pmatrix}$).
The initial state of the model encodes  $\rho^{\mathrm{SWD}}$ into $\ket{e_1}$ and $\ket{e_2}$. The state $\ket{g}$ represents the states in the $4^2\textrm{S}_{1/2}$ manifold. Next, an amplitude damping map is applied to each qutrit, where excited states decay to the ground state with a lifetime $\tau = 1.045$\SI{}{\second}, corresponding to the lifetime of the $3^2\textrm{D}_{5/2}$ manifold~\cite{James1998}.
The amplitude damping map has the following form
\begin{equation*}
    \mathcal{E}(\rho) = \sum_{K_1, K_2, K_3 \in \mathcal{K} }(K_1\otimes K_2\otimes K_3)\rho( K_1^{\dag}\otimes K_2^{\dag}\otimes K_3^{\dag}),
\end{equation*}
where the set $\mathcal{K}$ contains the following kraus operators respectively~\cite{Chessa2021}
\begin{equation*}
    \begin{pmatrix}
    \sqrt{1 - p} & 0 & 0\\
    0& \sqrt{1 - p}  & 0\\
    0 & 0 & 1\\
    \end{pmatrix},
    \hspace{1cm}
    \begin{pmatrix}
    0 & \sqrt{p} & 0\\
    0& 0  & 0\\
    0 & 0 & 0\\
    \end{pmatrix},
    \hspace{1cm}
    \begin{pmatrix}
    0 & 0 & 0\\
    0& 0  & 0\\
    \sqrt{p} & 0 & 0\\
    \end{pmatrix},
\end{equation*}
and the error probability is given by $p(t) = (1 - e^{-t/\tau})$.
Finally, the model applies an ideal $\pi$ pulse on the $\ket{g}\leftrightarrow\ket{e_1}$ transition, and extracts the fidelity with respect to the nearest GHZ state encoded in superpositions of the $\ket{g}\leftrightarrow\ket{e_2}$ states.
This model predicts a fidelity drop of $0.16(1)$. Here, the uncertainty is calculated by applying the channel separately to 200 Monte-Carlo-sampled (Section~\ref{tomo}) density matrices of the tomographically reconstructed state $\rho^{\mathrm{SWD}}$ in Figure 2d of the main text and computing the standard deviation of the resulting set of fidelities. The errorbar is centered around the value calculated from the raw data.

\subsection{Calibration of applied quadratic signal strengths}\label{appendix:calibration}

The main text states that seven different applied quadratic field strengths are applied to the sensors, each of which are separately calibrated to be $B^q_{\mathrm{cal}} = [ 0.0, 2.1, 5.0, 7.6, 9.5, 11.9, 15.2]~\SI{}{\pico\tesla\per\square\micro\meter}$, corresponding to laser-induced frequency shifts of $\ket{-2}$ by up to  $\SI{12.3}{\hertz}$. This section describes the experiments done to justify these statements by first detailing how a quadratic field is generated using a laser, and then providing an equation relating the effective quadratic field strength with parameters of the laser used to generate it. 

The quadratic field strength $B^q$ was defined in the main text such that the resulting magnetic field on three collinear sensors separated by a distance $d$ is given by $\bm{B} = \frac{B^qd^2}{2}\bm{f}^q$, where $\bm{f}^q = (1, 0, 1)$.
The field $\bm{f}^q$ is imprinted by applying a \SI{729}{\nano\meter} laser with nominally equal intensity on the three ions, and a frequency that is detuned from the $\ket{S}\leftrightarrow\ket{-2}$ transition by $\Delta_q = -700$\,kHz.
One contribution of the application of this laser is to shift the frequency of $\ket{-2}$ by $\delta_{\mathrm{AC}} = \frac{\Omega^2}{4\Delta_q}$ via an AC stark shift of the $\ket{S}\leftrightarrow\ket{-2}$ transition, where $\Omega$ is the corresponding Rabi frequency of the \SI{729}{\nano\meter} laser. 
When that laser is applied to the sensing states, this AC stark shift changes the frequency splitting of the middle ion-qubit by $-\delta_{\mathrm{AC}}$, since only the middle ion-qubit occupies $\ket{-2}$.
Within the Hilbert space spanned by the bold levels in Figure 1b of the main text, this frequency splitting imprints an effective quadratic field of the form $\bm{\tilde{f}}^q = (0, -1, 0) = \bm{f}^q - \bm{f}^c$, where $\bm{f}^c$ = (1,1,1). 
By design, the phase of the sensor states is insensitive to the $\bm{f}^c$ component.
We calculate the expected effective quadratic field strength from the \SI{729}{\nano\meter} laser to be $B^q =(C\frac{\Omega^2}{4\Delta_q})(\frac{1}{4\mu_Bg_D})(\frac{2}{d^2}) = \frac{C\Omega^2}{8\Delta_q\mu_Bg_Dd^2}$, where $g_D = 1.2$ is the Land\'e g factor of the $3^2\textrm{D}_{5/2}$ manifold, and $C$ is a constant factor accounting for AC stark shifts on all other dipole transitions, which we independently calibrate as described in the following paragraph. 

We now turn to describing the experiments that were done to measure and calibrate the applied quadratic field strengths ($B^q_{\mathrm{cal}}$).
The first experiment was performed with three ions, and the same trap confinement and cooling sequence as for the sensing experiments in the main text. 
The experiment was done to measure $\Omega$ on the $\ket{S}\leftrightarrow\ket{-2}$ transition for each of six different intensities of the \SI{729}{\nano\meter} laser. Those six different intensities were used to implement the six non-zero $B^q_{\mathrm{cal}}$ values given in the main text and at the beginning of this section.
The laser intensity was controlled through the RF power sent to an acousto-optic modulator in the beam's path.
In each case $\Omega$ was measured by applying a laser pulse for time $t_{\mathrm{pulse}}$ that is resonant with the $\ket{S}\leftrightarrow\ket{-2}$ transition. The excitation probabilities were measured as a function of $t_{\mathrm{pulse}}$ (Rabi flops), and were fit to a numerical model (Equation A.3 of~\cite{RoosThesis}).
This fit gave $\Omega = 2\pi\times[ 2.21(3), 3.39(1), 4.20(1), 4.67(1), 5.24(1), 5.92(2)]~\SI{}{\kilo\hertz}$.
The final task is to measure the value of $C$. This is done by extracting the frequency shifts on the relevant qubit for each ion due to dipolar AC stark shifts, and then calculating what quadratic field strength those frequency splittings correspond to. We measure a value $C = 0.979$.
Substituting the measured values of $\Omega$ into the equation above for $B^q$, one finds the seven aforementioned calibrated values of $B^q_{\mathrm{cal}}$.

\subsection{Phase offset of the sensor states}
\label{appendix:phi0}

In the absence of an applied quadratic field via the laser field, the measured parity oscillations in Figure 3a-b of the main text display a phase offset $\phi_0$, obtained using the fit function described in the main text. The values obtained for the entangled and separable states are $\phi_0 = $\SI{-0.18(2)}{\radian} and $\phi_0 = $ \SI{-0.19(8)}{\radian}, which are the same to within statistical uncertainty.

Re-measuring that phase offset for the entangled state in the absence of the noise fields yields no statistically significant difference: the noise fields do not thus generate a significant quadratic component.  
However, in the absence of noise and laser, the phase offset is found to increase linearly as a function of the wait time at a rate that is consistent with a background effective quadratic field strength of \SI{0.188(5)}{\hertz\per\micro\meter\squared}. 
That background quadratic field sets the measured values of $\phi_0$ above. Repeated measurements of $\phi_0$ taken on different days yield statistically consistent values, confirming that the background field is static. Its value can therefore be calibrated, causing no effect on the results of the main text. 
While we were not able to fully identify the physical origin of the background field, an independent investigation determined that it is not consistent with only the expected electric quadrupole shifts, quadratic Stark shifts and quadratic Zeeman shifts~\cite{Milena}. 

\subsection{Dynamic range and sensor performance as a function of shots}\label{appendix:dynamicrange}

\begin{figure}
    \centering
    \includegraphics[width=1\linewidth]{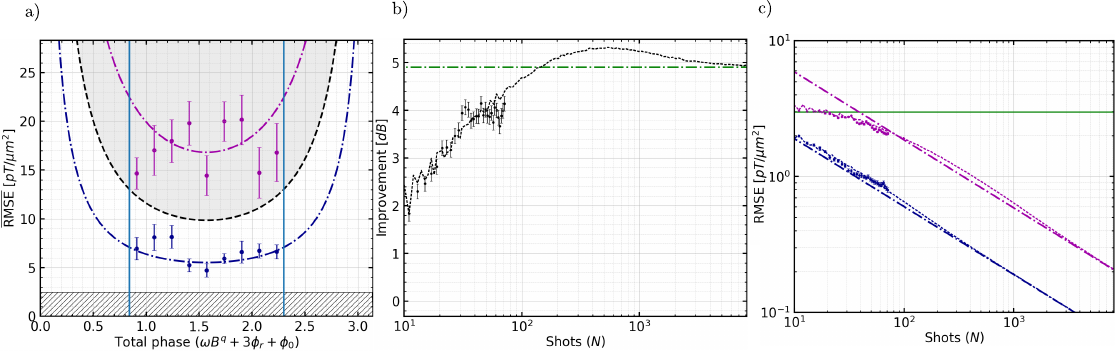}
    \caption{Dynamic range and performance scaling with shot number. In all panels, shapes with error bars show data. All data is for $n=72$ shots
    \textbf{(a)} Dynamic range of the estimation protocols. Blue circles show SWD protocol data. Magenta circles show separable protocol data. Striped area is beyond the Heisenberg limit and inaccessible. The grey area is achievable by all two-level separable estimation protocols. The magenta and blue colored dot-dashed lines show limits for $n\to \infty$ obtained via the classical Fisher information and the measured amplitudes $A$. The black dot-dashed line is similar, but for a perfect separable state ($A=0.25$).    
    Blue lines enclose the phase range $\frac{\pi}{2} \pm 0.73$. Data is binned into nine equally sized intervals. 
    \textbf{(b)} Improvement scaling with shot number: improvement is quantified by the ratio of the RMSE for the SWD protocol to the one of the separable protocol. The black dotted line shows simulated data. The green flat line shows infinite limit (ratio of magenta to blue dot-dashed lines in (a) for a phase of 1.5 radians). 
    \textbf{(c)} RMSE scaling with shot number. Blue circles show SWD protocol data. Magenta circles show separable protocol data. Color-matched dotted lines show simulations. Dot-dashed lines each show limits obtained via the classical Fisher information and the measured amplitudes $A$, which both have $1/\sqrt{n}$ scaling.
    The horizontal green line is the expected RMSE corresponding a uniform random guess of the phase between $\pm \pi/2$. 
    }
    \label{fig:dynamic-range}
\end{figure}

The minimum $\overline{\mathrm{RMSE}}$ (as bounded by the classical Fisher information in Equation \ref{equation:cramer-rao}) is only rigorously achieved in the limit of infinite shots, and when the total phase of the parity ($\Phi = \omega B^q + 3\phi_r + \phi_0$) is $\pi/2$. 
The total phase is not known exactly in advance, since the value of $B^q$ is yet to be measured. 
It is therefore desirable that a sensor not only achieves the minimum $\overline{\mathrm{RMSE}}$, but also maintains a small $\overline{\mathrm{RMSE}}$ over a range of total phase. 
The extent to which a sensor achieves those two properties is qualitatively referred to as the dynamic range of a sensor.
The total phase of the selected estimates $B^q_\mathrm{est}$ (as defined in the main text) are within the range $\mathrm{mod}_{\pi}\Phi\in\frac{\pi}{2} \pm 0.73$.
Figure \ref{fig:dynamic-range} shows that the measured $\overline{\mathrm{RMSE}}$ from both the separable and SWD protocol do not significantly vary over this phase range. The same figure also shows
that the implemented SWD protocol has a larger dynamic range than the separable protocol in the limit of infinite shots $n$. 

We now investigate how much larger the measured improvement (the ratio of the measured $\overline{\mathrm{RMSE}}$ of the SWD protocol and the separable protocol, as defined in the main text) could be, given more shots of the experiment.
First, we calculate the improvement in the limit of infinite shots using the Cram\'er-Rao bound (Equation~\ref{equation:cramer-rao}). 
This calculation yields an improvement of \SI{4.9(1)}{\decibel} (Figure \ref{fig:dynamic-range}b, dot-dash line), where the error bar originates from the uncertainty in the measured amplitude values A (given explicitly in the next paragraph).
The improvement we achieved for $n=72$ shots (as reported in the main text) is five standard deviations below that limit. 

Next, we simulate the expected performance of both protocols as a function of $n$. This is done by first  sampling parity outcomes from the parity function introduced in Equation 1 of the main text, using the measured amplitude values $A = 0.45(2)$ and $A = 0.146(9)$ (see main text). Finally, the simulated $\overline{\mathrm{RMSE}}$ is obtained following the same process as for the experimental data. 
The simulation results for the improvement are compared with the data in Figure \ref{fig:dynamic-range}b and show that two orders of magnitude more shots would be required for the improvement to converge to the value for infinite shots. 
The simulation results for the RMSE of each protocol separately (the absence of the overline indicating that we do not normalize by $\sqrt{n}$, so the expected $1/\sqrt{n}$ is observed) are shown in Figure \ref{fig:dynamic-range}c. 
While the measured RMSE from the SWD protocol reduces with $1/\sqrt{n}$ from $n\approx10$ shots, the measured RMSE from the separable protocol performs comparably to a random guess for $n<40$, and converges to the $1/\sqrt{n}$ scaling only after $n \approx 60$ shots (Figure \ref{fig:dynamic-range}c).

\section{Theoretical background of the SWD and separable protocols}

\subsection{General concepts and methods}

This section introduces a theoretical framework for sensing in the presence of correlated noise. Notation introduced in this section is used throughout this document.
In Section \ref{appendix:cramerrao}, the classical and quantum Fisher information are introduced, and it is shown how they bound the achievable $\overline{\mathrm{RMSE}}$. Section \ref{appendix:generalhamiltonian} generalizes the sensing Hamiltonian to arbitrary numbers of sensors and fields. In Section \ref{appendix:correlateddephasingnoise}, the effect of fluctuating noise fields on initial sensor states is derived and the concept of a decoherence-free subspace is formally defined. 
Section \ref{appendix:findoptimalstep} presents the method to find the sensor state that maximises the QFI for arbitrary fields. 
Finally, in \ref{appendix:mlegeneral}, the maximum likelihood estimator is generally defined and explicitly given for the case in hand.

\subsubsection{Bounding the achievable RMSE}\label{appendix:cramerrao}
This section introduces the classical and quantum Cram\'er-Rao bound and then applies them to the $\overline{\mathrm{RMSE}}$ used in the main text.
The classical Fisher information (CFI) quantifies the information that a probability distribution for an outcome $x$ ($p(x|\Theta)$) contains about a parameter $\Theta$.
To estimate $\Theta$ after observing an outcome $x$, an estimator $\hat{\Theta}$ assigns an estimation $\hat{\Theta}_x$ of $\Theta$  to each outcome $x$. A loss function quantifies the error of the estimator.
We use the root mean squared error $\mathrm{RMSE} = \sqrt{\sum_x p(x|\Theta)(\hat{\Theta}_x - {\Theta})^2 }$ as the loss function. 
The RMSE considers both the bias $(B=\langle \hat{\Theta}_x \rangle_{p(x|\Theta)} - \Theta)$ and statistical variance (Var) of the estimator as follows: $\mathrm{RMSE} = \sqrt{\mathrm{Var} + B^2}$.

We now introduce established bounds on the statistical variance for a quantum mechanical experiment, where the experimental outcome $x$ and associated probability distribution $p(x|\Theta)$ results from a measurement of a state $\rho_{\Theta}$ with a positive operator valued measurement (POVM) element $Q_x$. For unbiased estimators ($B = 0$), those bounds also extend to the RMSE.
The Cram\'er-Rao bound~\cite{Rao1992,CramerHarald1946Mmos} lower-bounds the achievable variance of any unbiased estimator for a given $Q_x$ by
\begin{equation}\label{equation:cramer-rao}
    \mathrm{Var}(\hat{\Theta}_x) \geq \frac{1}{\mathcal{F}_C},
\end{equation}
where $\mathcal{F}_C$ is the classical Fisher~\cite{fisherMathematicalFoundationsTheoretical1922} information defined by 
\begin{equation}\label{equ:fisher}
    \mathcal{F}_C=\sum_x \frac{(\partial_\Theta p(x|\Theta))^2}{p(x|\Theta)}.
\end{equation}
The quantum Cram\'er-Rao bound generalizes the Cram\'er-Rao bound to a parameterized quantum state $\rho_{\Theta}$. 
The quantum Cram\'er-Rao bound lower-bounds the achievable variance for any unbiased estimator and for any POVM by
\begin{equation}
    \mathrm{Var}(\hat{\Theta}_{x}) \geq \frac{1}{\mathcal{F}_Q(\rho_\Theta)},
\end{equation}
where $\mathcal{F}_Q$ is the quantum Fisher information (QFI)~\cite{helstrom1969quantum,parisQuantumEstimationQuantum2009}. The QFI is given by 
\begin{equation}
    \mathcal{F}_Q(\rho_\Theta) = \mathrm{Tr}(\rho_\Theta L_\Theta^2),
\end{equation}
where $L_\Theta$ is the symmetric logarithmic derivative (SLD) defined via 
\begin{equation}\label{equ:sld}
    \frac{L_\Theta \rho_\Theta + \rho_\Theta L_\Theta}{2} = \partial_\Theta\rho_\Theta.
\end{equation}
The CFI reaches the maximal value set by the QFI only when the optimal measurement POVM is used~\cite{parisQuantumEstimationQuantum2009}.   
For any pure states, where an arbitrary initial state $\ket{\psi(0)}$ is mapped to $\ket{\psi(\Theta)} = \exp(-\ii \Theta G)\ket{\psi(0)}$ via a Hermitian generator $G$, the quantum Fisher information simplifies to $\mathcal{F}_Q=4 \mathrm{Var}(G)_{\ket{\psi(0)}}$.

In the main text, the error in our estimates for $B^q$ is quantified by $\overline{\mathrm{RMSE}} = \sqrt{\frac{n}{m}\sum_{i=1}^m (B^q_{\mathrm{cal}}-B^q_{\mathrm{est},i})^2}$.
The maximum likelihood estimator used in the main text is evaluated on estimates from $n$ shots of the experiment (See Section~\ref{appendix:mlegeneral} for details).
The joint probability distribution for the outcomes $\bm{x} = (x_1, ..., x_N)$ of $n$ independent shots of the experiment is given by $p(\bm{x}|B^q) = \prod_i p(x_i|B^q)$.
The CFI for that joint probability distribution is simply n times the CFI for the single-outcome probability distribution $p(x|B^q)$.
The same is true for the QFI, resulting in a QFI for the joint probability distribution of $n\mathcal{F}_Q$.
$\overline{\mathrm{RMSE}}$ is an approximation of the expectation value $\overline{\mathrm{RMSE}}' = \sqrt{n\langle (\hat{B}^q(\bm{x}) - B^q)^2 \rangle_{p(\bm{x}|B^q)}}$ from $m$ independent estimates (as defined in the main text).

We therefore get the following chain of inequalities\footnote{For the case when $\hat{B}^q(\bm{x})$ is unbiased.}

\begin{equation}\label{equ:bound_rmse}
    \overline{\mathrm{RMSE}}\approx \overline{\mathrm{RMSE}}' \geq \sqrt{n \mathrm{Var}} \geq \frac{1}{\sqrt{\mathcal{F}_C}} \geq \frac{1}{\sqrt{\mathcal{F}_Q}}.
\end{equation}
Finally, we remark that the inequalities are tight in the limit of $n\rightarrow\infty$ \cite{parisQuantumEstimationQuantum2009}, and the approximation is
exact in the limit of $m\rightarrow\infty$ independent estimates.

\subsubsection{The sensing Hamiltonian generalized to more sensor positions and fields.}
\label{appendix:generalhamiltonian}

Recall the Hamiltonian for sensor $i$ as presented in the main text: $H(x_i) = \hbar\kappa\sum_{k = 1}^{n_i}B(x_i)s_k^i\ket{s_k^i}\bra{s_k^i}$, where $s_k^i$ labels the $n_i$ levels of the sensor.
The arbitrary field $B(x_i)$ is decomposed into the static signal field component and fluctuating noise components in the following way $B(x_i) = B_{\mathrm{signal}} \Tilde{f}_{\rm signal}(x_i) + \sum_jB^j_{\rm noise}(t)\Tilde{f}^j_{\rm noise}(x_i)$, 
where the functions with tildes are unit-less spatial functions of the noise or signal fields.
The task is to estimate the parameter $B_{\rm signal}$ with spatial profile $\Tilde{f}_{\rm signal}$ in the presence of fluctuating noise fields $B^j_{\rm noise}(t) \Tilde{f}^j$, where each $B^j_{\rm noise}(t)$ fluctuate independently. 

The total Hamiltonian for all sensors $i$ is given by
\begin{equation}\label{eq:htot}
    H_{\mathrm{tot}} = \sum_i H(x_i) = \underbrace{\sum_i\hbar\kappa B_{\mathrm{signal}}\Tilde{f}_{\mathrm{signal}}(x_i)D^i}_{H_\mathrm{signal}} + \underbrace{\sum_i\sum_j\hbar\kappa B^j_{\mathrm{noise}}(t)\Tilde{f}^j_{\mathrm{noise}}(x_i)D^i}_{H_\mathrm{noise}},
\end{equation}
where we introduce $D^i= \sum_{k = 1}^{n_i} s_k^i \ket{s_k^i}\bra{s_k^i}$ for convenience. 
The corresponding time-evolving unitaries are
\begin{align}\label{eq:signalham}
    U_{\mathrm{signal}} &= \exp{(-\ii H_{\mathrm{signal}}t/\hbar)} = \exp{(-\ii B_{\mathrm{signal}}G_{\mathrm{signal}}})\\ 
    U_{\mathrm{noise}} &= \exp{(-\ii \int dtH_{\mathrm{noise}}/\hbar)},
\end{align}
where $G_{\mathrm{signal}}$ is the signal generator is generally given by $G_{\mathrm{signal}}=\kappa t\sum_i \Tilde{f}_{\rm signal}(x_i) D^i $. 

Since $[U_{\mathrm{signal}}, U_{\mathrm{noise}}] = 0$, the Hamiltonian time evolution of any initial sensor state $\rho(0)$ factors into 
\begin{equation}\label{eq:signoise}
    \rho(t) = U_{\mathrm{signal}}U_{\mathrm{noise}}\rho(0)U_{\mathrm{noise}}^\dagger U_{\mathrm{signal}}^\dagger.
\end{equation}
Section~\ref{appendix:correlateddephasingnoise} extends Equation \ref{eq:signoise} to the stochastic time-evolution due to the noise fields.

In the sensing task considered in the main text, we estimate the spatially-quadratic field component $B^q$ with $\bm{\Tilde{f}}^q = ({\Tilde{f}}^q_1,{\Tilde{f}}^q_2,{\Tilde{f}}^q_3) = d^2 (1,0,1)/2$, where $d$ is the distance between neighboring ions. 
Substituting $B_{\rm signal} = B^q$ and $\Tilde{f}_{\rm signal}(x_i)={\Tilde{f}}^q_i$ gives signal generator and time-evolving unitary
\begin{equation}\label{equ:signalGeneratorD5half}
    G^q = \kappa t d^2 (D^1+D^3)/2\hspace{2.5cm} U_{B^q} = \exp(-\ii B^qG^q)
\end{equation}
with $D^i=J^i_z/\hbar - \frac{\id^j}{2}$ and $J^i_z$ is the z-component of the total angular momentum operator in the $3^2 D_{5/2}$ manifold of sensor $i$. 

\subsubsection{Spatially-correlated noise and the emergence of decoherence free subspaces}\label{appendix:correlateddephasingnoise}
This section models the effect of stochastic time-dependent spatially-correlated fields on the time evolution of general sensor states. 
The concept of a decoherence-free subspace (DFS) and overwhelming correlated noise is then introduced.
Finally, the DFS which protects our sensors from spatially constant and gradient noise is identified. 

According to Equation \ref{eq:signoise}, the time evolution of a sensor state in the presence of $H_{\mathrm{tot}}$ (Equation \ref{eq:htot}) can be computed by first time evolving under the noise fields, and then under the signal field.
Here, we consider the first step of time evolving under the noise fields.
We consider the sensors to be prepared in an arbitrary initial state $\ket{\psi(0)} = \sum_{\bm s}\braket{\bm s|\psi(0)} \ket{\bm s}$, expressed in the basis of the sensor levels $\ket{\bm{s}}=\bigotimes_{i=1}^{N_{\mathrm{sensors}}} \ket{s_i}$, where $N_{\mathrm{sensors}}$ is the number of sensors. 
The sensors are exposed to the Hamiltonian $H_{\mathrm{noise}}$ for a time $T$, mapping the initial state to $\ket{\psi(T)}=\sum_{\bm s}\braket{\bm s|\psi(T)} \ket{\bm s}$, which has elements given by
\begin{align}\label{equ:appendix:time_pure}
    \bra{\bm s}{\psi(T)}\rangle = \exp\left(-\ii \kappa \int_0^T  \sum_j \bm{s}\cdot \bm{\Tilde{f}}^j_{\mathrm{noise}} B_{\mathrm{noise}}^j(t)dt\right)\bra{\bm s}{\psi(0)}\rangle = \exp\left(-\ii \kappa   \sum_j \bm{s}\cdot \bm{\Tilde{f}}^j_{\mathrm{noise}} \Bar{B}_{\mathrm{noise}}^j\right)\bra{\bm s}{\psi(0)}\rangle ,
\end{align}
where $\bm{\Tilde{f}}_{\mathrm{noise}}^j= \bm{f}_{\mathrm{noise}}^j/B_{\mathrm{noise}}^j$ are  the noisy field vectors with unit field strength and $\Bar{B}_{\mathrm{noise}}^j=\int_0^Tdt B_{\mathrm{noise}}^j(t)$ are the integrated noise field strengths.

Stochastically varying noise fields lead to a distribution $p(\bar{B}_{\mathrm{noise}}^j)$ of integrated field strengths.
Then, the noise channel is given as a statistical mixture of the pure evolution, 
according to $p(\bar{B}_{\mathrm{noise}}^j)$.
Explicitly, the effect of the noise on the initial state $\rho(0)$ is given by the noise channel $\mathcal{N}(\rho(0))=\sum_{\bm s \bm s'} \ket{\bm s'} \bra{\bm{s}'}\mathcal{N}(\rho(0))\ket{\bm{s}}\bra{\bm s}$, with elements given by
\begin{equation}\label{eq:noise}
    \bra{\bm{s}'}\mathcal{N}(\rho(0))\ket{\bm{s}} = \left\langle\int e^{-\ii\kappa t \sum_j  (\bm{s}-\bm{s'})\cdot \bm{\Tilde{f}}^j_{\mathrm{noise}} B_{\mathrm{noise}}^j(t)} dt\right\rangle_{\{p(\bar{B}_{\mathrm{noise}}^j)\}}  \bra{\bm{s}'}\rho(0)\ket{\bm{s}},
\end{equation}
where the mixture $\langle Y\rangle_{p(x)}=\sum_x Y p(x)$ is done with respect to all integrated noise field strengths $\{\bar{B}_{\mathrm{noise}}^j\}$. 
For matrix elements where $(\bm{s} - \bm{s'})\cdot\bm{\Tilde{f}}_{\mathrm{noise}}^j$ is non-zero, $\bar{B}_{\mathrm{noise}}^j$ generates a complex phase $\kappa t(\bm{s} - \bm{s'})\cdot\bm{\Tilde{f}}_{\mathrm{noise}}^j\bar{B}_{\mathrm{noise}}^j(t)$. 
When the value of the complex phase fluctuates strongly between shots of the experiment, it is approximately equally distributed on the unit circle, and the corresponding matrix element will average to zero. 
We define a noise field $\bm{f}_{\mathrm{noise}}^j$ to be overwhelming when
its application alone results in matrix elements of $\mathcal{N}(\rho(0))$, where $(\bm{s} - \bm{s'})\cdot\bm{\Tilde{f}}_{\mathrm{noise}}^j \neq 0$, being zero.
There are two cases when the exponent in Equation \ref{eq:noise} is zero: for diagonal elements ($\bm{s}=\bm{s'}$), and for decoherence-free subspaces (DFSs) where $(\bm{s}-\bm{s'}) \perp \bm{f}_{\mathrm{noise}}^j$ for all noise fields $j$. 
For these elements $
\mathcal{N}(\rho(0))_{\bm{s},\bm{s'}} = \rho(0)_{\bm{s},\bm{s'}}$, meaning that the elements are protected from the noise. The constraint $\bm{s} \cdot \bm f_{\mathrm{noise}}^j = 0$ in the main text is a special case with $\bm{s}=- \bm{s'}$ and sufficient to find the optimal SWD state.
In general there can be multiple DFSs, which we identify by the real numbers $\bm{\eta} = (\eta_1, ... )$ where $\eta_j = \bm{f}_{\mathrm{noise}}^j\cdot\bm{s}=\bm{f}_{\mathrm{noise}}^j\cdot\bm{s'}$  are the \enquote{energies} in the different noise fields.
Formally each DFS is defined by
\begin{equation}\label{equation:DFS}
    \mathrm{DFS}_{\bm{\eta}}= \mathrm{span}(\left\{\ket{\bm{s}}\Big\vert \bm{f}_{\mathrm{noise}}^j\cdot\bm{s} = \eta_j \right\}).
\end{equation}
A DFS has to have a dimension of at least two: the individual diagonal elements $\ket{\bm{s}}$ that are not in a DFS are summarized in $\perp$. 
The resulting noise channel for overwhelming noise is 
\begin{equation}\label{equ:correlated_noise}
    \mathcal{N}(\rho) = \bigoplus_{\bm{\eta}}p_{\bm{\eta}}\rho_{\bm{\eta}} + p_\perp \rho_\perp,
\end{equation}
where $p_{\bm{\eta}} = \tr(\Pi_{\bm{\eta}}\,\rho)$, $\rho_{\bm{\eta}} = \Pi_{\bm{\eta}} \rho \Pi_{\bm{\eta}} /p_{\bm{\eta}}$, 
\begin{equation}\label{eq:proj}
\Pi_{\bm{\eta}} = \sum_{\{\bm{s} \vert \bm{f}^j\cdot\bm{s} = \eta_j\}} \ket{\bm{s}}\bra{\bm{s}}
\end{equation} and $\rho_{\perp}= \frac{1}{p_\perp}\bigoplus_{\bm{s}\notin DFS} \rho_{\bm{s}\bm{s}}\ket{\bm{s}}\bra{\bm{s}}$. The latter are the diagonal elements orthogonal to any DFS with the normalization $p_\perp=1-\sum_{\bm{\eta}}p_{\bm{\eta}}=\sum_{\bm{s}\notin DFS} \bra{\bm{s}}\rho\ket{\bm{s}}$.

\begin{figure}
    \centering
    \includegraphics[width=0.33\linewidth]{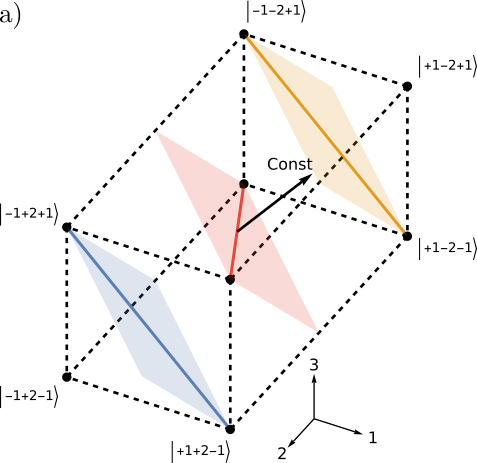}\includegraphics[width=0.33\linewidth]{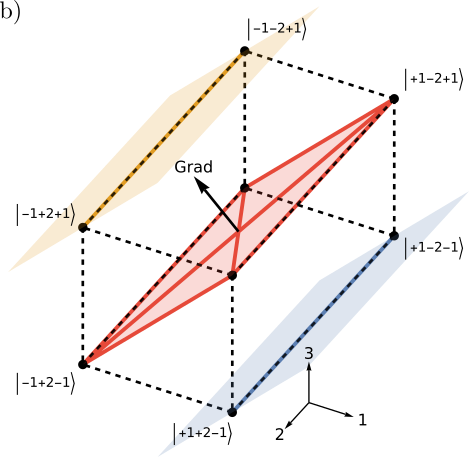}
    \includegraphics[width=0.33\linewidth]{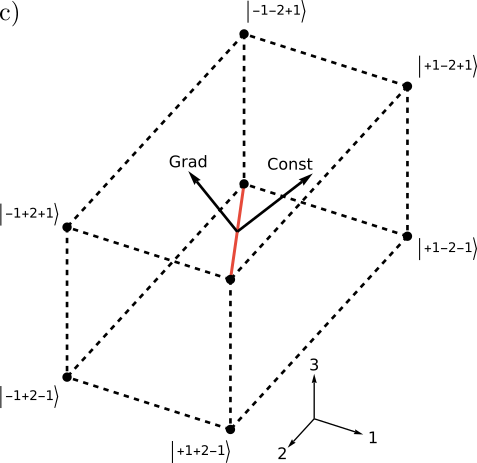}
    \caption{Visualizing DFSs protected from \textbf{a}) constant fields, \textbf{b}) gradient fields, and \textbf{c}) both constant and gradient fields. The eight sensor eigenstates within the reduced Hilbert space ($\mathrm{span}\left(\ket{\pm1,\pm2,\pm1}\right)$, bold levels in Figure 1 of the main text) are visualized as black points that form a three-dimensional grid, where axis 1, 2 and 3 (as labeled in the insets) represent the sensitivity of sensor 1, 2 and 3 respectively. Colored areas represent equipotential planes with respect to the corresponding fields. Solid lines highlight states within the same DFS. Yellow, red and blue colors represent DFSs with energies $\eta=-2,0,2$ respectively.}
    \label{fig:affineDFS}
\end{figure}
Figure~\ref{fig:affineDFS} visualizes all DFSs that are found in the scenario investigated in the main text. The sensors are prepared within the reduced Hilbert space given by the bold levels ($\mathrm{span}\left(\ket{\pm1,\pm2,\pm1}\right)$ in Figure 1 of the main text. 
In the case of only constant noise, Equation \ref{equation:DFS} identifies the following three DFSs
\begin{align}
    \mathrm{DFS}^c_{0} &= \mathrm{span}\left(\ket{1,-2,1},\ket{-1,2,-1}\right)\\
    \mathrm{DFS}^c_{\pm2} &= \mathrm{span}\left(\ket{\pm1,\pm2,\mp1},\ket{\mp1,\pm2,\pm1}\right).
\end{align}
In the case of only gradient noise, we obtain the following three DFSs 
\begin{align}
    \mathrm{DFS}^l_{0} &= \mathrm{span}\left(\ket{1,-2,1},\ket{1,2,1},\ket{-1,-2,-1},\ket{-1,2,-1}\right)\\
    \mathrm{DFS}^l_{\pm2} &= \mathrm{span}\left(\ket{\pm1,2,\mp1},\ket{\pm1,-2,\mp1}\right).
\end{align}
Finally, in the case of constant and gradient noise, the intersection of $\mathrm{DFS}^c_0$ and $\mathrm{DFS}_0^l$ gives the combined DFS
\begin{align}\label{eq:dfsexp}
    \mathrm{DFS}_{\mathrm{exp}} &= \mathrm{DFS}_{0,0} =\mathrm{span}\left(\ket{1,-2,1},\ket{-1,2,-1}\right).
\end{align}
which is protected from both spatially constant and gradient noise. 
$\mathrm{DFS}_{\mathrm{exp}}$ is the one used in this work.

The noise channel in the case of both noise fields is 
\begin{equation}\label{appendix:noiseideal}
\mathcal{N}(\rho) = p_{\mathrm{DFS}}\rho_{\mathrm{DFS}} + p_{\perp} \rho_\perp,
\end{equation}
where  $p_{\mathrm{DFS}}=\tr( \Pi_{\mathrm{DFS}}\,\rho)$, $\Pi_{\mathrm{DFS}}$ is the projector on $\mathrm{DFS}_{\mathrm{exp}}$  and $\rho_{\mathrm{DFS}} = \Pi_{\mathrm{DFS}}\rho \Pi_{\mathrm{DFS}}/p_{\mathrm{DFS}}$. That channel generates a block-diagonal structure for any initial state.

Now the effect of overwhelming correlated noise in the general case (Equation~\ref{equ:correlated_noise}) on the QFI, with respect to the signal field $\Tilde{f}_{\rm signal}$ (as defined in Section \ref{appendix:generalhamiltonian}), is presented.
The QFI is given by
\begin{align}
    \mathcal{F}_Q &= \tr(\mathcal{N}(\rho) L^2),
    \intertext{ Where $L$ is the SLD (Equation~\ref{equ:sld}) with respect to the signal field strenth $B_{\mathrm{signal}}$ (defined in Section \ref{appendix:generalhamiltonian}). That equation simplifies to}
    \mathcal{F}_Q &= \sum_{\bm\eta} p_{\bm\eta} \tr(\rho_{\bm{\eta}} L^2) + p_\perp \tr(\rho_\perp L^2).\\
    \intertext{The SLD within any DFS (labeled $\mathrm{DFS}_{\bm\eta}$) is given by the projection $L_{\bm\eta} = \Pi_{\bm{\eta}} L \Pi_{\bm{\eta}}$ into the $\mathrm{DFS}_{\bm\eta}$. Thereby the QFI}
    \mathcal{F}_Q &= \sum_{\bm\eta} p_{\bm\eta} \underbrace{\tr(\rho_{\bm{\eta}} L_{\bm{\eta}}^2)}_{\mathcal{F}_Q^{\bm\eta}}\label{equation:QFI:DFS}
\end{align}
simplifies to a convex combination of the QFI in the DFSs $\mathcal{F}_Q^{\bm\eta}$. The maximal QFI can always be obtained within a single DFS with the maximal spectral range $\Delta_{\bm \eta}$ (as introduced in Section~\ref{appendix:findoptimalstep}).
In the investigated scenario of overwhelming constant and linear noise with the chosen three two-level sensors, $\mathrm{DFS}_{\mathrm{exp}}$ is the only DFS, and the QFI is $p_{\mathrm{DFS}} \mathcal{F}_Q^{\mathrm{DFS}}$, where $\mathcal{F}_Q^{\mathrm{DFS}}=\tr(\rho_{\mathrm{DFS}} L^2)$ is the QFI in $\mathrm{DFS}_{\mathrm{exp}}$.

\subsubsection{Finding the sensor state with maximum QFI}\label{appendix:findoptimalstep}

The action of the signal field (with field strength $B_{\mathrm{signal}}$ as defined in Section \ref{appendix:generalhamiltonian}) onto any pure initial state $\ket{\psi_0}$ results in the state $\ket{\psi(B_{\mathrm{signal}})} = \exp(-\ii B_{\mathrm{signal}} G_{\mathrm{signal}})\ket{\psi_0}$, where the signal generator $G_{\mathrm{signal}}$ is introduced in Equation \ref{eq:signalham}.
The QFI is then given by $4\mathrm{Var}(G_{\mathrm{signal}})_{\ket{\psi_0}}$.
The state that maximizes the QFI is given by the superposition $\ket{\psi_0}=\left(\ket{\mathrm{max}}+\ket{\mathrm{min}}\right)/\sqrt{2}$, where $\ket{\mathrm{max}}$ is the eigenstate of $G_{\mathrm{signal}}$ with the maximal eigenvalue $\lambda_\mathrm{max}$, and $\ket{\mathrm{min}}$ has the minimal $\lambda_\mathrm{min}$.
This state reaches the maximum QFI of $\Delta^2$, where $\Delta=\lambda_\mathrm{max}-\lambda_\mathrm{min}$ is the spectral range of $G_{\mathrm{signal}}$~\cite{parisQuantumEstimationQuantum2009,giovannettiAdvancesQuantumMetrology2011}.

Having found the state which maximizes the QFI for a given generator, we apply this to the scenario of overwhelming correlated noise (as introduced in Section~\ref{appendix:correlateddephasingnoise}). In the presence of overwhelming spatially-correlated noise only coherences within a DFS are available.
The state which maximizes the QFI
of $B_{\mathrm{signal}}$ within any one $\mathrm{DFS}_{\bm{\eta}}$ is given by an equal superposition of the maximal and minimal eigenvectors of the projected generator $\hat{G}_{\bm\eta}=\Pi_{\bm{\eta}} \hat{G}_{\mathrm{signal}} \Pi_{\bm\eta}$, where the projectors $\Pi_{\bm\eta}$ are defined in Equation \ref{eq:proj}. 
From the convexity shown in Equation~\ref{equation:QFI:DFS} it follows that a DFS with the largest spectral range (which we label by $\bm{\eta} = \bm{\eta}_{\mathrm{opt}}$) contains a state with maximum QFI.
That state is given by $(\ket{\bm{s}_{\mathrm{max}}} + \ket{\bm{s}_{\mathrm{min}}})/\sqrt{2}$
, where $\ket{\bm{s}_{\mathrm{max}}}$ and $\ket{\bm{s}_{\mathrm{min}}}$ are the eigenvectors of the corresponding projected generator within that DFS with maximal and minimal eigenvalues.

We find the maximal/minimal eigenvalues by the optimization
\begin{equation}\label{equ:spectral_range}
    \lambda_{\mathrm{max}}^{\bm{\eta}} = \max_{\ket{\bm{s}}} \bra{\bm s}\hat{G}_{\bm{\eta}_{\mathrm{opt}}}\ket{\bm s} = \kappa t\max_{\ket{\bm{s}} \in \mathrm{DFS}_{\bm{\eta}_{\mathrm{opt}}}} \bm{\Tilde{f}}_{\mathrm{signal}}\cdot\bm{s} \quad\text{and}\quad \lambda_{\mathrm{min}}^{\bm{\eta}} = \kappa t\min_{\ket{\bm{s}} \in \mathrm{DFS}_{\bm{\eta}_{\mathrm{opt}}}} \bm{\Tilde{f}}_{\mathrm{signal}}\cdot\bm{s}
\end{equation}
where the sensor-field interaction strength $\kappa$ and the sensing time $t$ are introduced in the main text, and we introduce the signal field vector $\bm{\Tilde{f}}_{\mathrm{signal}}$, the elements of which are the values of the function $\Tilde{f}_{\mathrm{signal}}$ (as defined in Section \ref{appendix:generalhamiltonian}) at the position of the corresponding sensor.
The corresponding QFI is given by $\Delta^2_{\bm{\eta}}$, where we define $\Delta_{\bm{\eta}} = \lambda_{\mathrm{max}}^{\bm{\eta}} - \lambda_{\mathrm{min}}^{\bm{\eta}}$.
The sensitivity of the $i$-th sensor $s_i$ is typically constrained by the system to a maximal value.
There are two cases of the optimization. First, $s_i$ might only be allowed to take discrete values within the set of available sensitivities ($\mathcal{S}$) in the system; then, the optimization is discrete and on a grid (a minimal example is shown in Figure \ref{fig:affineDFS}). The second case, where we can modify $s_i$ values continuously within $\mathcal{S}$, provides an upper bound to the optimization result for the discrete case. For the second case, it has been shown that $\bm{\eta}_{\mathrm{opt}}=0$  and $\bm{s}_{\mathrm{max}}=-\bm{s}_{\mathrm{min}}$ is optimal~\cite{sekatskiOptimalDistributedSensing2020}. 
This result was used in the main text to identify the SWD state $\ket{\psi^{\mathrm SWD}} = (\ket{1,-2,1}-\ket{-1,2,-1})/\sqrt{2}$.

It has been shown that the SWD sensor state, found by the procedure detailed above, achieves a quantum Fisher information ($\mathcal{F}_Q$) that scales with Heisenberg scaling ($\mathcal{F}_Q\propto N^2$) in the number of sensors $N$\cite{hamannApproximateDecoherenceFree2022}.

\subsubsection{The maximum likelihood estimator}\label{appendix:mlegeneral}

To estimate $B^q$ after observing an outcome $x$ given a probability distribution for observing that outcome $p(x|B^q)$ (as introduced in Section \ref{appendix:cramerrao}), we use the maximum likelihood estimator (MLE)
\begin{equation}
    \hat{B}^q = \mathrm{argmax}_{B^q} p(x|B^q).
\end{equation}
The MLE is proven to minimize the $\overline{\mathrm{RMSE}}$ and saturates the Cram\'er-Rao bound (Equation \ref{equation:cramer-rao}) in the limit of infinite shots $n\to\infty$~\cite{parisQuantumEstimationQuantum2009}. 
After $n$ shots of the experiment, outcomes $\pm$ (as introduced in Section \ref{appendix:parity_measurement}) are observed $n_{\pm}$ times.
The total likelihood is given by $p(B^q|n_+,n_-)=p(+|B^q)^{n_+}p(-|B^q)^{n_-}$, where $p(\pm|B^q) = 1/2\pm A \cos(\omega B^q + 3\phi_r + \phi_0)/2$ is the probability of measuring outcome $\pm$ given a field $B^q$. That probability extends the one defined in Equation~\ref{eq:sepprob} by the inclusion of an amplitude $A$ that accounts for imperfections in state preparation, and by a phase offset $\phi_0$ that is determined by the phase of the initial state (Section \ref{appendix:phi0}). 
The $\hat{B}^q$ estimator which maximizes the total likelihood is given by the following analytical expression
\begin{equation}\label{equation:MLE}
    \hat{B}^q(P) = \frac{\pi-3\phi_r-\phi_0+2\pi \lfloor\mu/2\rfloor}{\omega}+\frac{(-1)^\mu}{\omega}\cdot\begin{cases}
			\arccos(P/A)\\
            \pi, & \text{if $\frac{P}{A} < -1$}\\
            0 & \text{if $\frac{P}{A}>1$}\\ 
		 \end{cases}.
\end{equation}
Here $P=(n_+-n_-)/n$ is the parity estimate, and $\mu = \lfloor\frac{\omega B^q_\mathrm{cal} + 3 \phi_r + \phi_0}{\pi}\rfloor$ denotes the expected fringe of the cosine, which we assume to be known.

\subsection{Proving the SWD protocol minimizes the achievable $\mathrm{\overline{RMSE}}$}\label{appendix:SWDoptimal}
Using the initial state and measurement of the SWD protocol it is possible to achieve the minimum $\mathrm{\overline{RMSE}}$ over all states and measurements in the $3^2 D_{5/2}$ manifold that are insensitive to the noise fields, as set by the Heisenberg limit.
Here we prove this statement. 
The classical and quantum Cram\'er-Rao bound presented in Section~\ref{appendix:cramerrao} lead to the following three steps for identifying the state and measurement that minimize the $\mathrm{\overline{RMSE}}$.
First, we choose the state that maximizes the QFI. Second, we identify a measurement that, when paired with the chosen state, maximizes the CFI. Finally, the minimal $\mathrm{\overline{RMSE}}$ is achieved by the maximum likelihood estimator in the infinite limit $n\to \infty$. The first two of those steps are presented separately in the following two subsections. The final step requires no further explanation.

\subsubsection{The SWD state maximizes the QFI}\label{appendix:chooses}

According to the inequality in Equation \ref{equ:bound_rmse}, states that minimize the achievable $\mathrm{\overline{RMSE}}$ also maximize the QFI\footnote{If the quantum Cram\'er-Rao bound is saturable, which it is for the scenario considered in the paper (\ref{appendix:cramerrao}). }.
The procedure for finding the state with the maximal QFI is given in Section \ref{appendix:findoptimalstep}. As a reminder:  one finds a DFS with the maximal spectral range and then prepares the state $(\ket{\bm{s}_{\mathrm{max}}}+\ket{\bm{s}_{\mathrm{min}}})/\sqrt{2}$, where the states $\ket{\bm{s}_{\mathrm{max}}}$ and $\ket{\bm{s}_{\mathrm{min}}}$ have the maximal and minimal energy within the DFS, respectively.

We first identify the DFS with the maximal spectral range within the $3^2 D_{5/2}$ manifold of the three sensors that is protected from spatially constant and linear dephasing.
The manifold is spanned by $6^3$ basis states $\ket{s_1,s_2,s_3}$, where $s_i$ can take integer values from $-2$ to $3$.
Using Equation \ref{equation:DFS}, we identify all $68$ DFSs, $32$ of which have a maximal spectral range of $\Delta=\omega$, where $\omega$ is the frequency of the parity oscillations introduced in Equation~1 of the main text. $\mathrm{DFS}_{\mathrm{exp}}$ --- the DFS used in the main text, as defined in Equation \ref{eq:dfsexp} --- attains the maximal spectral range.

We next find $\ket{\bm{s}_{\mathrm{max}}}$ and $\ket{\bm{s}_{\mathrm{min}}}$ within $\mathrm{DFS}_{\mathrm{exp}}$, using the optimization in Equation \ref{equ:spectral_range}, yielding $\bm{s}_{\mathrm{max}} = - \bm{s}_{\mathrm{min}} = (1, -2, 1)$.
We have now identified the state with the overall maximal QFI for the scenario of this work: $(\ket{1, -2, 1} + \ket{-1, 2, 1})/\sqrt{2}$. We have thus recovered $\ket{\psi_{(1, -2, 1)}^{\mathrm{SWD}}}$ as the optimal state, which achieves the maximum of 
\begin{equation}
    \mathcal{F}_Q^{\mathrm{SWD}} = \omega^2.
\end{equation}
This state saturates the Heisenberg limit, which corresponds to the maximum achievable QFI in the presence of the correlated dephasing noise. 

\subsubsection{The parity measurement maximizes the CFI}\label{appendix:parity_measurement}

According to the inequality in Equation \ref{equ:bound_rmse}, measurements that minimize the achievable $\mathrm{\overline{RMSE}}$ also maximize the CFI.
This section verifies that the parity measurement used in the experiment --- paired with the SWD state found in Section \ref{appendix:chooses} --- maximizes the CFI.
The parity measurement is implemented by the measurement operator $M = U_{\phi_r}^\dagger \sigma^1_z\sigma^2_z\sigma^3_z U_{\phi_r}=\sigma^1_{\phi_r}\sigma^2_{\phi_r}\sigma^3_{\phi_r} $, where $\sigma^i_{\phi_r}=\cos{\phi_r}\sigma^i_y-\sin{\phi_r}\sigma^i_x$, $\sigma^i_k$ is the Pauli $k$ matrix acting on the $i$-th sensor and $\phi_r$ is the measurement phase defined in the main text.
One can explicitly compute the outcome probability distribution 
\begin{equation}\label{eq:sepprob}
    p(\pm|B^q) = \bra{\psi_{(1, -2, 1)}^{\mathrm{SWD}}}U_{B^q}^\dagger\Pi_\pm U_{B^q}\ket{\psi_{(1, -2, 1)}^{\mathrm{SWD}}} = \frac{1 \pm \cos(\omega B^q +3\phi_r)}{2},
\end{equation}
for an ideal implementation, where $\Pi_\pm$ are the projectors onto the $\pm$ eigenspaces of $M$, and the unitary $U_{B^q}$ is defined in Equation \ref{equ:signalGeneratorD5half}. 
In the experiment, the parity estimation is extracted from the outcome probabilities via $\braket{M} = p(+|B^q) - p(-|B^q)$. From Equation \ref{eq:sepprob}, one finds $p(+|B^q) - p(-|B^q) = \cos(\omega B^q +3\phi_r)$,
which recovers the form of Equation 1 in the main text, apart from an extra phase $\phi_0$ which is determined by the phase of the initial state (Section \ref{appendix:phi0}), and an amplitude $A$ accounting for imperfections in state preparation.
From the outcome probability distribution $p(\pm|B^q)$, one obtains a CFI of 
\begin{equation}\label{appendix:CFISWD}
    \mathcal{F}_C^{\mathrm{SWD}} = \omega^2,
\end{equation}
in the case in which $\omega B^q +3\phi_r = \pi/2$, which is equal to the QFI of $\omega^2$ found in Section \ref{appendix:chooses}, and is therefore the maximum achievable CFI.

In the case $A \leq 1$, the CFI given in Equation \ref{appendix:CFISWD} becomes $\mathcal{F}_C^{\mathrm{SWD}} = (A\omega)^2$. The state fidelity ($F$) of the SWD state is bounded by $F \leq \frac{1 + A}{2}$, with the inequality being saturated when all imperfections are caused by dephasing. The CFI can then be written as the inequality $\mathcal{F}_C^{\mathrm{SWD}} \leq (\omega(2F - 1))^2$. The RMSE is bounded by the CFI via Equation \ref{equ:bound_rmse}. 

\subsection{Minimizing the $\mathrm{\overline{RMSE}}$ for separable protocols}

For separable protocols, we introduce the restriction that both the state and the measurement are separable and therefore don't utilize entanglement.
Restricting the measurement to be separable means that the QFI is not an appropriate general metric, as the bound provided by the QFI assumes that entangled measurements are available.
Therefore, the three steps used to minimise the $\mathrm{\overline{RMSE}}$ for the SWD protocol in Section \ref{appendix:SWDoptimal} don't apply. 
Here, an appropriate metric is the CFI, which lower-bounds the achievable $\mathrm{\overline{RMSE}}$ for a given state and measurement by Equation \ref{equ:bound_rmse}.
From that equation, states and measurements that minimize the achievable $\mathrm{\overline{RMSE}}$ also maximize the CFI.
Maximizing the CFI is achieved by jointly maximizing over both the initial state and measurement.

A full numerical optimization for the scenario of this work --- three noise-protected sensors in the $3^2 D_{5/2}$ manifold --- is given in Section \ref{appendix:ideal_sep_six} and yields the six-level separable protocol.
In Section \ref{appendix:ideal_sep_two}, we use the same numerical optimization of the CFI to verify that the separable protocol is optimal when restricting to two-level separable states and separable projective measurements on the $3^2 D_{5/2}$ manifold.

\subsubsection{The six-level separable state and measurement}\label{appendix:ideal_sep_six}
We found another separable protocol that utilizes all six sensor levels in the $3^2 D_{5/2}$ manifold, which outperforms the ideal two-level protocol but necessarily falls short of the ideal SWD protocol.
The measured SWD protocol $\mathrm{\overline{RMSE}}$ of \SI{6.7(3)}{\pico\tesla\per\micro\meter\squared} does not outperform the value of \SI{5.42}{\pico\tesla\per\micro\meter\squared} achieved by an ideal implementation of that six-level separable protocol. Outperforming that protocol would require achieving $A>0.454$, which is achievable in our system with improvements to the phase stability of our \SI{729}{\nano\meter} laser.
In this section, we explicitly show the methods used to find the six-level separable protocol. We don't experimentally realize the six-level separable protocol for comparison here; the control techniques to allow that are only currently being developed~\cite{ringbauerUniversalQuditQuantum2022}.

The six-level separable protocol is obtained by optimizing the classical Fisher information (CFI) jointly over separable states and measurements, which utilize the entire $3^2D_{5/2}$ manifold, which we now describe in detail. 
The initial state is given by $\ket{\psi}=\ket{\psi_1}\otimes\ket{\psi_2}\otimes\ket{\psi_3}$, 
where each $\ket{\psi_i}=\frac{\bm{a}^i + \ii\bm{b}^i}{\sqrt{|\bm{a}^i + \ii\bm{b}^i|}} \in \mathbb{C}^6$ is specified by vectors $\bm{a}^i = (a^i_1,...,a^i_6)$ and 
$\bm{b}^i=(b^i_1,...,b^i_5,b^i_6=0)$ of real numbers (yielding 33 real parameters in total). 
The constraint $b^i_6=0$ fixes the physically irrelevant global phase of each state. The measurements are given by the eigenvectors $\ket{m}$ of the observable $O^1\otimes O^2\otimes O^3$, where $O^i$ are defined by $6\times6$ Hermitian matrices
\begin{equation}
    O_i = \left(
\begin{array}{cccccc}
 o_{i11} & o_{i12}+\ii o^c_{i12} & o_{i13}+\ii o^c_{i13} & o_{i14}+\ii o^c_{i14} & o_{i15}+\ii o^c_{i15} & o_{i16}+\ii o^c_{i16} \\
 o_{i12}-\ii o^c_{i12} & o_{i22} & o_{i23}+\ii o^c_{i23} & o_{i24}+\ii o^c_{i24} & o_{i25}+\ii o^c_{i25} & o_{i26}+\ii o^c_{i26} \\
 o_{i13}-\ii o^c_{i13} & o_{i23}-\ii o^c_{i23} & o_{i33} & o_{i34}+\ii o^c_{i34} & o_{i35}+\ii o^c_{i35} & o_{i36}+\ii o^c_{i36} \\
 o_{i14}-\ii o^c_{i14} & o_{i24}-\ii o^c_{i24} & o_{i34}-\ii o^c_{i34} & o_{i44} & o_{i45}+\ii o^c_{i45} & o_{i46}+\ii o^c_{i46} \\
 o_{i15}-\ii o^c_{i15} & o_{i25}-\ii o^c_{i25} & o_{i35}-\ii o^c_{i35} & o_{i45}-\ii o^c_{i45} & o_{i55} & o_{i56}+\ii o^c_{i56} \\
 o_{i16}-\ii o^c_{i16} & o_{i26}-\ii o^c_{i26} & o_{i36}-\ii o^c_{i36} & o_{i46}-\ii o^c_{i46} & o_{i56}-\ii o^c_{i56} & o_{i66} \\
\end{array}
\right)
\end{equation} parameterized by 36 real numbers. 
This parametrization leads to a total of $108$ real parameters for the measurement observable ($36$ for each $O^i$). 
For the optimization of the CFI, we choose the case $B^q=0$; this can be done without loss of generality as the optimal CFI value is the same for all $B^q$ by transforming the optimal state via the unitary  $U_{B^q}^\dagger$ (which is defined in Equation \ref{equ:signalGeneratorD5half}).
According to Equation \ref{equ:fisher}, the CFI is computed by 
\begin{align}
    \mathcal{F}_C &= \sum_{m} \frac{(\partial_{B^q} p(m|B^q))^2}{p(m|B^q)}\Bigg\vert_{B^
    q=0}\\
    \intertext{with outcome probabilities $p(m|B^q) = \bra{m}U_{B^q}\mathcal{N}\big(\ket{\psi}\hspace{-1mm}\bra{\psi}\big)U_{B^q}^\dagger\ket{m}$, where the noise channel $\mathcal{N}$ is introduced in Section \ref{appendix:correlateddephasingnoise}. The equation above can be re-expressed as}
    \mathcal{F}_C &=\sum_{m} \frac{\left(-2 \mathrm{Im}\left(\bra{m}G^q\,\mathcal{N}\big(\ket{\psi}\hspace{-1mm}\bra{\psi}\big)\ket{m}\right)\right)^2}{\bra{m}\mathcal{N}\big(\ket{\psi}\hspace{-1mm}\bra{\psi}\big)\ket{m}},  
\end{align}

\noindent where the generator $G^q$ is defined in Equation \ref{equ:signalGeneratorD5half}.

Numerical optimization of the CFI over all $141$ parameters provides $\mathcal{F}_C = 3.3\frac{\omega^2}{16}$. 
In order to verify the robustness of the optimization, three additional runs were performed. Each run replaces $p(m|B^q)$ with $(1-\epsilon) p(m|B^q) +\frac{\epsilon}{6^3}$, which mimics the effect of depolarizing noise, where $\epsilon\in(0.01, 0.05, 0.1)$ is the depolarization probability. 
The results confirm the noise-free optimization.
Using the corresponding MLE, the six-level protocol obtains the $\overline{\mathrm{RMSE}}\geq(\sqrt{3.3}\frac{\omega}{4})^{-1}=\SI{5.4}{\pico\tesla\per\square\um}$ in the limit of infinite shots ($n\to\infty$). Therefore, the ideal six-level protocol outperforms the ideal two-level separable protocol ($\mathrm{\overline{RMSE}} \geq$  \SI{9.847}{\pico\tesla\per\square\um}) but not the ideal SWD protocol ($\mathrm{\overline{RMSE}} \geq$  \SI{2.462}{\pico\tesla\per\square\um}).

\subsubsection{The two-level separable protocol}\label{appendix:ideal_sep_two}
In the main text, we write: \enquote{The initial state and measurement of the separable protocol can achieve the minimum $\mathrm{\overline{RMSE}}$ over all noise-protected separable states restricted to occupying two levels of the $3^2 D_{5/2}$ manifold and projective measurements that don't utilize entanglement}. This section presents how this result was obtained. 

Our approach to minimize the $\mathrm{\overline{RMSE}}$ is to maximize the classical Fisher information (CFI), which are related via Equation~\ref{equ:bound_rmse}. The CFI is maximized by constraining the state of the optimization in Sec.~\ref{appendix:ideal_sep_six} to the bold levels of Figure 1 in the main text.
The optimization yields a maximum CFI of 
\begin{equation}\label{appendix:CFIsep}
    \mathcal{F}_C^{\mathrm{sep}} = \frac{\omega^2}{16},
\end{equation}
obtained by the initial state $\ket{\psi_{(1, -2, 1)}^{\mathrm{sep}}}$ and measurement of the separable protocol. Here $\omega$ is proportional to the sensor-field interaction as defined in Equation 1 of the main text. 
The action of the full noise channel on the initial separable state is to partially project it into the two-dimensional DFS$_{\mathrm{exp}}$ (defined in Equation \ref{eq:dfsexp}) which has a spectral range of $\Delta=\omega$ (introduced in Section \ref{appendix:findoptimalstep}).

A full numerical optimization over all possible two-level sensors is computationally impractical. However, such optimization would only yield states with an equal or poorer CFI of $\mathcal{F}_C\leq \frac{\omega^2}{16}$. We come to this conclusion by analyzing the effect of our full noise channel for all possible combinations of two-level sensors. Each combination has an associated eight-dimensional Hilbert space that falls into one of three categories:
\begin{enumerate}
    \item The Hilbert space fully dephases, and therefore any state and measurement have a CFI $\mathcal{F}_C=0$.
    \item The Hilbert space is isomorphic to the bold levels in Figure 1, and the state is partially projected to a DFS that has the same spectral range $\Delta=\omega$ as DFS$_{\mathrm{exp}}$. The optimization thus yields the same CFI of $\mathcal{F}_C=\frac{\omega^2}{16}$.
    \item The Hilbert space is isomorphic to the bold levels, but the states are partially projected into a DFS that has half the spectral range $\Delta=\omega/2$ as DFS$_{\mathrm{exp}}$. Here, the optimization yields a CFI of $\mathcal{F}_C=\frac{\omega^2}{64}$.
\end{enumerate}
In conclusion, the initial state $\ket{\psi_{(1, -2, 1)}^{\mathrm{sep}}}$ and measurement of the separable protocol is the optimal two-level protocol with projective measurements in the $3^2 D_{5/2}$ manifold that doesn't utilize entanglement. 
Combined with the Cramer-Rao bound (\ref{equ:bound_rmse}), this two-level protocol reaches $\mathrm{\overline{RMSE}} \geq  \SI{9.847}{\pico\tesla\per\square\um}$.
\\
On applying the ideal noise channel (defined in Equation \ref{appendix:noiseideal}) to the separable state with $A\leq1$, its fidelity can be expressed by $F = \frac{4}{5}(1 + A)$. 
The CFI given in Equation \ref{appendix:CFIsep} becomes $\mathcal{F}_C^{\mathrm{sep}} = (A\omega)^2 = ((5F - 4)\omega/4)^2$. 

\subsubsection{Intuition for the two-level separable protocol}
\label{appendix:theseparableprotocol}
The intuition for the two-level separable protocol is that it approximates the SWD protocol in the following way. When overwhelming spatially constant and gradient noise is applied to the separable state $\ket{\psi_{(1, -2, 1)}^{\mathrm{sep}}}$, the result is equivalent to preparing the state $\ket{\psi_{(1, -2, 1)}^{\mathrm{SWD}}}$ with success probability 1/4; the highest success probability over all separable states. With probability 3/4, states that live outside the DFS$_{\mathrm{exp}}$ are prepared. 
The SWD protocol is therefore successfully implemented for $1/4$ of the shots, and the unsuccessful $3/4$ provide a random outcome. 
These random outcomes commensurately reduce the amplitude of the sinusoidal parity $A$ (as defined in the main text) to $1/4$. This results in a CFI that is a factor $A^2 = 1/16$ smaller. 

\subsection{Exponential advantages of the SWD protocol}
\label{appendix:exp}

In the main text, we write: \enquote{Consider the task of sensing the highest order of an $M$ order Taylor expansion, whilst being insensitive to fluctuations in all lower orders, using the minimum number of sensors, $N_{\mathrm{min}}$. 
That task requires a minimum of $N_{\mathrm{min}} = M+1$ sensors all at different locations. We find that the SWD protocol achieves a lower $\mathrm{\overline{RMSE}}$ compared to the best two-level separable protocol by an amount that grows exponentially in $M$. 
Moreover, we prove that this exponential advantage extends to arbitrary spatial field distributions}.
This exponential advantage is now proven for arbitrary spatial field distributions. 
After dealing with arbitrary spatial field distributions, we finally prove the case of an $M$ order Taylor expansion explicitly. 
The proof employs two-level sensors with levels $\ket{\pm s_i}$\footnote{Since the transformation $s_{1, 2}' \rightarrow s_{1, 2} + c$ of the two sensitivities $s_1$ and $s_2$, where $c$ is a real number, corresponds to a physically irrelevant global phase, any two-level sensor can always be expressed in this symmetric form $\ket{\pm s_i}$.} that have tunable sensitivity values $s_i$, given by $0\leq s_i \leq s_{\mathrm{max}}$.
Section \ref{appendix:tunesensitivity} summarises how such tunable sensitivities could either be directly engineered or approximated. The proof makes no assumptions about the size of fluctuations of the noise fields, however it is predicated on the condition that the sensing protocols are insensitive to those fluctuations. In the presence of overwhelming noise in all the noise fields, the exponential advantage exists even without requiring the sensing protocol to be insensitive to those fluctuations. Quantifying the advantage between the SWD protocol and the best two-level separable protocol when the number of sensors used is not minimal remains an open question.

First, we define the conditions under which the necessary DFS exists that is protected from a set of $k$ noise fields. 
Consider $l$ sensors that can each be placed at distinct positions, $x_i$.  
A DFS of those sensors, that is protected from $k$ noise fields, with continuous spatial functions $\Tilde{f}_{\rm noise}^j$ for $1\leq j\leq k$\footnote{As in Section \ref{appendix:generalhamiltonian}, spatial functions with tildes are unit-less.}, exists iff
\begin{equation}\label{eq:generaldfscondition}
    \exists x_1,...,x_l \in X \Big | \mathrm{rank}(Q) < l,
\end{equation}
where $X$ is the set of the available sensor locations and we have introduced the noise matrix $Q$ with elements $Q_{ji}= \Tilde{f}^j(x_i)_{\rm noise}$.
We define $N_{\mathrm{min}}$ as the minimal number of distributed sensors $l$ that fulfill Equation \ref{eq:generaldfscondition}.
In the worst case, when all $k$ noise vectors (rows of Q) are linearly independent --- which is mathematically equivalent to $\mathrm{rank}(Q) = k$ --- then one obtains $N_{\mathrm{min}} = k + 1$.
The kernel of Q for that minimal number of sensors is one-dimensional.

Next, we explicitly find the protected DFS for sensors of minimal size $N_{\mathrm{min}}$, making use of the following property. 
Two distinct states within the same DFS ($\ket{\bm{s}},\ket{\bm s'}\in \mathrm{DFS} $) cannot share the same local sensitivity (for all sensors $i$, $s_i \neq s_i'$), since the DFS would still exist when the corresponding sensor (sensor $i$) is traced out, violating the condition that the sensors are of minimal size. Since each sensor has levels of the form $\ket{\pm s_i}$, those two distinct states must therefore be of the form $\ket{\bm{s}}, \ket{-\bm{s}}$. 
The energy of such a DFS is given by $\bm{\eta} = Q \bm{s} = Q(-\bm{s})$, where $\bm{\eta}$ is defined in Equation \ref{equation:DFS}. There is only one solution, $\bm{\eta} = 0$, for non-zero $\bm s$. 
Any possible DFS therefore has the form
\begin{align}\label{equ:DFS0}
    \mathrm{DFS}_{\bm{0}}=\mathrm{span}\left\{\ket{\bm{s}}\Big| Q\bm{s} = 0\right\} = \mathrm{span}\left\{\ket{\alpha\bm{\Tilde{s}}_\perp},\ket{-\alpha\bm{\Tilde{s}}_\perp}\right\},
\end{align}
where we introduce $\alpha$ as a free scaling parameter, and $\bm{\Tilde{s}}_{\perp}$ is chosen to have a maximal absolute component of one.
The optimal DFS, which is the one with the largest spectral range $\Delta$  (as proven in Section \ref{appendix:findoptimalstep}), is obtained by choosing $\alpha = s_{\mathrm{max}}$.

So far we have established the conditions for a minimal set of sensors that possess a DFS that is protected from the noise fields, and that such a DFS is two-dimensional. Now, we compare the maximal QFI for separable and entangled states of the sensors, with respect to a signal field $\Tilde{f}_{\rm signal}$ that is linearly independent to all noise fields.  The approach is to follow the general procedure for finding the state which optimizes QFI within a given DFS in Section \ref{appendix:findoptimalstep}. 

The state within $\mathrm{DFS}_{\bm{0}}$ that has the maximal QFI with respect to the signal vector $\bm{\Tilde{f}}^{\mathrm{signal}}$ (with elements $\bm{\Tilde{f}}^{\mathrm{signal}}_i= \Tilde{f}_{\mathrm{signal}}(x_i)$) is given by $(\ket{\bm{s}_\perp} +\ket{-\bm{s}_\perp})/\sqrt{2}$, where $\bm{s}_\perp=s_{\mathrm{ max}} \bm{\Tilde{s}}_\perp$. The QFI is given by the squared spectral range $\Delta^2_{\bm{0}} = (2\kappa t\bm{\Tilde{f}}_{\mathrm{signal}}\cdot\bm{s}_\perp)^2$, where the sensor-field interaction $\kappa$ and sensing time $t$ are introduced in the main text.
The initial state of any product strategy is given by
$\ket{\psi} = \bigotimes_i( a_i\ket{s_i} +b_i\ket{-s_i})$, where $a_i, b_i$ are complex numbers with the normalization constraint $|a_i|^2 + |b_i|^2 = 1$. 
Applying Equation \ref{equation:QFI:DFS} to the product state $\ket{\psi}$ and the only existing DFS ($\mathrm{DFS}_{\bm 0}$ as shown above) results in a QFI of
\begin{align}\label{eq:fisherexponential}
    \mathcal{F}_Q &= p_{\bm{0}} \mathrm{Var}(G_{\bm 0})_{\ket{\psi_{\bm{0}}}}\\
    &= \Delta_{\bm{0}}^2 \left(p_{\bm{0}} - \frac{p(\bm{s})-p(-\bm{s})}{p_{\bm{0}}}\right),
\end{align}
where $\bm{s} = (s_1,...)$, $p_{\bm{0}} = p(\bm{s})+p(-\bm{s})$, $p(\bm{s}) = \prod_i |a_i|^2 $, $p(-\bm{s}) = \prod_i |b_i|^2 $ and $G_{\bm 0}$ is the generator of the signal field within $\mathrm{DFS}_{\bm{0}}$ (as defined in Section Section~\ref{appendix:findoptimalstep}). 
From Equation \ref{eq:fisherexponential} the maximal QFI is $2^{1-N_{\mathrm{min}}}\Delta_{\bm 0}^2$, obtained in the case $\bm{s}=\bm{s}_\perp$ and $a_i = b_i = \frac{1}{\sqrt{2}}$, which decreases exponentially in $N_{\mathrm{min}}$.
Finally, the quantum Cram\'er-Rao bound (as defined in Equation \ref{equ:bound_rmse}) relates the QFI with the $\mathrm{\overline{RMSE}}$.
We have therefore found that the SWD protocol achieves an exponentially lower $\mathrm{\overline{RMSE}}$ as compared to the best two-level separable protocol for arbitrary spatial field distributions, when the minimum number of sensors are used. 

Finally, we explicitly show how the proof applies to the case of sensing the highest order of an $M$ order Taylor expansion, whilst being insensitive to fluctuations in all lower orders. In this case, there are $k = M$ linearly independent noise fields, $\Tilde{f}_{\rm noise}^j(x) = x^{j-1}/(j-1)!$ where $j$ takes on integer values in the range $1\leq j\leq M$. The signal field is given by $\Tilde{f}_{\rm signal}^j(x) = x^M/M!$. 
Given $l$ distributed sensors, where each sensor $l$ is at a distinct position $x_l$, the noise matrix has elements $Q_{jl} = (x_l)^{j-1}/(j-1)!$, and the signal vector has elements $\bm{\Tilde{f}}^{\mathrm{signal}}_l=x_l^{M}/M!$. 
The matrix Q has the form of a Vandermonde matrix~\cite{matrixAnalysis}, therefore its rank is $\mathrm{rank}(Q) = \min(M,l)$. Condition~\ref{eq:generaldfscondition} is therefore satisfied by a minimum of $N_{\mathrm{min}} = M + 1$ distinct sensor positions.
The QFI of the SWD and separable protocols then follows the expressions given in the paragraph above, substituting the noise matrix $Q$ and spectral range $\Delta_0$ for this considered task. This confirms that the QFI for the separable protocol is smaller than that for the SWD protocol by an amount that is exponential in $M$. The QFI bounds the achievable $\mathrm{\overline{RMSE}}$ by Equation \ref{equ:bound_rmse}. We have proven that for the task of sensing the highest order of an $M$ order Taylor expansion, whilst being insensitive to fluctuations in all lower orders using the minimum number of sensors, the SWD protocol achieves a lower $\mathrm{\overline{RMSE}}$ compared to the best two-level separable protocol by an amount that grows exponentially in $M$.
More examples of exponential advantages, for specific fields distributions, are provided in~\cite{sekatskiOptimalDistributedSensing2020}.

\subsection{Methods for realizing arbitrary sensitivities in the sensor state}
\label{appendix:tunesensitivity}

In this section, we present two detailed methods for tuning the sensitivity of a two-level sensor to a field, where the second method was briefly presented in \cite{sekatskiOptimalDistributedSensing2020}. We consider a sensor-field interaction which causes shifts in energies of the two-level sensors. 
The first method achieves an arbitrary reduction in the sensitivity of each sensor by implementing the Hamiltonian $H = H_{\mathrm{sensor}} + H_{\mathrm{field}}$, where $H_{\mathrm{sensor}} = \frac{\hbar\Delta}{2}\sigma_z + \frac{\hbar\Omega}{2}\sigma_x$, $H_{\mathrm{field}} = \hbar\delta\sigma_z$, and $\hbar\delta$ is the energy shift caused by the field to be sensed. 
$H_{\mathrm{sensor}}$ could for example be implemented by a qubit in our system encoded in superpositions spanning the $4^2\textrm{S}_{1/2}$ and $3^2\textrm{D}_{5/2}$ manifolds. 
The application of a \SI{729}{\nano\meter} laser with a Rabi frequency of $\Omega$ and a frequency of $\omega_L$ implements $H$, where $\Delta = \omega_0 - \omega_L$ and $\omega_0$ is the frequency difference between the ion-qubit states in the absence of the laser.
This method is valid when the energy shift caused by the field to be sensed is perturbatively small as given by the expression $(\delta\Delta/(\Omega^2 + \Delta^2))^2 << 1$.
The dressed eigenstates of $H_{\mathrm{sensor}}$ are given by $\ket{+_d} = \sin{\theta}\ket{+1} + \cos{\theta}\ket{-1}$ and $\ket{-_d} = \cos{\theta}\ket{+1} - \sin{\theta}\ket{-1}$, where $\tan{2\theta} = -\Omega/\Delta$ and $\ket{\pm1}$ are the eigenstates of $\sigma_z$. The corresponding eigenenergies are $E_{\pm_d} = \mp\hbar\sqrt{\Omega^2 + \Delta^2}/2$.
To first order in $\delta$, the total Hamiltonian $H$ has the same eigenstates $\ket{\pm_d}$, but different corresponding energies of $E'_{\pm_d} = E_{\pm_d}\mp\hbar\delta\Delta/(2\sqrt{\Delta^2 + \Omega^2})$.
The resulting s values are $s = \mp\Delta/(2\sqrt{\Delta^2 + \Omega^2})$.
This scheme can be separately applied to each sensor-qubit to tune their respective s values, as required by the SWD protocol. 
 
We now present the second method, which is briefly discussed in~\cite{sekatskiOptimalDistributedSensing2020}.
For the case in which the fields are constant over the  total sensing time (but not necessarily from shot to shot), the sensitivity of the $j$-th sensor-qubit can be reduced by factor $F^j_{\mathrm{red}}$ by applying a spin echo to it after time $F^j_{\mathrm{red}}t_{\mathrm{echo}}/2$, where $t_{\mathrm{echo}}$ is set to the total sensing time ($t_s$), and a final spin echo after time $t_{\mathrm{echo}}$. For the case in which the fields fluctuate over the total sensing time, the solution is to perform spin echoes often enough such that a quasistatic situation is realised. Specifically, the above echo sequence should be concatenated $k$ times with $t_{\mathrm{echo}} = t_s/k$, where $k$ is large enough such that the field can be treated as static over the time $t_s/k$.
Here, we assume that the duration of the spin echo pulses is negligible compared to $t_s/k$.


%